\documentclass[mnsc,nonblindrev]{informs1}
\OneAndAHalfSpacedXI %
\usepackage{rotating}
\usepackage{bbding}
\usepackage[left=1in,top=1in,right=1in,bottom=1in]{geometry}
\usepackage{amssymb,amsmath,amsxtra,amstext,xfrac,mathtools}
\usepackage{graphicx,booktabs,enumerate,paralist,mdwlist}
\usepackage{dsfont,verbatim,latexsym}
\usepackage{mathrsfs}
\usepackage{bbm}
\usepackage{multirow}
\usepackage[table]{xcolor}
\newcolumntype{C}[1]{>{\centering\let\newline\\\arraybackslash\hspace{0pt}}m{#1}}
\usepackage{longtable,multirow,threeparttable}
\usepackage{float}
\usepackage{tikz,qtree}
\usetikzlibrary{arrows,positioning,decorations.pathreplacing}
\usepackage{mwe}
\usepackage{makecell}
\usepackage{eurosym}
\usepackage[english]{babel}
\usepackage[utf8]{inputenc}
\usepackage{fancyhdr}
\usepackage{color}
\usepackage{array}
\usepackage{subcaption}
\usepackage{geometry}
\usepackage{natbib}
\usepackage{url}
\newlength{\regionplotwidth}
\usepackage[colorlinks=TRUE,citecolor=MyDarkBlue,urlcolor=MyDarkBlue,linkcolor=MyDarkBlue]{hyperref}
\definecolor{MyDarkBlue}{RGB}{158,0,0}

\usepackage{natbib}
\bibpunct[, ]{(}{)}{,}{a}{}{,}%
\def\bibfont{\small}%
\TheoremsNumberedThrough
\EquationsNumberedThrough

\makeatletter
\long\def\FIGURE#1#2#3{%
  \long\def\testfignote{#3}%
  \long\def\itis@empty{}%
  \ifx\testfignote\itis@empty
    \long\def\fignote##1{\relax}%
  \else
    \long\def\fignote##1{\par\noindent
      \FigureNoteStyle\HD{11}{0}{\it\FigureNoteName}\enskip ##1\endgraf}%
  \fi
  \setbox4=\hbox to \hsize{\vbox{%
    \begin{minipage}{\hsize}
      \hbox to\hsize{\hss{#1}\hss}%
      \vspace{0.5em}%
      \caption[]{#2}%
      \fignote{#3}%
    \end{minipage}}}%
  \box4}
\makeatother
\hypersetup{pdftitle={Artificial Shoppers and Attention Competition},
  pdfauthor={Xiaomeng Guo, Fasheng Xu, Xiaoli Zhang}}
\begin{document}

\RUNAUTHOR{Guo, Xu, and Zhang}
\RUNTITLE{Artificial Shoppers and Attention Competition}
\TITLE{Artificial Shoppers and Attention Competition}
\ARTICLEAUTHORS{%
	\AUTHOR{Xiaomeng Guo$^1$ ~~~ \href{https://www.fashengxu.com/}{\textcolor{black}{Fasheng Xu}}$^2$ ~~~ Xiaoli Zhang$^1$ \\[-2mm]}
	\AFF{$^1$Department of Logistics and Maritime Studies, Hong Kong Polytechnic University}
 \AFF{$^2$School of Business, University of Connecticut}
}
\ABSTRACT{%
We study how AI shopping interfaces affect price competition, platform profit, and consumer welfare in a market with three products of different quality.
A platform hosts two products, while the third is sold off platform.
Without AI, consumers compare the hosted pair under salience distortion
and overweight whichever attribute, price or quality, stands out.
We distinguish two AI modes.
Under \textit{assistance}, an \textit{AI assistant} recommends the two products with the highest true utility, leaving consumers to make the same distorted comparison.
Under \textit{delegation}, an \textit{AI agent} selects the best product directly%
; consumers thus bypass comparison but retain purchase approval.
Both modes access all three products and know consumers' preferences; firms set prices anticipating these choice rules.
First, assistance can soften competition despite expanded product access.
The middle-quality firm may forgo low-valuation consumers because attracting them requires cutting the price paid by higher-valuation consumers as well.
Relative to no AI with the same assortment, assistance lowers both hosted prices under the high- and low-quality pair but can raise both under the high- and middle-quality pair.
Second, we identify an \textit{alignment paradox}: some consumers can become worse off than without AI even though their best option remains recommended and becomes cheaper.
Salience can induce a switch to a lower-quality product whose quality loss exceeds the saving relative to the original purchase.
The assistant ranks products accurately but does not optimize the shortlist for consumers' realized welfare.
Third, assistance can raise both platform profit and aggregate consumer surplus with the high- and middle-quality pair; delegation can do so with the high- and low-quality pair.
Fourth, among AI designs, the platform weakly prefers delegation with the high- and middle-quality pair wherever all compared pricing equilibria exist.
Yet this preferred design can yield lower platform profit and aggregate consumer surplus than the no-AI optimum.
}%

\KEYWORDS{AI-mediated shopping, AI agent, salience distortion, price competition, platform assortment }
\HISTORY{Current version: November, 2026}
\maketitle
\vspace{-8mm}
\section{Introduction}
AI shopping systems are changing which products consumers can access and how they choose among them.
For retail platforms, this creates a joint design problem: whether AI should help consumers compare products or select on their behalf, and which products to host when AI can reach outside sellers.
Google describes this shift as an ``agentic commerce era'' and has announced the Universal Commerce Protocol (UCP) to standardize interactions among AI surfaces, merchants, and payment providers \citep{googleCloudAgenticCommerce2026, googleDevelopersBlogUCP2026}.
Adobe Analytics reports a 693.4\% year-over-year increase in traffic from generative AI tools to U.S. retail websites during the 2025 holiday season, albeit from a modest base \citep{adobeHolidayShopping2026}.

An \textit{AI assistant} curates options while leaving the final product choice to the consumer; we call this mode \textit{assistance}.
Amazon introduced Rufus to answer product questions, summarize information, and recommend alternatives \citep{amazonRufus2024}; ChatGPT's shopping research provides buying guides tailored to consumers' needs and preferences \citep{openai2025shoppingresearch}.
An \textit{AI agent} instead selects a product on the consumer's behalf, subject to the consumer's purchase approval; we call this mode \textit{delegation}.
Amazon's Alexa for Shopping illustrates the movement toward delegated shopping by combining personalized recommendations and product comparisons with automated cart building and routine purchases \citep{amazonAlexaShopping2026}.
Meta's Muse can research products and make purchases with consumers' approval \citep{metaMuseShopping2026}.
A service can support both functions; our distinction concerns who selects the product.

\begin{figure}[h]
\FIGURE
{%
\begin{minipage}{\linewidth}
\centering
\begin{minipage}[t]{0.18\linewidth}
\centering
\includegraphics[trim=0bp 70bp 0bp 40bp,clip,width=\linewidth,height=5.55cm,keepaspectratio]{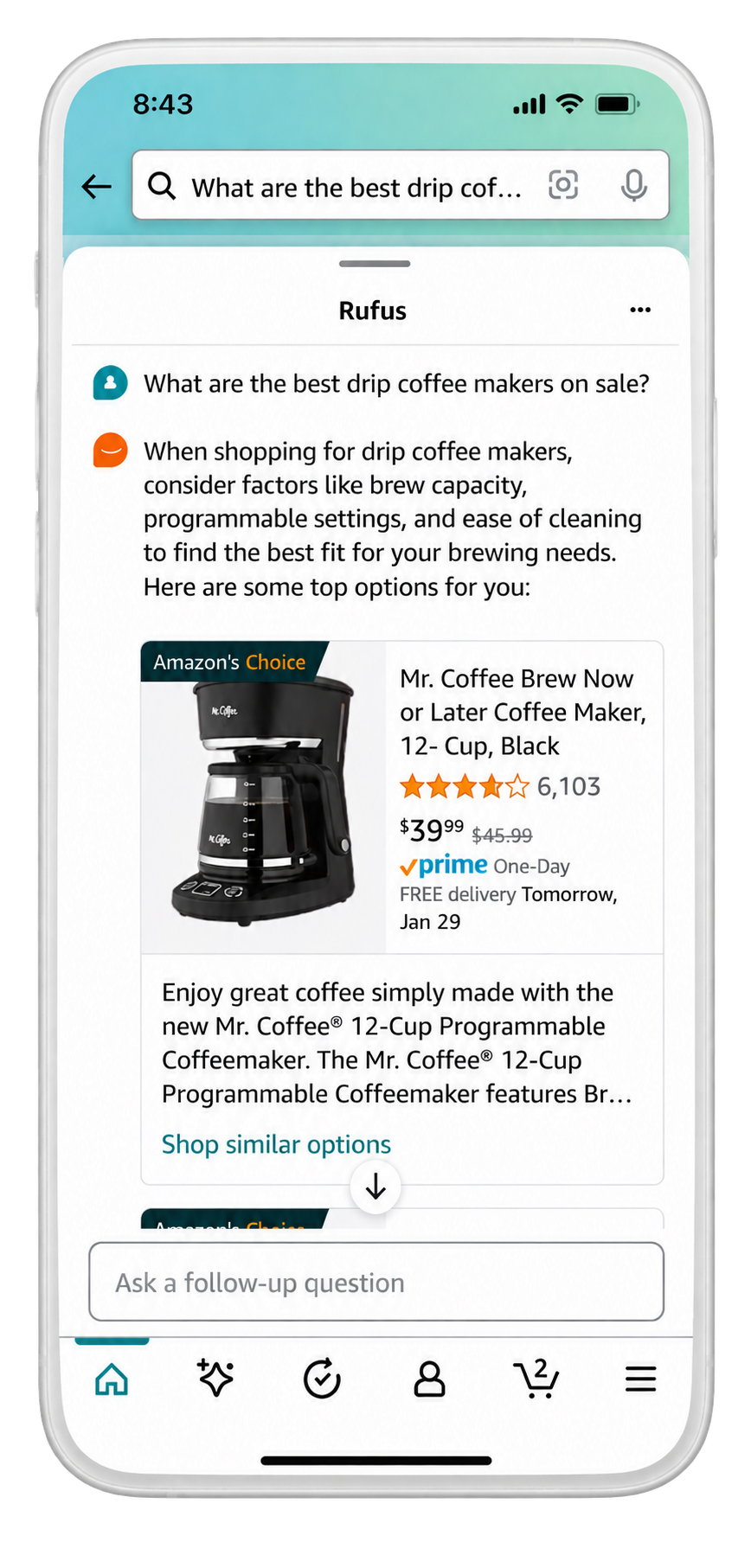}
\par\smallskip
{\scriptsize\textbf{(a) AI assistant}\\Rufus}
\end{minipage}\hfill
\begin{minipage}[t]{0.375\linewidth}
\centering
\includegraphics[trim=0bp 35bp 0bp 50bp,clip,width=\linewidth,height=5.55cm,keepaspectratio]{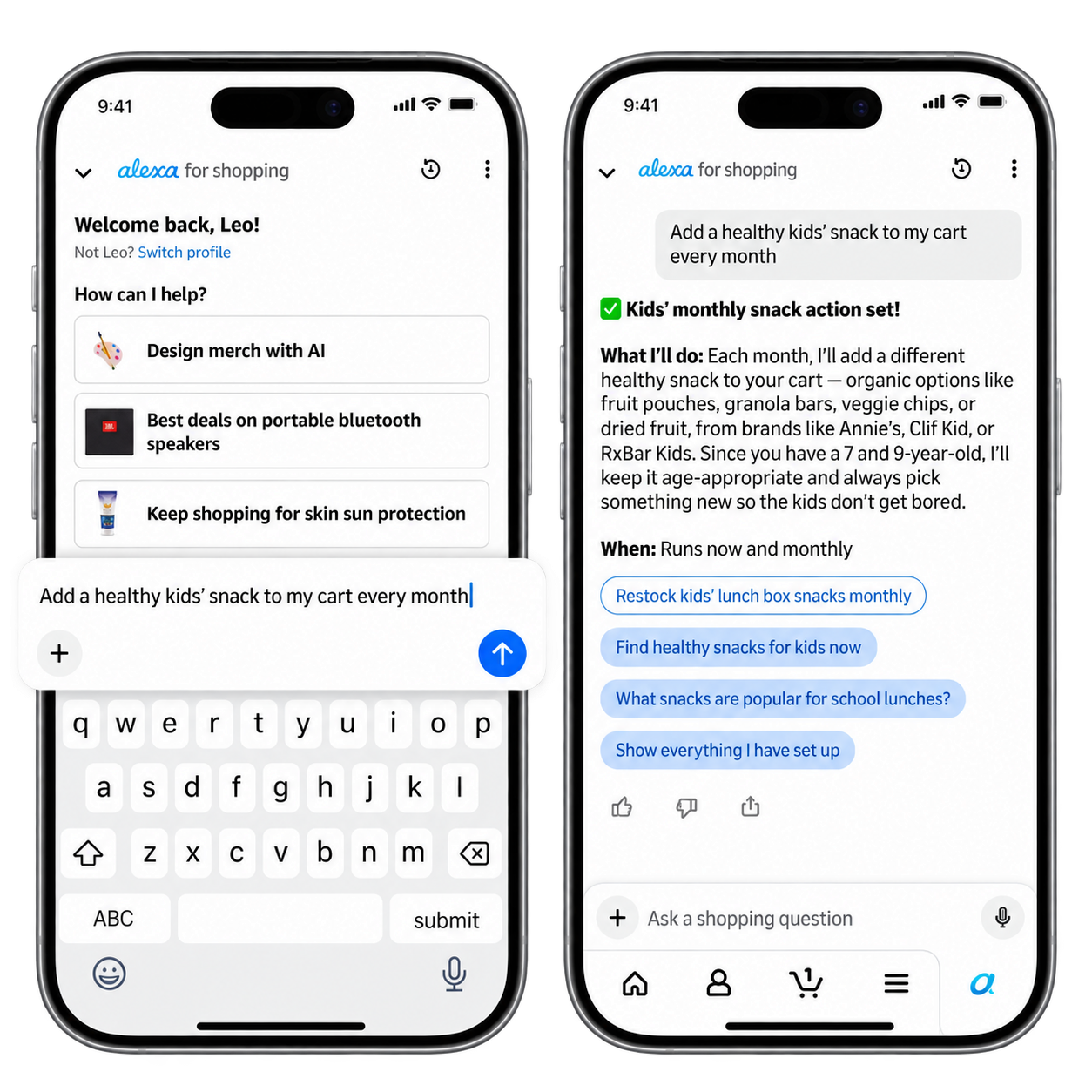}
\par\smallskip
{\scriptsize\textbf{(b) AI agent}\\Alexa for Shopping}
\end{minipage}\hfill
\begin{minipage}[t]{0.42\linewidth}
\centering
\includegraphics[trim=270bp 10bp 270bp 10bp,clip,width=\linewidth,height=5.55cm,keepaspectratio]{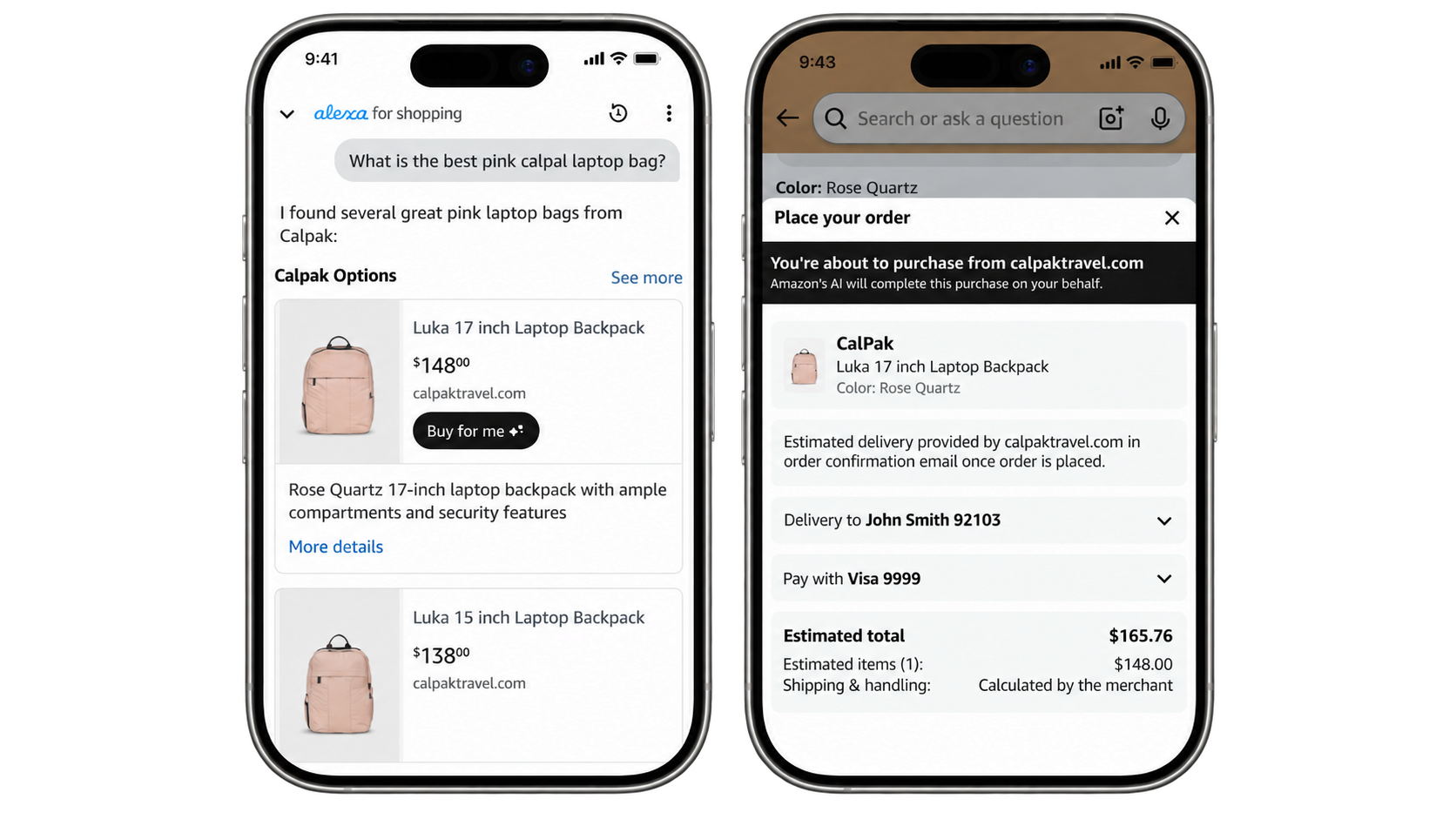}
\par\smallskip
{\scriptsize\textbf{(c) Cross-platform shopping}\\Buy for Me}
\end{minipage}
\end{minipage}
}
{Examples of AI-Mediated Shopping Interfaces\label{fig:ai-shopping-examples}}
{Sources: \citet{amazonNews2024rufusVideo} for panel (a) and \citet{amazonAlexaShoppingHowTo2026} for panels (b) and (c).}
\end{figure}

AI-mediated shopping can also span platform boundaries. Amazon's \textit{Buy for Me} helps consumers discover products from external merchants and, upon request, completes purchases on their behalf \citep{amazon2025buyforme}. Amazon has expanded this cross-platform access through Shop Direct, which now includes more than 100 million products from over 400,000 external merchants, with tens of millions eligible for Buy for Me \citep{amazonShopDirect2026}. By making off-platform products accessible to consumers, AI intermediation extends competition beyond the platform’s own assortment. Figure~\ref{fig:ai-shopping-examples} shows examples of AI assistance, delegated shopping, and cross-platform purchasing.

The consequences of AI-mediated shopping depend not only on \textit{which} products consumers discover but also on \textit{how} they evaluate them. Even when consumers know their preferences, comparing products across price, quality, and other attributes can be cognitively demanding \citep{payne1976task}, and the alternatives considered together can shape which attributes attract attention \citep{bettman1998constructive}. The Williams--Sonoma bread-maker example illustrates this contextual effect: after the retailer introduced a larger, more expensive model, sales of its existing model increased although its price and quality were unchanged \citep{lee2013williams}. More generally, adding an alternative can shift choices among existing products, as in the decoy and compromise effects \citep{tversky1993context}. The salience framework of \citet{bordalo2013} offers a unified account: attributes that stand out relative to the comparison set receive greater weight, so changing the set can change how a product is evaluated. Retailers have long shaped comparison contexts through assortment and product presentation. AI alters this process in two ways: An assistant can personalize that context by curating a shortlist; an agent can remove the consumer comparison stage by selecting directly. AI thus affects both what consumers consider and how---or whether---they compare alternatives.

Retailers already modify website content to improve their visibility in AI recommendations \citep{reutersAIShoppingTraffic2026}.
Pricing introduces a further strategic response: when AI uses posted prices, a seller's price affects both its inclusion in a shortlist and the purchases that shortlist generates.
The seller must therefore account for the alternatives against which consumers compare its product.
Platforms face a related trade-off: access to outside sellers can improve consumers' options while diverting purchases away from hosted products that generate commission revenue.
The platform needs to evaluate its AI design together with its product assortment, anticipating how firms price and consumers choose. We therefore ask two questions:

\textit{How do assistance and delegation change price competition and the division of surplus among consumers, firms, and the platform?}
\textit{Which combinations of AI mode and assortment can benefit both the platform and consumers, and which does the platform prefer?}

We study these questions in a market with three vertically differentiated products and consumers who differ in willingness to pay for quality.
The platform chooses two products to host and earns commissions on hosted sales; the third  is sold off platform. When AI is available, it also chooses between assistance and delegation.
Firms then simultaneously set a single price for each product, anticipating the applicable intermediation rule and consumer choices.
Without AI, consumers compare only the hosted pair.
Both AI modes access all three products and observe consumer types.
The assistant ranks products by true utility and recommends the top two for the consumer to compare; the agent selects the true-utility-maximizing product, which the consumer buys if its utility is nonnegative.
Consumer comparisons under no AI and assistance are subject to salience distortion \citep{bordalo2013}.
We first compare each AI mode with no AI for a fixed assortment, then let the platform choose its preferred design with and without AI.

Under assistance, the central mechanism links personalized comparison sets to a seller's single price across consumers.
The middle-quality product can appear alongside a higher-quality rival for some consumers and a lower-quality rival for others.
A price cut to attract the latter group also reduces revenue from each sale to the former.
The seller may therefore forgo low-valuation demand to preserve revenue from high-valuation consumers, so greater visibility need not translate into more aggressive competition.
Under delegation, the agent instead selects directly. Because both modes can access the same products and observe each consumer's preferences, comparing them shows how leaving the choice to consumers changes prices and welfare.

For the fixed-assortment comparisons, we focus on a \textit{broad-spectrum assortment} hosting the high- and low-quality products, $A$ and $C$, and a \textit{specialized-premium assortment} hosting the high- and middle-quality products, $A$ and $B$.
The former has the widest on-platform quality gap; the latter has a narrower gap.
Three results and a joint-design implication follow.

First, whether AI intensifies or softens competition relative to no AI depends on the platform's assortment.
On a broad-spectrum platform, access to the middle-quality product, $B$, introduces an intermediate substitute and reduces both hosted prices under the assistant.
On a specialized-premium platform, the assistant can instead raise both prices when firm $B$ specializes in high-valuation demand.
Delegation can also raise hosted prices in this assortment.
Without AI, consumers underweight quality differences, restraining prices.
An agent values those differences fully, which can soften competition when the newly accessible low-quality product, $C$, is a weak substitute.
Within either AI mode, all products are accessible and proportional commissions only rescale firms' profits, leaving prices independent of which pair is hosted.
The assortment changes the no-AI benchmark against which these price effects are measured.

Second, we identify an \textit{alignment paradox}: on a broad-spectrum platform, an assistant that ranks products according to consumers' true utility can leave some consumers worse off even as both hosted prices fall.
These consumers previously chose \(A\) from \(\{A,C\}\), but switch to \(B\) when the assistant recommends \(\{A,B\}\).
\(A\) remains available and becomes cheaper, but its price premium relative to its quality advantage is larger in the new pair.
Price salience can therefore induce a switch whose quality loss exceeds the price saving.
Here, alignment concerns the ranking of individual products; the assistant does not optimize shortlist composition for the consumer's eventual salience-based choice.
These individual losses coexist with higher aggregate consumer surplus.

Third, either AI mode can raise both platform profit and aggregate consumer surplus, but under different assortments.
With the specialized-premium assortment, niche specialization under the assistant can raise platform profit. In some of these cases, consumer gains from moving to a higher-quality product and from purchases of $C$ by previously unserved consumers outweigh  losses from higher hosted prices and downward switching.
With the broad-spectrum assortment, joint gains under the agent require a sufficiently wide quality range and a middle-quality product that benefits consumers without eroding too much of the premium product's quality advantage.
In the joint-gain region, undistorted quality valuation allows the premium firm's additional profit to outweigh the low-quality firm's loss.
The other two mode--assortment pairings cannot deliver a strict win--win relative to their respective no-AI benchmarks.

Finally, we allow the platform to choose its assortment and, when AI is available, its AI mode.
Without AI, it prefers the broad-spectrum assortment because the wider quality gap softens competition between the hosted products.
With AI, it weakly prefers the specialized-premium assortment with an agent wherever all compared pricing equilibria exist.
Hosting $A$ and $B$ then captures the two larger profit streams without changing competition within that mode.
Delegation yields weakly higher  platform profit than assistance in the characterized equilibria.
The no-AI optimum hosts $A$ and $C$, so moving to this AI design changes both how products are selected and which sellers' sales generate commissions.
Relative to that optimum, the preferred AI design can benefit both the platform and consumers, hurt both, or benefit either party at the other's expense.

For \textit{platform operators}, the central task is to coordinate how AI selects products with which sales the platform monetizes.
Once AI makes all products accessible, hosting determines the commission base, whereas the AI mode shapes consumer choice and price competition.
A design can perform best among AI alternatives yet earn less than the no-AI optimum.
Platform evaluations therefore need to account for firms' price responses and purchases diverted to outside sellers.

For \textit{retailers and brands}, the value of recommendation exposure depends on the rivals displayed alongside a product and the profitability of the consumers it reaches.
When a common price serves several recommendation groups, attracting additional low-valuation consumers requires extending the price cut to higher-valuation consumers as well.
A seller may therefore prefer to maintain a premium position even as AI exposes its product to a broader audience.
For firms on a specialized-premium platform, delegation can also support higher prices when the newly accessible low-quality alternative is a sufficiently weak substitute.
For a fixed assortment, the off-platform seller moves from zero sales to positive profit under either AI mode in the characterized equilibria.

For \textit{policymakers and regulators}, evaluating an interface requires distinguishing accurate product rankings from beneficial purchase decisions.
An assistant can include the best option and rank it correctly, yet induce a harmful switch through the alternative displayed beside it.
Evaluations should therefore consider both which products consumers choose and whether the resulting welfare gains survive firms' price adjustments.
Delegation removes the  comparison distortion, but its equilibrium welfare effects still depend on those adjustments.

\section{Related Literature}
\label{sec:literature}

Our paper relates to three streams of research: AI intermediation in markets, salience and context-dependent choice, and consideration-set formation and recommendation design.

\subsection{AI Intermediation in Markets}
A growing literature studies the role of artificial intelligence (AI) in market interactions.
Some studies use large language models (LLMs) to approximate consumer responses and preferences in market-research settings \citep{brand2023using, li2024frontiers, goli2024frontiers}.
Other work examines AI as a firm-side technology.
AI assistance can improve customer sentiment in interactions with human service agents \citep{zhang2025engaging}.
AI-generated personalized advertising and sponsored-search content can improve engagement and conversion \citep{kapoor2025frontiers, reisenbichler2025applying}.

Work on consumer-side intermediation ranges from search assistance to autonomous agents.
Voice-based shopping assistants can reduce information acquisition costs and increase consumer spending \citep{sun2025effect}.
AI-enhanced search can increase purchases by translating consumers' underlying needs into effective queries \citep{zheng2025generative}, while query recommenders can expand consideration sets and consumption diversity \citep{zheng2025frontiers}.
Research also examines  autonomous buyer agents' responses to product positioning \citep{allouah2025your} and AI agents' use in seller-side bargaining \citep{kong2025fishbargain} and supply chain bargaining \citep{liang2026negotiate}.

Recent theoretical work examines how AI acquires preference information before recommending products.
\citet{kumar2026much} show how platform objectives shape preference elicitation: conversion can favor little elicitation, whereas commission revenue can favor more elicitation  because it raises equilibrium prices.
These price responses can offset consumers' gains from improved matching.
\citet{cao2026solicit} characterize how deeper preference elicitation can substitute for larger recommendation sets in improving product matches.
\citet{dong2026preference} study how consumers' costly preference refinement shapes agentic search, voluntary adoption, and platforms' fidelity choices.

In our setting, AI already observes each consumer's willingness to pay for quality.
We distinguish two AI modes: an assistant recommends the two products with the highest true utility for the consumer to compare, while an agent selects the best product directly.
We study how retaining or bypassing salience-based comparison affects consumer choices and firms' pricing incentives.

\subsection{Salience and Context-Dependent Choice}
Standard rational-choice models assume that a product's evaluation is independent of surrounding alternatives. Yet decoy, compromise, and related context effects show that changes in the choice set can shift demand even when products' own attributes remain unchanged \citep{tversky1993context}. Salience theory attributes such effects to consumers overweighting attributes that stand out relative to the choice-set average \citep{bordalo2013}. Related models imply opposite range effects: larger cross-option differences attract more attention under focusing \citep{kHoszegi2013model}, whereas a given difference receives less weight when the surrounding range is larger under relative thinking \citep{bushong2021model}.

Extending salience theory to competitive markets, \citet{bordalo2016} show how firms compete for attention through price--quality positioning. Subsequent work studies attention-shaping strategies such as loss leaders and decoys \citep{apffelstaedt2021competition}, selective emphasis on attributes to limit attention dilution \citep{zhu2017prominent}, and probabilistic selling that shifts attention from price to quality \citep{zheng2019probabilistic}. Salience effects have also been documented in crowdfunding \citep{weinmann2023attraction} and portfolio choice \citep{dertwinkel2025salient}.

We build on the insight that firms' pricing incentives depend on which attributes receive attention.
Under assistance, an intermediary  assigns consumers to personalized two-product comparisons using preferences and posted prices.
A seller can therefore appear in different comparisons while charging a single price.
This links its incentives across recommendation groups and can induce it to forgo low-valuation demand to protect its margin in a higher-valuation group.
We compare these outcomes with those under delegation, where the agent bypasses salience-based comparison.

\subsection{Consideration-Set Formation and Recommendation Design}
Consumer choice is often modeled as a two-stage process in which consumers first form a consideration set and then choose among the products considered \citep{howard1969theory, payne1976task}. \citet{hauser1990evaluation} and \citet{roberts1991development} formalize this process as a trade-off between the benefit of evaluating an additional brand and its evaluation cost. Whereas these models treat consideration-set formation as consumer-driven, digital platforms can actively shape which products consumers consider through search design, ranking, filtering, sponsored placement, and self-preferencing \citep{dukes2016online, dinerstein2018consumer, zhong2023platform, jiang2020consumer, chu2020position, long2024platform, long2024self}.

Recommendations also affect whether consumers begin searching and what suppliers choose to offer.
In a field experiment, \citet{long2025choice} find that larger recommendation sets can reduce purchases, partly by discouraging consumers from starting a search.
\citet{zou2025quality} model how recommendation rules shape creators' entry and quality choices and the resulting variety of content.

The closest analytical studies already connect recommendation design to strategic pricing.
\citet{li2018recommender} show how recommendation objectives and precision alter manufacturers' price competition and platform profit.
\citet{zhou2024precision} show that greater profiling accuracy can first lower and then raise prices as sellers compete for recommendations; their main competitive mechanisms also persist when recommendations are based on consumer surplus.

We build on this strategic interaction by allowing recommendations to determine which products consumers compare on price and quality.
The assistant includes the consumer's highest-utility product, yet the comparison with the second recommended product can induce the consumer to choose the lower-utility option.
Both the no-AI comparison and the assistant's shortlist contain two products, allowing us to examine the role of comparison-set composition at a fixed size.
Cross-platform access shapes how these outcomes compare with the platform's restricted no-AI assortment, while delegation  bypasses the consumer comparison stage, clarifying its role.

\section{Model Setup}\label{sec:model}

We consider a market with a retail platform, three firms selling vertically differentiated products, and a unit mass of heterogeneous consumers. The platform chooses two products to host, while the third remains available off platform. AI availability is exogenous. In the no-AI scenario, consumers compare only the hosted products. In the AI scenario, the platform can choose between two AI modes: assistance through an AI assistant and delegation through an AI agent. Both modes access all three products and observe each consumer's willingness to pay for quality. The assistant recommends a personalized two-product shortlist for the consumer to compare and choose from, whereas the agent directly selects the product that maximizes the consumer's true utility. When consumers compare products themselves, their choices are subject to salience distortion \citep{bordalo2013}. Firms set prices anticipating the applicable intermediation rule and consumer behavior. Superscripts \(m\in\{n,s,g\}\) denote the no-AI case, assistance, and delegation, respectively. The main notation is summarized in Table~\ref{tab:notation} of Appendix~\ref{sec: technical details}.

\subsection{Firms and Platform}

Three firms each sell one vertically differentiated product \(i\in\{A,B,C\}\), with qualities \(q_A>q_B>q_C>0\). Each firm sets a price \(p_i\in[0,p^{\max}]\) to maximize profit. Marginal costs are normalized to zero.\footnote{Allowing quality-dependent marginal costs \(c_A>c_B>c_C>0\) would alter equilibrium prices and thresholds and may change the comparison regions. We leave that extension for future work.}
The platform chooses a two-product assortment \(S_P\subset\{A,B,C\}\) for its shopping interface. The remaining product \(k\notin S_P\) is available off platform but not easily discovered or purchased through the platform's conventional interface. The full product set is \(\mathcal I\equiv S_P\cup\{k\}\).

We interpret hosting as inclusion in the platform's conventional shopping interface and commission arrangement.
Sales of hosted products generate platform revenue, whereas purchases from the off-platform seller do not.
The platform chooses among available hosting arrangements, taking seller participation as given; entry, commission-setting, and the negotiation of hosting terms are outside the model.

We focus on two assortments with different on-platform quality gaps.\footnote{The platform-design comparison in Section~\ref{sec:platform design} also includes the third assortment \(S_P=\{B,C\}\).} Under \(S_P=\{A,C\}\), the platform hosts the high- and low-quality products, creating the maximal on-platform quality gap; we call this the \textit{broad-spectrum assortment}. Under \(S_P=\{A,B\}\), it hosts the high- and middle-quality products, creating a narrower quality gap; we call this the \textit{specialized-premium assortment}.

\enlargethispage{\baselineskip}
\subsection{Consumers}

There is a unit mass of consumers whose types are distributed as \(\theta\sim U[0,\Theta]\), where \(\Theta>0\) is the upper bound of the type distribution and a higher \(\theta\) indicates a greater willingness to pay for quality. Consumers know their own types, whereas the platform and firms know only the type distribution when making platform-design and pricing decisions; prices are not type contingent. A type-\(\theta\) consumer's true utility from purchasing product \(i\) is $u_i(\theta)=\theta q_i-p_i$, while the utility from not purchasing is normalized to zero. We call \(u_i(\theta)\) ``true utility'' to distinguish it from the salience-distorted perceived utility introduced in Section~\ref{sec:salience_choice}.

\subsection{Salience-Based Choice}\label{sec:salience_choice}

In the no-AI scenario and under assistance, consumers compare two products and make the product choice themselves. Let \(T=\{i,j\}\) denote their comparison set; its composition is specified below. Although consumers know their willingness to pay for quality, comparing products can be cognitively demanding \citep{payne1976task}, and their evaluations may depend on the alternatives considered together \citep{tversky1993context}. We capture this context dependence using the salience framework of \citet{bordalo2013}, which offers a unified account of many documented context effects, including the decoy and compromise effects, and has been applied across diverse settings and received experimental support \citep{zheng2019probabilistic,dertwinkelkalt2017demand}.

Following \citet{bordalo2012salience,bordalo2016}, we define the reference values for quality and price as the comparison-set averages
\[
\bar q(T)=\frac{q_i+q_j}{2},
\qquad
\bar p(T)=\frac{p_i+p_j}{2},
\]
and measure how strongly an attribute value \(a\) stands out relative to its reference \(\bar a\) by
\[
\sigma(a,\bar a)=\frac{|a-\bar a|}{\bar a},
\qquad a\geq0,\quad \bar a>0,
\]
with \(\sigma(0,0)=0\). The function captures ordering and diminishing sensitivity: moving either \(a\) or \(\bar a\) away from the other weakly increases salience, whereas moving both upward by the same amount weakly decreases it.

For each product \(i\in T\), we compare quality salience, \(\sigma(q_i,\bar q(T))\), with price salience, \(\sigma(p_i,\bar p(T))\).
The larger value identifies the salient attribute.
With the pair averages above, both products have the same quality salience, \(|q_i-q_j|/(q_i+q_j)\), and the same price salience, \(|p_i-p_j|/(p_i+p_j)\); price salience is zero when both prices are zero.
Thus, the attribute on which the two products differ proportionally more is salient, and both products share the same salient attribute.
For \(q_i>q_j>0\) and \(p_i>p_j>0\), this comparison is equivalent to comparing quality-to-price ratios:
\[
\sigma(q_i,\bar q(T))\gtrless\sigma(p_i,\bar p(T))
\quad\text{if and only if}\quad
\frac{q_i}{p_i}\gtrless\frac{q_j}{p_j}.
\]
On this ordered-price domain, quality is salient when \(q_i/p_i>q_j/p_j\), price is salient when the inequality is reversed, and neither attribute is overweighted at equality.
We use this equivalence to characterize equilibrium. As shown in Appendix~\ref{sec:proof}, when the higher-quality product costs no more, every consumer weakly prefers it to the lower-quality product, regardless of which attribute is salient.
Hereafter, we abbreviate ``if and only if'' as ``iff.''
Our analysis uses this specification; salience functions that do not preserve the stated equivalence are outside its scope.

We model salience distortion with \(\delta\in[0,1]\): the salient dimension retains unit weight, while the non-salient dimension receives weight \(\delta\). At \(\delta=1\), perceived utility coincides with true utility; lower \(\delta\) implies stronger salience distortion. Thus, product \(i\)'s perceived utility in \(T\) is
\[
\tilde u_i(\theta\mid T)=
\begin{cases}
\theta q_i-\delta p_i, & \text{if }\sigma(q_i,\bar q(T))>\sigma(p_i,\bar p(T)),\\
\delta\theta q_i-p_i, & \text{if }\sigma(q_i,\bar q(T))<\sigma(p_i,\bar p(T)),\\
\theta q_i-p_i, & \text{if }\sigma(q_i,\bar q(T))=\sigma(p_i,\bar p(T)).
\end{cases}
\]
The consumer purchases the product in \(T\) with the highest perceived utility if that utility is nonnegative; otherwise, the consumer does not purchase.

We focus on moderate salience distortion, $\delta\geq \frac{1}{2}$, so the non-salient dimension retains at least half the weight of the salient dimension. This range is consistent with the benchmark parameterization in \citet{bordalo2012salience} and subsequent experimental estimates in \citet{konigsheim2019salience}.  

\subsection{Shopping Scenarios and AI Modes}
\label{sec:ai_intermediation}

We consider two shopping scenarios: no AI and AI. Within the AI scenario, we distinguish two modes by the degree of decision authority delegated to AI: assistance through an AI assistant and delegation through an AI agent. We model the AI interface as platform-provided, so the platform can choose between these modes when AI is available. Our subsequent analysis for a given assortment and AI mode also applies if the interface is instead provided by a third party.

\noindent\textbf{No-AI scenario.} Cross-platform search frictions make it difficult for consumers to discover off-platform products. We capture this restriction with the following assumption.
\begin{assumption}[Baseline consideration]\label{ass:baseline_consider}
In the no-AI scenario, consumers consider only the hosted products, so their comparison set is \(T^n=S_P\).
\end{assumption}

Assumption~\ref{ass:baseline_consider} captures the platform interface's role in determining which products consumers consider, consistent with evidence that consideration sets are small and shaped by the search and display environment \citep{hauser2014consideration,dinerstein2018consumer}.
Consumers apply the salience-based decision rule in Section~\ref{sec:salience_choice} to the hosted pair and may choose either product or not purchase.

\noindent\textbf{AI scenario.}
AI reshapes product discovery and consumer choice. The AI interface can access the full product set \(\mathcal I\), including the off-platform product \(k\). Amazon's Buy for Me, for example, surfaces products from merchants outside its marketplace and allows consumers to purchase them \citep{amazon2025buyforme}. This cross-platform access exposes the off-platform product to new consumers and makes it a strategic competitor to on-platform firms.

Beyond product access, AI can use information about a consumer's preferences to personalize product recommendations.
We capture this capability by assuming that, at the intermediation stage after firms have set prices, the AI observes each consumer's type \(\theta\) and ranks products according to true utility, \(u_i(\theta)=\theta q_i-p_i\).
The ranking therefore depends on both the consumer's willingness to pay and the posted prices.
Firms anticipate this rule when pricing but do not observe individual consumer types.

Platform revenue, click-through, and conversion do not enter this ranking, and we call the AI \textit{consumer-aligned} in this ranking sense.
For the assistant, alignment concerns how individual products are ranked; it does not mean that the recommended pair maximizes the consumer's realized true utility under salience-based choice.
The assistant recommends the two highest-ranked products without adjusting the shortlist for that behavioral response, whereas the agent selects the highest-ranked product directly, subject to the consumer's purchase decision.

\paragraph{AI assistant.}
Under assistance, the AI evaluates the products in \(\mathcal I\) according to each consumer's true utility, breaking ties in favor of higher quality. Let \(i_\theta^1\) and \(i_\theta^2\) denote the first- and second-ranked products. The assistant recommends the personalized two-product shortlist
\[
T_\theta^s=\{i_\theta^1,i_\theta^2\}\subseteq\mathcal I,
\qquad |T_\theta^s|=2.
\]
Consumers compare the recommended products under the salience-based decision rule and decide whether to purchase and, if so, which product to choose.
The shortlist always contains the product that maximizes true utility over \(\mathcal I\).
Nevertheless, the alternative displayed alongside it can change the consumer's evaluation and purchase.
Because the ranking depends on posted prices, a firm's price affects not only its demand within a shortlist but also which consumers' shortlists it enters, and firms take both effects into account when pricing.

Relative to no AI, the assistant changes which products are compared while keeping the comparison-set size at two.
This keeps the number of alternatives fixed while allowing shortlist composition and equilibrium prices to change.
A two-product shortlist also captures the filtering role of AI interfaces that present small, curated selections \citep{Diaz2025AmazonAITool,Biyani2025GPTAITool}.
The effects of larger catalogs and shortlist sizes are left for future work.

\paragraph{AI agent.}
Under delegation, consumers delegate product selection to the AI while retaining the final purchase decision.\footnote{Amazon's Buy for Me asks customers to confirm their order details before AI completes the purchase \citep{amazon2025buyforme}. Our results would remain unchanged if the agent also decided whether to purchase based on the consumer's true utility.} Compared with human decision makers, AI agents can evaluate alternatives more systematically and are less constrained by limited attention and cognitive capacity \citep{balasubramanian2022substituting,montard2026where}. As under assistance, the AI evaluates all products in \(\mathcal I\) according to each consumer's true utility. Instead of recommending a shortlist, it directly selects the first-ranked product: \(i_\theta^g=i_\theta^1\). The consumer observes the selected product, and we model this approval decision as undistorted: the consumer buys it iff its true utility is nonnegative, \(u_{i_\theta^g}(\theta)\geq0\).
The agent thus removes the comparison between products, and the retained approval step introduces no salience distortion.

The two modes therefore affect choice through different channels. The assistant changes which products consumers compare while retaining salience-based choice; the agent selects a product directly, bypassing the comparison stage. Firms anticipate these choice rules when setting prices, so the two modes can have distinct effects on competition and welfare.

\subsection{Timing}\label{sec:timing_equil}

The shopping scenario (no AI or AI) is exogenously given. The game then proceeds in four stages.
\begin{enumerate}[(1)]
    \item \textbf{Platform choice.} The platform chooses its two-product assortment \(S_P\). In the AI scenario, it also chooses between assistance and delegation.

    \item \textbf{Pricing.} Observing the assortment and, in the AI scenario, the mode, firms simultaneously set prices.

    \item \textbf{Consideration and intermediation.} In the no-AI scenario, consumers' comparison set is \(T^n=S_P\). In the AI scenario, the interface observes each consumer's type \(\theta\) and, at the posted prices, recommends \(T_\theta^s\) under assistance or selects \(i_\theta^g\) under delegation.

    \item \textbf{Consumer choice.} In the no-AI scenario and under assistance, consumers compare products in \(T\) and make purchase decisions based on perceived utility. Under delegation, consumers buy the selected product iff its true utility is nonnegative.
\end{enumerate}

For each assortment, we first characterize the pure-strategy pricing equilibria under no AI, assistance, and delegation, and compare outcomes wherever the relevant equilibria exist.
We then compare platform designs on the common equilibrium-existence domain where every pricing subgame entering the comparison has a characterized pure-strategy equilibrium.
If an assistant pricing equilibrium does not exist in pure strategies, our analysis does not determine the platform's optimal AI mode at those parameter values.

\subsection{Outcomes and Welfare}\label{sec:welfare}

Given an assortment \(S_P\), let \(D_i^m\) denote product \(i\)'s equilibrium demand under \(m\in\{n,s,g\}\), so that \(p_i^mD_i^m\) is its sales revenue before commissions.
The platform collects a fixed, uniform share \(\tau\in(0,1)\) of hosted sales revenue and receives no revenue from off-platform sales.
With zero marginal costs, firm \(i\)'s profit and the platform's profit are, respectively,
\[
\pi_i^m=
\begin{cases}
(1-\tau)p_i^mD_i^m, & i\in S_P,\\
p_i^mD_i^m, & i\notin S_P,
\end{cases}
\qquad
\Pi^m=\tau\sum_{i\in S_P}p_i^mD_i^m.
\]
Thus, commissions divide hosted sales revenue between firms and the platform.
The main text uses these actual profits; Appendix~\ref{sec:proof} introduces normalized profits to simplify the proofs.

Let \(x^m(\theta)\in\mathcal I\cup\{0\}\) denote the choice of a type-\(\theta\) consumer under \(m\), where \(0\) denotes no purchase and yields zero utility. Consumer surplus is
\[
CS^m=\frac{1}{\Theta}\int_0^\Theta u_{x^m(\theta)}(\theta)\,d\theta.
\]
Following \citet{bernheim2009beyond,bordalo2013}, we evaluate consumer surplus using true utility. Salience may change how consumers weigh quality and price when choosing, but their realized welfare depends on the quality they receive and the price they pay.

Within either AI mode, all three products are accessible regardless of \(S_P\), so the hosted assortment does not affect demand at a given price vector.
With zero marginal costs, paying commissions multiplies a firm's payoff at every price by the same positive factor \(1-\tau\), leaving its best response unchanged.
Thus, within a given AI mode, equilibrium prices, demands, and consumer surplus are independent of the hosted assortment.
Hosting determines which firms pay commissions.

Without AI, consumers compare only the hosted pair, so changing \(S_P\) also changes the no-AI outcome.
Consequently, the  effects of a given AI mode on prices and consumer surplus relative to no AI vary across assortments only because the no-AI benchmark changes.
The platform's profit comparison also depends on which products' sales generate commissions.

\subsection{No-AI Benchmark}\label{sec:benchmark}

The no-AI benchmark provides the reference point for evaluating AI intermediation. By Assumption~\ref{ass:baseline_consider}, consumers compare only the hosted products, so \(T^n=S_P\). The off-platform product therefore has zero demand, and its price does not affect equilibrium outcomes; we leave it unspecified. The two hosted firms set prices simultaneously, anticipating salience-based consumer choice.

\begin{lemma}[No-AI benchmark]\label{lem:no AI_equil}
Fix any two-product assortment $S_P=\{i,j\}\subset\{A,B,C\}$ with $q_i>q_j$.
In the no-AI benchmark, the two hosted firms' pricing game has a unique pure-strategy equilibrium.
At this equilibrium, price is salient and the prices are
\begin{equation*}
    (p_i^{n},p_j^{n})
    =
    \left(
    \frac{2 \delta \Theta q_i(q_i-q_j)}{4q_i-q_j},
    \frac{\delta \Theta (q_i-q_j) q_j}{4q_i-q_j}
    \right).
\end{equation*}
The corresponding equilibrium profits and consumer surplus are characterized in Appendix~\ref{sec: technical details}.
\end{lemma}

Price salience reflects the lower-quality firm's need to attract consumers. If quality were salient, or the two attributes equally salient, the higher-quality product would be priced no higher per unit of quality than its rival, so no consumer would buy the lower-quality product; the proof in Appendix~\ref{sec:proof} shows that every such price profile admits a profitable deviation.

With price salient, stronger salience distortion (lower $\delta$) reduces the weight consumers place on quality relative to price. This narrows perceived vertical differentiation and intensifies competition, and both equilibrium prices fall in proportion to $\delta$. Because a price-salient consumer values each product at \(\delta\theta q-p\), this proportional price cut leaves unchanged both the type that is indifferent between the two products and the lowest type that buys: consumers buy the same products at lower prices. The resulting gain in consumer surplus is a one-for-one transfer from the hosted firms and the platform, with no change in the allocation.

\section{Assistance: The AI Assistant}\label{sec:assistant}

We first analyze assistance through the AI assistant.
The assistant personalizes the pair each consumer compares and can include an off-platform product, while leaving salience-based evaluation intact.
For fixed product qualities, its pricing equilibrium is the same under either platform assortment.
We characterize this equilibrium and then compare prices, profits, and consumer welfare with the no-AI benchmark for each assortment.

\subsection{Equilibrium Characterization}

We begin with the firms' pricing game. Firms set prices simultaneously, accounting for how their prices affect the assistant's recommendations and consumers' purchase decisions. To describe the equilibrium regions and the comparative statics that follow, we reparametrize product qualities as
\[
Q\equiv q_A-q_C,
\qquad
\lambda\equiv\frac{q_B-q_C}{q_A-q_C}\in(0,1).
\]
Equivalently, \(q_A=q_C+Q\) and \(q_B=q_C+\lambda Q\).

Here, \(Q\) measures the quality span between \(A\) and \(C\), while \(\lambda\) locates \(B\) within that span. A higher \(\lambda\) places \(B\) closer to \(A\) and farther from \(C\). Accordingly, under \(S_P=\{A,C\}\), \(\lambda\) captures the quality position of the off-platform product \(B\) between the platform's products; under \(S_P=\{A,B\}\), it captures both the proximity of the platform's two products and \(B\)'s quality advantage over the off-platform product \(C\). Using this notation, the following lemma summarizes equilibrium existence, the assistant's recommendation rule, and the resulting purchase outcomes.

\begin{lemma}\label{lem: assis_equil}
Under the AI assistant:
\begin{enumerate}[\upshape(i)]
\item \textbf{Existence.} The parameter space is partitioned into \(\mathcal R_1\), \(\mathcal R_{2a}\), and \(\mathcal R_{2b}\), each with a unique pure-strategy pricing equilibrium, and \(\mathcal R_{\varnothing}\), with no such equilibrium.

\item \textbf{Recommendations and salience.} In each equilibrium region, the assistant recommends \(\{B,C\}\) to consumers with \(\theta<\theta^{\mathrm{rec}}\) and \(\{A,B\}\) otherwise, where \(\theta^{\mathrm{rec}}\equiv(p_A^s-p_C^s)/Q\). Price is salient in both recommended groups.

\item \textbf{Demand for the middle product.} In \(\mathcal R_1\), firm \(B\) serves both groups. In \(\mathcal R_{2a}\) and \(\mathcal R_{2b}\), it serves only the \(\{A,B\}\) group, and its purchase threshold satisfies \(z_B<\theta^{\mathrm{rec}}\) and \(z_B=\theta^{\mathrm{rec}}\), respectively, where \(z_B\equiv p_B^s/(\delta q_B)\).
\end{enumerate}
The formal region conditions and closed-form prices, profits, and consumer surplus are provided in Appendix~\ref{sec: technical details}.
\end{lemma}

The two thresholds describe different decisions.
At \(\theta^{\mathrm{rec}}\), products \(A\) and \(C\) give the same true utility, and the assistant changes which of them it recommends alongside \(B\).
By contrast, \(z_B\) is the type at which \(B\)'s perceived utility becomes nonnegative when price is salient.
A consumer can obtain nonnegative perceived utility from \(B\) and still prefer the other recommended product.

\begin{figure}
	   \FIGURE
	   {%
       \begin{minipage}{\linewidth}
       \centering
       \begin{subfigure}[t]{\regionplotwidth}
         \centering
         \includegraphics[width=\linewidth]{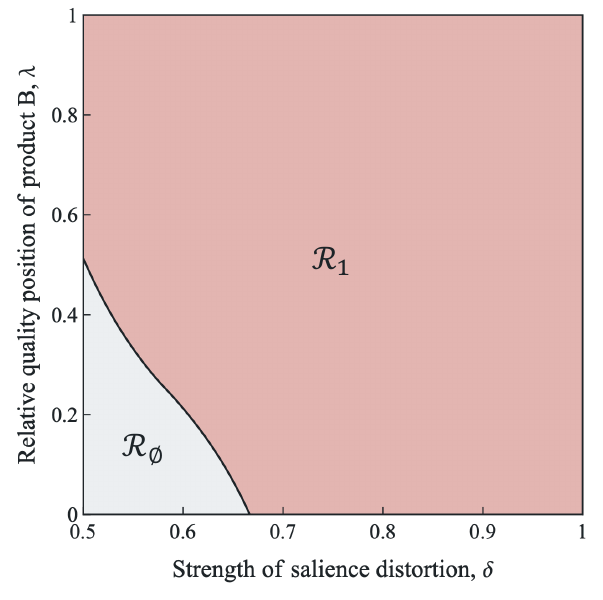}
         \caption{Low \(Q\)}
         \label{fig: assis_equil_a}
       \end{subfigure}
       \begin{subfigure}[t]{\regionplotwidth}
         \centering
         \includegraphics[width=\linewidth]{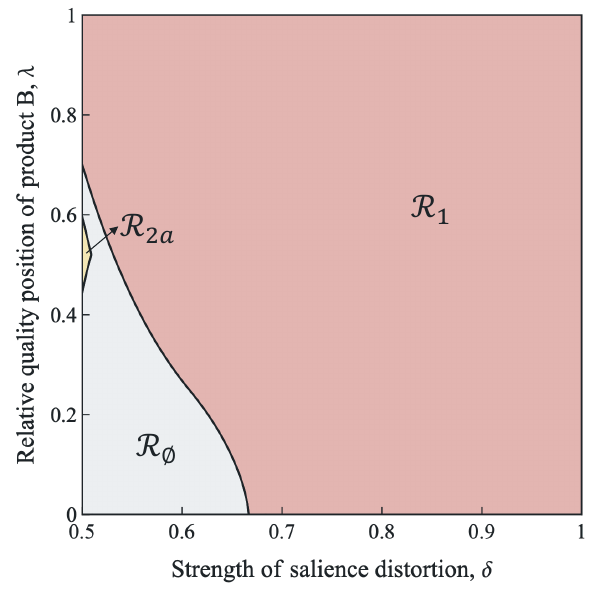}
         \caption{Medium \(Q\)}
         \label{fig: assis_equil_b}
       \end{subfigure}
       \begin{subfigure}[t]{\regionplotwidth}
         \centering
         \includegraphics[width=\linewidth]{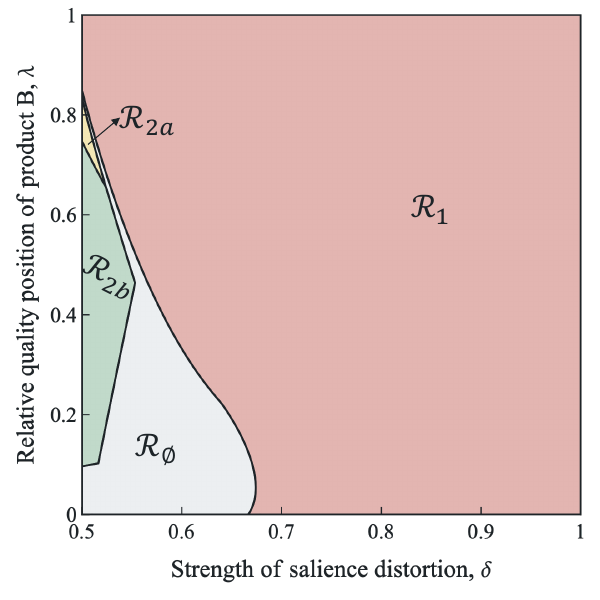}
         \caption{High \(Q\)}
         \label{fig: assis_equil_c}
       \end{subfigure}

       \begin{subfigure}{0.78\linewidth}
         \vspace*{0.7em}
         \centering
         \includegraphics[width=\linewidth]{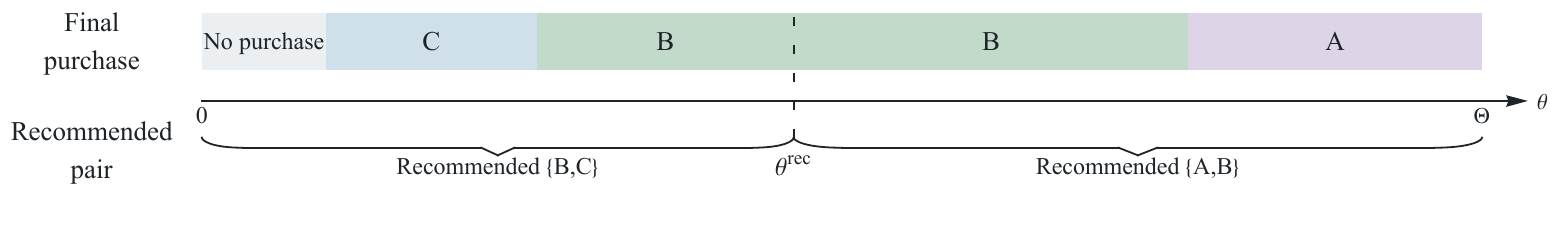}
         \caption{\(\mathcal R_1\): Market-sharing equilibrium}
         \label{fig: assis_equil_d}
       \end{subfigure}

       \begin{subfigure}{0.78\linewidth}
         \vspace*{0.7em}
         \centering
         \includegraphics[width=\linewidth]{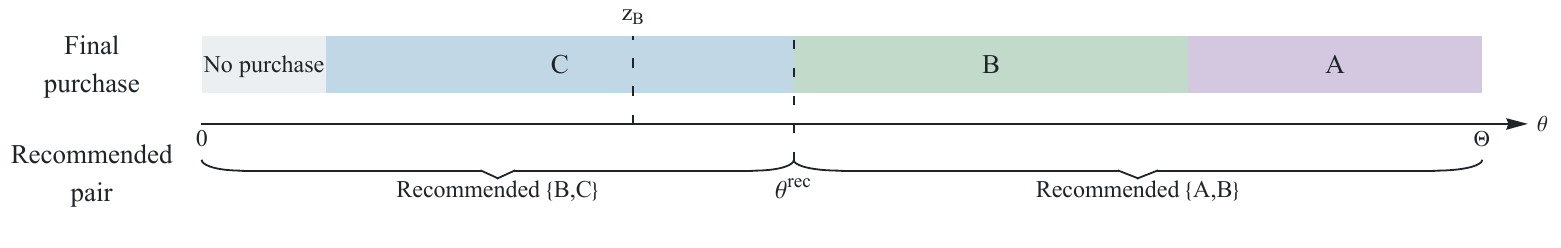}
         \caption{\(\mathcal R_{2a}\): Interior strategic-niche equilibrium}
         \label{fig: assis_equil_e}
       \end{subfigure}

       \begin{subfigure}{0.78\linewidth}
         \vspace*{0.7em}
         \centering
         \includegraphics[width=\linewidth]{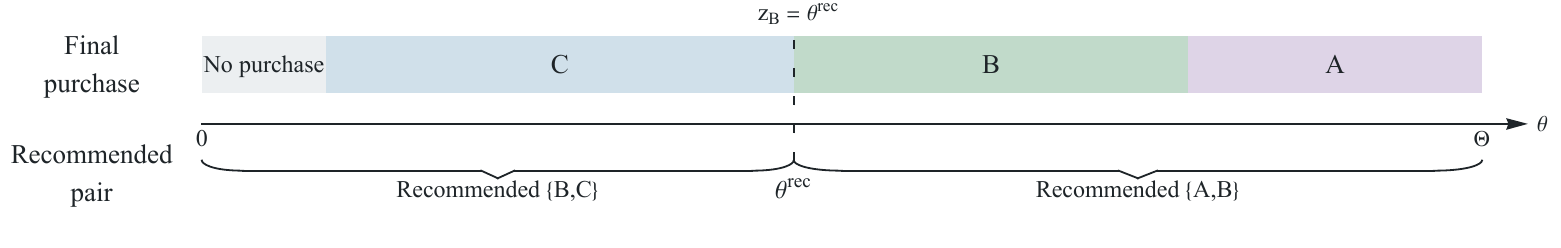}
         \caption{\(\mathcal R_{2b}\): Boundary strategic-niche equilibrium}
         \label{fig: assis_equil_f}
       \end{subfigure}
       \end{minipage}%
     }
	   {Equilibrium with AI Assistant\label{fig: assis_equil}}
	   {\scriptsize \(\Theta=2\); \(q_C=0.5\); \(Q=0.5\) in Figure~\ref{fig: assis_equil}(a), \(Q=1.3\) in Figure~\ref{fig: assis_equil}(b), and \(Q=3\) in Figure~\ref{fig: assis_equil}(c). Panels (a)--(c) plot the equilibrium regions in the \((\delta,\lambda)\) plane. Panels (d)--(f) illustrate the assistant's recommendations and purchase outcomes along consumer type \(\theta\) in \(\mathcal R_1\), \(\mathcal R_{2a}\), and \(\mathcal R_{2b}\), respectively. The dashed lines mark the two thresholds in Lemma~\ref{lem: assis_equil}, \(\theta^{\mathrm{rec}}\) and \(z_B\). In panel (d), \(B\) serves both groups. In panels (e) and (f), \(B\) cedes low-valuation demand to \(C\) and specializes in the \(\{A,B\}\) group.}
    \end{figure}
In \(\mathcal R_1\), \(B\) sells to consumers in both recommendation groups, competing with \(A\) in the premium segment and with \(C\) in the low-valuation segment.
We call this the \textit{market-sharing equilibrium}.
Because \(B\) charges one price, a cut intended to attract buyers from \(C\) also lowers its margin on premium sales and the price against which \(A\) competes.
Competition in the low-valuation group therefore carries over to the premium group.

In \(\mathcal R_{2a}\) and \(\mathcal R_{2b}\), \(B\) remains in both recommended pairs but sets a price at which buyers in the \(\{B,C\}\) group choose \(C\).
It serves only the high-valuation \(\{A,B\}\) group; we call these \textit{strategic-niche equilibria}.
Forgoing low-valuation sales lets \(B\) preserve its premium margin instead of cutting its common price to compete with \(C\).
The two niche regions differ at the lowest type receiving the premium shortlist.
In the interior region \(\mathcal R_{2a}\), this type obtains positive perceived utility from \(B\), because \(z_B<\theta^{\mathrm{rec}}\).
Consumers in the low-valuation group above \(z_B\) would also buy \(B\) rather than nothing, but still prefer \(C\).
In the boundary region \(\mathcal R_{2b}\), the lowest type receiving the premium shortlist is indifferent between buying \(B\) and not buying, because \(z_B=\theta^{\mathrm{rec}}\).

Figure~\ref{fig: assis_equil} displays the equilibrium regions for three quality spans and the corresponding recommendation and purchase patterns.
The distinction between being recommended and making a sale is central to the niche outcomes: the assistant continues to display \(B\) to low types even when \(B\) no longer serves them.

\subsection{Pricing and Profit Implications}

Under \(S_P=\{A,C\}\), the assistant introduces \(B\) between the two hosted products; under \(S_P=\{A,B\}\), it brings the low-quality product \(C\) into consideration.
Because the assistant equilibrium is the same under either assortment, its price  effects differ across assortments through the no-AI benchmark it replaces.
Its firm-profit effects also depend on whether a firm is hosted, while its effect on platform profit  depends on  which sales generate commissions.
Proposition~\ref{prop: assis_price and profits} gives these comparisons, and Table~\ref{tab:assistant_price_profit_summary} in Appendix~\ref{sec: technical details} summarizes them.
Here and below, superscript \(sr\), where \(r\in\{1,2a,2b\}\), denotes the
assistant outcome in equilibrium region \(\mathcal R_r\); underlined and
overlined threshold symbols denote lower and upper thresholds, respectively.
On thresholds, subscripts \(P\) and \(CS\) indicate platform profit and consumer surplus, respectively.

\begin{proposition}\label{prop: assis_price and profits}
Relative to the no-AI benchmark, the AI assistant affects prices and
profits as follows:
\begin{enumerate}[\upshape(i)]
\item Under \(S_P=\{A,C\}\), in each equilibrium region, both on-platform firms' prices and
profits, as well as platform profit, decrease.
\item Under \(S_P=\{A,B\}\), in \(\mathcal R_1\), both on-platform firms' prices and
profits, as well as platform profit, decrease. In \(\mathcal R_{2a}\) and
\(\mathcal R_{2b}\), both on-platform firms' prices, firm \(A\)'s profit, and platform
profit increase. Firm \(B\)'s profit increases in \(\mathcal R_{2a}\) iff
\(\lambda<\bar\lambda_B^s\) and in \(\mathcal R_{2b}\) iff
\(\lambda>\underline\lambda_B^s\); otherwise, it weakly decreases.
\end{enumerate}
No comparison applies in \(\mathcal R_{\varnothing}\). The thresholds are
defined in Appendix~\ref{sec: technical details}.
\end{proposition}

\begin{figure}[h]
	   \FIGURE
	   {%
       \begin{minipage}{\linewidth}
       \centering
       \includegraphics[width=0.96\linewidth]{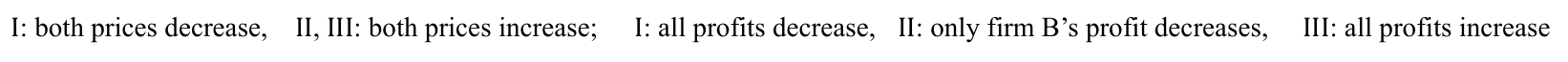}\par
       \vspace{0.35em}
       \begin{subfigure}[t]{\regionplotwidth}
         \centering
         \includegraphics[width=\linewidth]{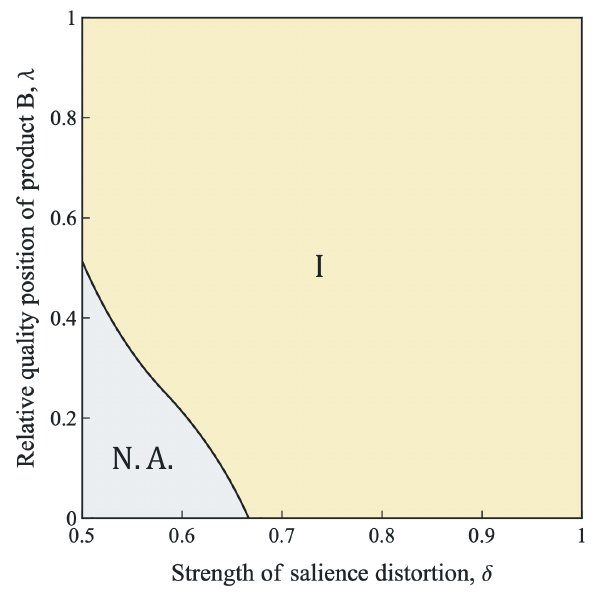}
         \caption{Low \(Q\)}
         \label{fig: assis_price_profits_a}
       \end{subfigure}
        \begin{subfigure}[t]{\regionplotwidth}
         \centering
         \includegraphics[width=\linewidth]{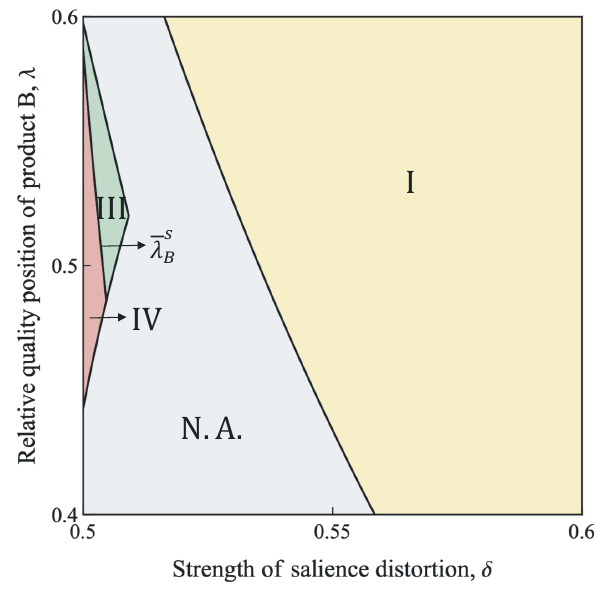}
         \caption{Medium \(Q\) (enlarged view)}
         \label{fig: assis_price_profits_b}
       \end{subfigure}
       \begin{subfigure}[t]{\regionplotwidth}
         \centering
         \includegraphics[width=\linewidth]{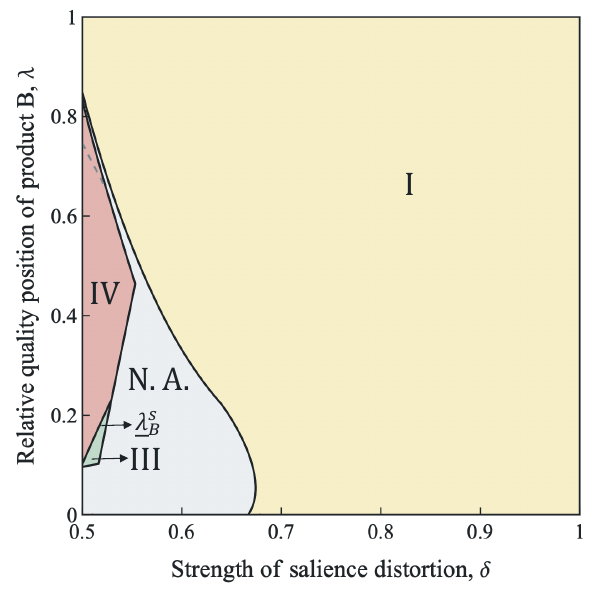}
         \caption{High \(Q\)}
         \label{fig: assis_price_profits_c}
       \end{subfigure}
       \end{minipage}%
     }
	   {Effects of the AI Assistant on Prices and Profits under \(S_P=\{A,B\}\)\label{fig: assis_price and profits}}
	   {\scriptsize \(\Theta=2\); \(q_C=0.5\); \(Q=0.5\) in Figure~\ref{fig: assis_price and profits}(a), \(Q=1.3\) in Figure~\ref{fig: assis_price and profits}(b), and \(Q=3\) in Figure~\ref{fig: assis_price and profits}(c). Panel (b) is displayed over \(\delta\in[0.5,0.6]\) and \(\lambda\in[0.4,0.6]\); all other panels use the full parameter range.}
    \end{figure}
Proposition~\ref{prop: assis_price and profits} shows how the platform's assortment shapes competition wherever a pure-strategy assistant equilibrium exists.
Under the broad-spectrum assortment \(S_P=\{A,C\}\), the no-AI benchmark pairs products with the largest quality gap.
The assistant introduces \(B\) into both recommended pairs: \(A\) faces a closer substitute in \(\{A,B\}\), while \(C\) faces a closer, higher-quality rival in \(\{B,C\}\).
Even in the niche equilibria, where \(B\) makes no sales in the low-valuation group, consumers in that group compare \(C\) with \(B\) rather than with \(A\).
As a result, both hosted firms charge less than in the no-AI benchmark, and their profits and the platform's profit fall.

Under the specialized-premium assortment \(S_P=\{A,B\}\), \(B\) is the lower-quality hosted product  and serves lower-valuation consumers without AI.

In the market-sharing equilibrium, adding \(C\) pushes \(B\) to compete for low-valuation consumers, and its common price carries this pressure into the premium group; both hosted prices and profits fall.
In either strategic-niche equilibrium, \(B\) instead cedes low-valuation sales to \(C\) and sets its price for the premium group alone.
This retreat leaves \(A\) facing a less aggressive rival than without AI, allowing both hosted firms to raise their prices.
Thus, the same niche outcome remains more competitive than the broad-spectrum benchmark while being less competitive than the specialized-premium benchmark.
Consistent with this comparison, Lemma~\ref{lem:no AI_equil} implies that \(A\)'s no-AI price is lower against \(B\) than against \(C\), holding product qualities fixed.
Figure~\ref{fig: assis_price and profits} illustrates the specialized-premium comparisons.

In both strategic-niche equilibria, \(A\) benefits from the higher price, while \(B\)'s profit depends on whether its higher margin compensates for the loss of low-valuation sales.
In the interior region \(\mathcal R_{2a}\), \(B\) gains iff \(\lambda<\bar\lambda_B^s\), as illustrated in Figure~\ref{fig: assis_price and profits}(b).
Holding \(q_C\), \(Q\), and \(\delta\) fixed within this region, a lower \(\lambda\) supports a larger proportional increase in \(B\)'s price relative to its no-AI price.
This price gain helps offset the forgone sales.

The boundary region \(\mathcal R_{2b}\) has the opposite cutoff: \(B\) gains iff \(\lambda>\underline\lambda_B^s\), as illustrated in Figure~\ref{fig: assis_price and profits}(c).
Here, \(B\)'s price makes the consumer at the recommendation boundary just willing to buy.
Within this region, a higher \(\lambda\) supports a larger proportional price gain and a smaller proportional sales loss relative to no AI, making specialization more favorable to \(B\).
Although \(B\) can lose under either niche regime, \(A\)'s gain more than offsets that loss, so platform profit rises.
The welfare analysis next examines whether consumers also benefit from these changes.

\subsection{Welfare Implications}\label{sec:assis welfare}

The price comparisons do not  determine consumer welfare because the assistant also changes which products consumers buy. We distinguish its effect on aggregate consumer surplus from the gains and losses of individual consumer types. Accurate  rankings and lower incumbent prices can coexist with consumer losses when induced purchases yield lower true utility.

\begin{proposition}\label{prop: assis_CS}
Relative to the no-AI benchmark, the AI assistant affects consumer
surplus as follows:
\begin{enumerate}[\upshape(i)]
\item Under \(S_P=\{A,C\}\), in each equilibrium region, consumer surplus
increases.
\item Under \(S_P=\{A,B\}\), in \(\mathcal R_1\), consumer surplus increases;
in \(\mathcal R_{2a}\), it decreases. In \(\mathcal R_{2b}\), it increases
iff \(\lambda<\bar\lambda_{CS}^s\); otherwise, it weakly decreases.
\end{enumerate}
No comparison applies in \(\mathcal R_{\varnothing}\). The threshold is
defined in Appendix~\ref{sec: technical details}.
\end{proposition}

\begin{figure}
	   \FIGURE
	   {%
       \begin{minipage}{\linewidth}
       \centering
       \begin{subfigure}[t]{\regionplotwidth}
         \vspace{0pt}
         \centering
         \includegraphics[width=\linewidth]{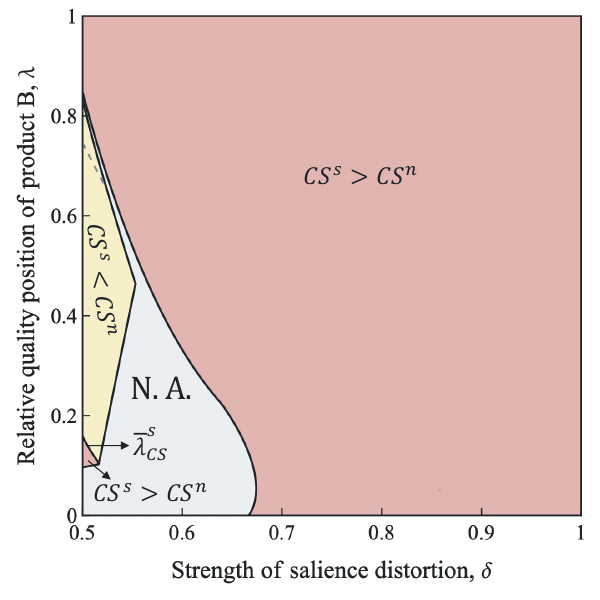}
         \caption{\(S_P=\{A,B\}\): high \(Q\)}
         \label{fig: assis_CS_a}
       \end{subfigure}%
       \hfill
       \begin{minipage}[t]{0.675\linewidth}
         \vspace{0pt}
         \centering
         \begin{subfigure}{\linewidth}
           \centering
           \includegraphics[width=\linewidth]{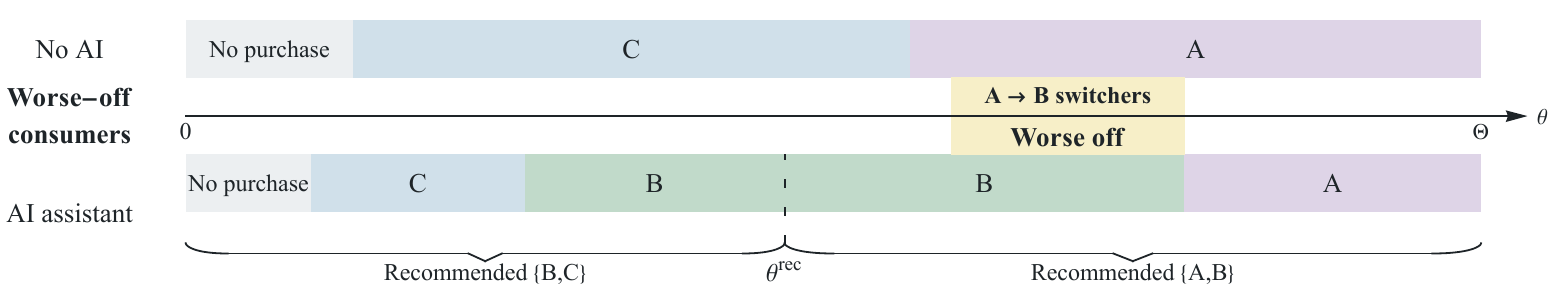}
           \caption{Individual consumer purchase and welfare changes under
             \(S_P=\{A,C\}\)}
           \label{fig: assis_CS_b}
         \end{subfigure}
         \begin{subfigure}{\linewidth}
           \vspace*{0.8em}
           \centering
           \includegraphics[width=\linewidth]{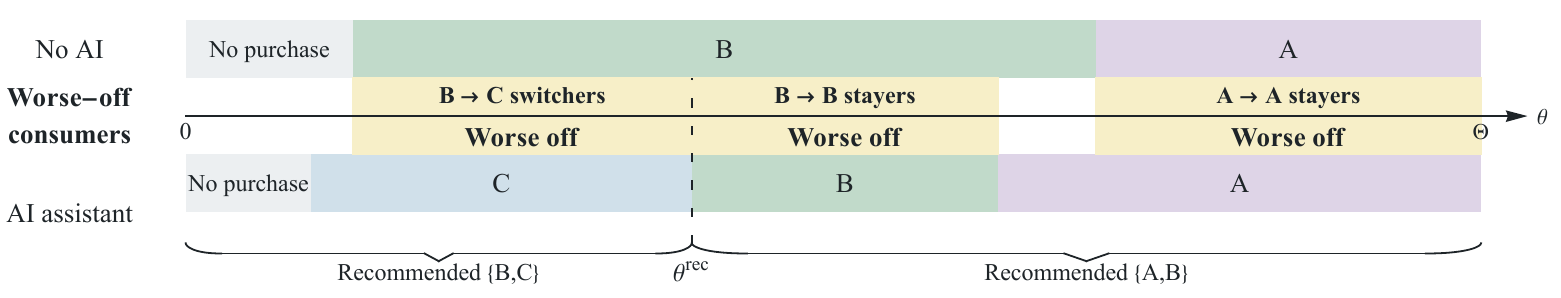}
           \caption{Individual consumer purchase and welfare changes under
             \(S_P=\{A,B\}\)}
           \label{fig: assis_CS_c}
         \end{subfigure}
       \end{minipage}
       \end{minipage}%
     }
	   {Effects of the AI Assistant on Consumer Surplus\label{fig: assis_CS}}
	   {\scriptsize \(\Theta=2\); \(q_C=0.5\); \(Q=3\) in Figure~\ref{fig: assis_CS}(a). Figures~\ref{fig: assis_CS}(b) and (c) illustrate the AI assistant's effects on individual purchases and welfare relative to the no-AI benchmark. Outcomes above the axis correspond to the no-AI benchmark, while those below the axis show the assistant's recommended pairs and induced purchases. The middle part identifies consumers who become worse off under the AI assistant. We call consumers who change their purchase decisions switchers and those who retain their choices stayers.}
    \end{figure}
Under the broad-spectrum assortment \(S_P=\{A,C\}\), aggregate consumer surplus rises in every equilibrium region.
Consumers who continue buying \(A\) or \(C\) benefit from lower prices, while access to \(B\) lets some former \(C\) buyers move to a higher-quality product.
These gains can coexist with losses for other consumers, so the aggregate comparison does not imply that every type benefits.

Figure~\ref{fig: assis_CS}(b) illustrates the \textit{alignment paradox}, which can arise in the market-sharing region \(\mathcal R_1\).
Some consumers who previously chose \(A\) from \(\{A,C\}\) receive the shortlist \(\{A,B\}\) and switch to \(B\).
Price is salient in both the original and recommended pairs, so a consumer chooses \(A\) only when its quality advantage, discounted by \(\delta\), covers its price premium.
For these switchers, \(A\)'s price premium per unit of quality advantage is higher against \(B\) at assistant prices than it was against \(C\) at no-AI prices.
The salience-based comparison can therefore favor \(B\) even when \(A\) still yields the highest true utility.
For the harmed switchers, the quality loss from moving to \(B\) exceeds the price saving relative to their original purchase of \(A\).
The assistant thus retains the best product in the shortlist and ranks it correctly, yet the consumer's comparison leads to a worse realized outcome.

The paradox distinguishes alignment of product rankings from welfare-maximizing design of comparison sets.
Here, alignment refers to the assistant's true-utility ranking, not to optimizing shortlist composition for the consumer's subsequent salience-based choice.
For interface design, ranking accuracy and the quality of induced decisions are therefore distinct evaluation criteria.

Under the specialized-premium assortment \(S_P=\{A,B\}\), the market-sharing equilibrium also raises aggregate consumer surplus, while the alignment paradox can still arise for individual switchers.
Both hosted prices fall, and access to \(C\) expands purchase opportunities.
The individual comparisons for this case are provided in Appendix~\ref{sec:proof}.

The strategic-niche equilibria instead raise both hosted prices and create four groups of affected consumers, illustrated in Figure~\ref{fig: assis_CS}(c).
Consumers who continue buying \(A\) or \(B\) lose because they pay more for the same product.
Consumers who move from \(B\) to \(C\) also lose: their quality reduction outweighs the price saving.
By contrast, consumers who move from \(B\) to \(A\) gain enough quality to offset their higher expenditure, and some former nonbuyers gain by purchasing \(C\).

In the interior niche region \(\mathcal R_{2a}\), the losses dominate.
In the boundary region \(\mathcal R_{2b}\), aggregate consumer surplus rises iff \(\lambda<\bar\lambda_{CS}^s\), as shown in Figure~\ref{fig: assis_CS}(a).
For fixed \(q_C\) and \(Q\), a lower \(\lambda\) makes \(B\) closer in quality to \(C\) and farther from \(A\): the quality loss from downward switching is smaller, while the gain from upward switching is larger.
The threshold \(\bar\lambda_{CS}^s\) incorporates these quality changes together with equilibrium prices and the sizes of the affected consumer groups.
Below it, gains from upward switching and new purchases outweigh the losses even though both hosted prices rise.

Figure~\ref{fig: assis_CS}(a) illustrates a gain--loss--gain pattern in the assistant's consumer-surplus effect as salience distortion weakens, that is, as \(\delta\) increases.
In this numerical example, the consumer-surplus threshold \(\bar\lambda_{CS}^s\) falls as \(\delta\) rises within the boundary niche region. At a fixed \(\lambda\), the consumer-surplus effect can therefore switch from positive to negative, while both hosted prices remain above their no-AI levels.
At higher \(\delta\), where the market-sharing equilibrium exists, the assistant lowers prices and again raises consumer surplus.
The comparison excludes intervening parameter values in \(\mathcal R_{\varnothing}\), where no pure-strategy assistant equilibrium exists.
The first welfare reversal thus need not involve a reversal in the price effect; the later gain accompanies a change in equilibrium region.

\begin{figure}[h]
    \FIGURE
    {%
      \centering
      \includegraphics[width=\regionplotwidth]{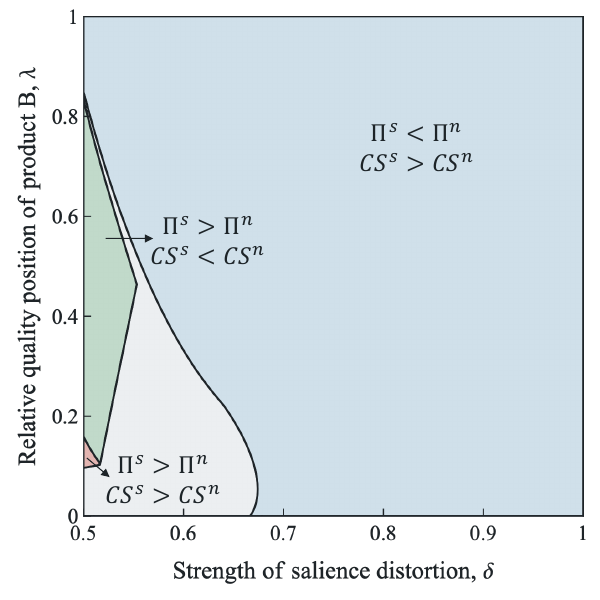}%
    }
    {Joint Effects of the AI Assistant on the Platform and Consumers under \(S_P=\{A,B\}\)\label{fig: assis_joint}}
    {\scriptsize \(\Theta=2\); \(q_C=0.5\); \(Q=3\).}
\end{figure}

We next ask whether platform profit and aggregate consumer surplus can both be strictly higher than in the no-AI benchmark for the same assortment.
Combining Propositions~\ref{prop: assis_price and profits} and~\ref{prop: assis_CS} yields the following result; Appendix~\ref{sec: technical details} provides the corresponding parameter conditions.
The off-platform firm earns zero without AI and positive profit at the characterized assistant equilibria, so the remaining question is whether the platform and consumers also gain.

\begin{proposition}\label{prop: assis_joint}
Where a pure-strategy assistant equilibrium exists, the AI assistant can generate a strict platform--consumer win--win relative to the no-AI benchmark under \(S_P=\{A,B\}\), but not under \(S_P=\{A,C\}\). Specifically:
\begin{enumerate}[\upshape(i)]
\item Under \(S_P=\{A,C\}\), it benefits consumers but hurts the platform.
\item Under \(S_P=\{A,B\}\), it may benefit both parties or benefit one party while hurting the other.
\end{enumerate}
\end{proposition}

Under \(S_P=\{A,C\}\), every assistant equilibrium raises aggregate consumer surplus at the platform's expense.
Under \(S_P=\{A,B\}\), market sharing has the same distributional pattern, while the interior niche equilibrium benefits the platform and hurts consumers in aggregate.
A strict win--win arises in the boundary niche region \(\mathcal R_{2b}\) when \(\lambda<\bar\lambda_{CS}^s\), as illustrated in Figure~\ref{fig: assis_joint}.
The hosted firms' higher combined profits benefit the platform, while upward switching and new purchases of \(C\) raise consumer surplus enough to offset the losses from higher hosted prices and downward switching.
Some consumers still lose, so the win--win concerns platform profit and aggregate consumer surplus.

\section{Delegation: The AI Agent}\label{sec:agent}
We now turn to delegation through the AI agent.
Both AI modes access the full product set, but the agent selects the product with the highest true utility instead of leaving the consumer to evaluate a shortlist.
The consumer buys the selected product iff its true utility is nonnegative.
We first characterize the pricing equilibrium and then compare prices, profits, and consumer welfare with the no-AI benchmark, holding the platform assortment fixed.

\subsection{Equilibrium Characterization}

Under the agent, firms set prices simultaneously, anticipating that it will directly select the true-utility-maximizing product from the full product set. The following lemma characterizes the pricing equilibrium.
\begin{lemma}[AI agent equilibrium]\label{lem: agent_equil} Under the AI agent, the firms' unique pure-strategy equilibrium prices are
\[ (p_A^{g},p_B^{g},p_C^{g})
=
\frac{\Theta(1-\lambda)Q}{6q_C+2(4\lambda-\lambda^2)Q}
\Big( (\lambda+3)q_C+4\lambda Q,\;
2\lambda(q_C+\lambda Q),\;
\lambda q_C
\Big). \]
\end{lemma}

At the prices in Lemma~\ref{lem: agent_equil}, all three firms have positive demand.
As willingness to pay rises, consumers move from no purchase to buying \(C\), then \(B\), and finally \(A\).
The agent allocates purchases by true quality--price tradeoffs, and neither its equilibrium prices nor its allocation depends on \(\delta\).

At \(\delta=1\), the assistant and agent induce the same purchases at any common price vector, apart from utility ties of measure zero.
The assistant's shortlist contains a product that maximizes true utility over \(\mathcal I\), and consumers evaluate the shortlisted products using that same utility.
The two modes therefore generate the same demand functions and pricing game at this limit.
For \(\delta<1\), only the assistant retains salience-based choice; any dependence on \(\delta\) in the agent's comparison with no AI comes from the benchmark.

\subsection{Pricing and Profit Implications}

Relative to no AI, delegation both opens access to the off-platform product and removes the salience discount on quality.
The first change introduces a competitor; the second makes selection depend on the full quality advantage relative to the price premium.
Firms adjust prices in response to both changes.
Proposition~\ref{prop: agent_price and profits} characterizes the price and profit effects, and Table~\ref{tab:agent_price_profit_summary} in Appendix~\ref{sec: technical details} summarizes the results.
On firm-specific thresholds, subscripts \(1\) and \(2\) refer to price and profit comparisons, respectively.

\begin{proposition}\label{prop: agent_price and profits}
Relative to the no-AI benchmark, the AI agent affects prices and
profits as follows:
\begin{enumerate}[\upshape(i)]
\item Under \(S_P=\{A,C\}\), firm \(C\)'s price and profit decrease. Firm \(A\)'s
price and profit increase iff \(\lambda<\overline{\lambda}_{A1}^{g}\) and
\(\lambda<\overline{\lambda}_{A2}^{g}\), respectively, and platform profit
increases iff \(Q>\underline{Q}^{g}\) and
\(\lambda<\overline{\lambda}_{P}^{g}\). Otherwise, each of these outcomes
weakly decreases. The \(\lambda\)-thresholds weakly decrease in \(\delta\).
\item Under \(S_P=\{A,B\}\), firm \(A\)'s and firm \(B\)'s prices increase iff
\(\lambda>\underline{\lambda}_{A1}^{g}\) and
\(\lambda>\underline{\lambda}_{B1}^{g}\), respectively; their profits increase
iff \(\lambda>\underline{\lambda}_{A2}^{g}\) and
\(\lambda>\underline{\lambda}_{B2}^{g}\), respectively; and platform profit
increases iff \(\lambda>\underline{\lambda}_{P}^{g}\). Otherwise, each of these
outcomes weakly decreases. The \(\lambda\)-thresholds weakly increase in
\(\delta\).
\end{enumerate}
All thresholds are defined in Appendix~\ref{sec: technical details}.
\end{proposition}

\begin{figure}[h]
	   \FIGURE
	   {%
       \begin{minipage}{\linewidth}
       \centering
       \includegraphics[width=0.83\linewidth]{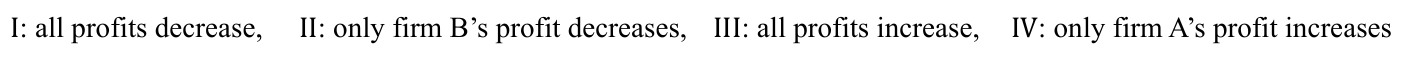}\par
       \vspace{0.35em}
       \begin{subfigure}[t]{\regionplotwidth}
         \centering
         \includegraphics[width=\linewidth]{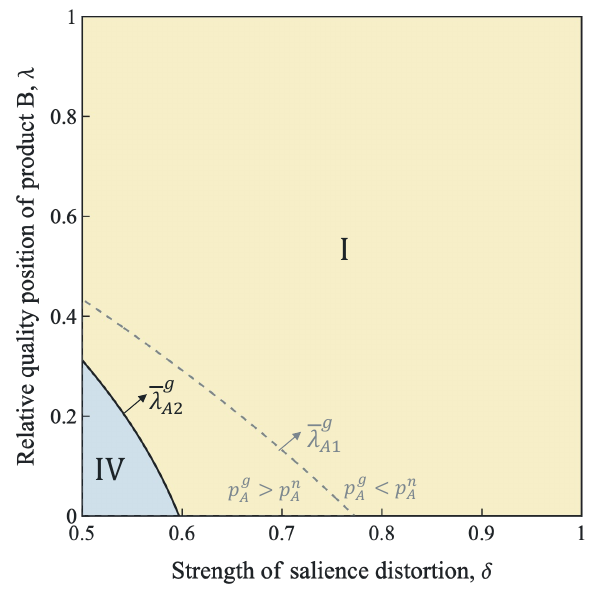}
         \caption{\(S_P=\{A,C\}: Q\leq\underline Q^g\)}
         \label{fig: agent_price_profits_a}
       \end{subfigure}%
       \hfill
       \begin{subfigure}[t]{\regionplotwidth}
         \centering
         \includegraphics[width=\linewidth]{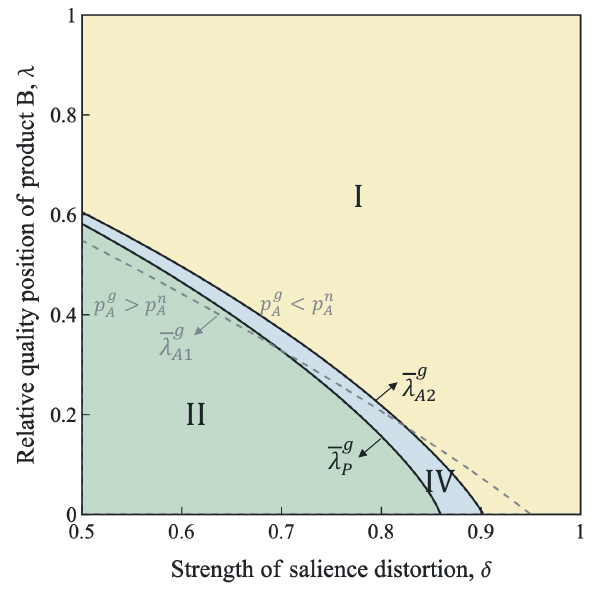}
         \caption{\(S_P=\{A,C\}: Q>\underline Q^g\)}
         \label{fig: agent_price_profits_b}
       \end{subfigure}%
       \hfill
       \begin{subfigure}[t]{\regionplotwidth}
         \centering
         \includegraphics[width=\linewidth]{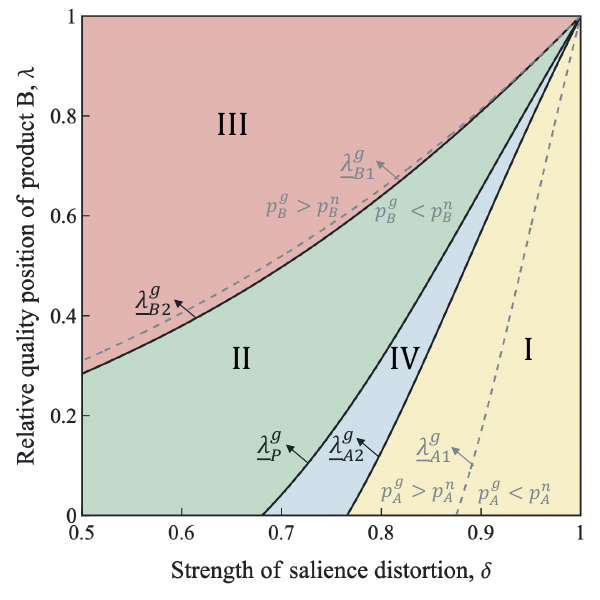}
         \caption{\(S_P=\{A,B\}\)}
         \label{fig: agent_price_profits_c}
       \end{subfigure}
       \end{minipage}%
     }
	   {Effects of the AI Agent on Prices and Profits\label{fig: agent_price and profits}}
	   {\scriptsize \(\Theta=2\); \(q_C=0.5\); \(Q=0.05\) in Figure~\ref{fig: agent_price and profits}(a), \(Q=2\) in Figure~\ref{fig: agent_price and profits}(b), and \(Q=0.5\) in Figure~\ref{fig: agent_price and profits}(c). Solid black curves represent thresholds for firm profits and platform profit, whereas gray dashed curves represent price thresholds. Price-increase and price-decrease regions relative to the no-AI benchmark are labeled inside the figures.}
    \end{figure}
Under the broad-spectrum assortment \(S_P=\{A,C\}\), \(C\) always charges less and earns less than without AI.
Access to \(B\) gives consumers a closer, higher-quality alternative to \(C\).
By eliminating salience-distorted choice, the agent also prevents price salience from masking \(C\)'s quality disadvantage relative to \(A\).
For \(A\), these forces work in opposite directions: adding \(B\) introduces a closer rival, but the agent weighs \(A\)'s full quality advantage against its price premium.
When \(\lambda\) is low, \(B\) remains far below \(A\) in quality, so the second force can dominate.
The separate cutoffs \(\overline\lambda_{A1}^g\) and \(\overline\lambda_{A2}^g\) identify when \(A\)'s price and profit, respectively, rise.

The platform gains only if \(A\)'s additional profit exceeds \(C\)'s loss.
For fixed \(q_C\) and \(\delta\), the quality-span condition \(Q>\underline Q^g\) identifies when such a gain is possible; \(\lambda<\overline\lambda_P^g\) then identifies the quality positions that deliver it.
A sufficiently wide span gives \(A\) room to earn a larger premium, provided \(B\) does not erode that advantage too much.
Figure~\ref{fig: agent_price and profits}(a) and (b) illustrate the two sides of the quality-span condition.

Under the specialized-premium assortment \(S_P=\{A,B\}\), price salience in the no-AI benchmark compresses the perceived quality gap between \(A\) and \(B\) and intensifies their competition.
Delegation removes this discount on quality differentiation but also brings \(C\) into the market.
For fixed \(q_C\) and \(Q\), a higher \(\lambda\) moves \(B\) closer to \(A\) and farther from \(C\), making \(C\) a more distant substitute for the hosted pair.
When competition from \(C\) is sufficiently weak, undistorted valuation supports higher hosted prices and profits.
These increases occur when \(\lambda\) exceeds the respective  thresholds in Proposition~\ref{prop: agent_price and profits}, which need not coincide.
Figure~\ref{fig: agent_price and profits}(c) illustrates these comparisons.
This argument turns on undistorted quality valuation, whereas the assistant's niche explanation turns on \(B\)'s withdrawal from sales to the low-valuation recommendation group.

The effects of salience distortion in both assortments come from the no-AI benchmark.
Lower \(\delta\) reduces no-AI prices and profits while leaving the agent's outcomes unchanged.
It therefore weakly expands the parameter regions in which delegation raises a hosted firm's price or profit, or platform profit.
This explains why the upper \(\lambda\)-thresholds under \(S_P=\{A,C\}\) weakly decrease with \(\delta\), whereas the lower thresholds under \(S_P=\{A,B\}\) weakly increase.

\subsection{Welfare Implications}\label{sec:agent welfare}
At any given prices, the agent selects the best available product by true utility and the consumer can decline to buy.
This removes the comparison-induced alignment paradox: the original purchase remains available, and the selected outcome gives at least as much true utility.
Once firms adjust prices, however, removing the salience discount on quality can ease price competition and make the original purchase more expensive.
The welfare comparison therefore turns on how the gains from access and undistorted selection interact with the pricing response.

\begin{proposition}\label{prop: agent_CS}
Relative to the no-AI benchmark, the AI agent affects consumer
surplus as follows:
\begin{enumerate}[\upshape(i)]
\item Under \(S_P=\{A,C\}\), consumer surplus increases iff
\(\lambda>\underline{\lambda}_{CS}^{g}\); otherwise, it weakly decreases. The
threshold \(\underline{\lambda}_{CS}^{g}\) weakly decreases in \(\delta\).
\item Under \(S_P=\{A,B\}\), consumer surplus increases iff
\(\lambda<\overline{\lambda}_{CS}^g\); otherwise, it weakly decreases. The
threshold \(\overline{\lambda}_{CS}^g\) weakly increases in \(\delta\).
\end{enumerate}
The thresholds are defined in Appendix~\ref{sec: technical details}.
\end{proposition}

\begin{figure}[h]
	   \FIGURE
	   {%
       \begin{minipage}{\linewidth}
       \centering
       \begin{subfigure}[t]{\regionplotwidth}
         \centering
         \includegraphics[width=\linewidth]{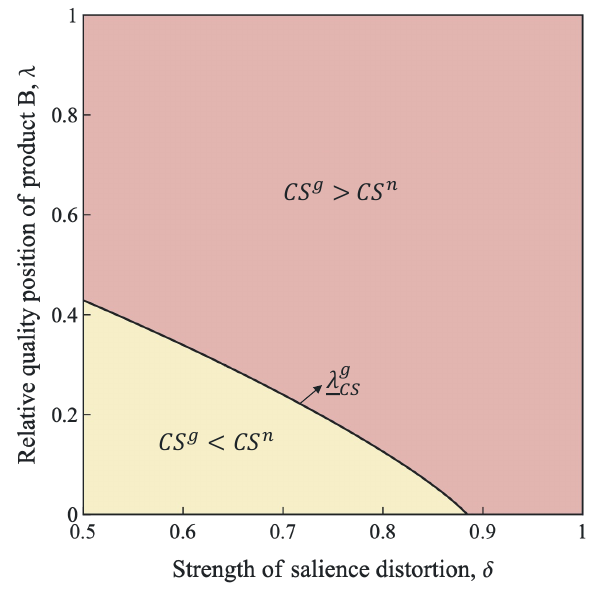}
         \caption{\(S_P=\{A,C\}\)}
         \label{fig: agent_CS_a}
       \end{subfigure}%
       \hspace{0.030\textwidth}%
       \begin{subfigure}[t]{\regionplotwidth}
         \centering
         \includegraphics[width=\linewidth]{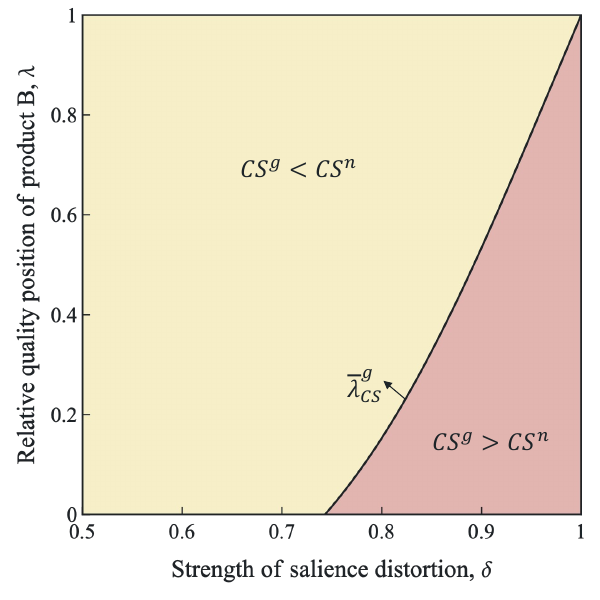}
         \caption{\(S_P=\{A,B\}\)}
         \label{fig: agent_CS_b}
       \end{subfigure}

       \begin{subfigure}{0.78\linewidth}
         \vspace*{0.7em}
         \centering
         \includegraphics[width=\linewidth]{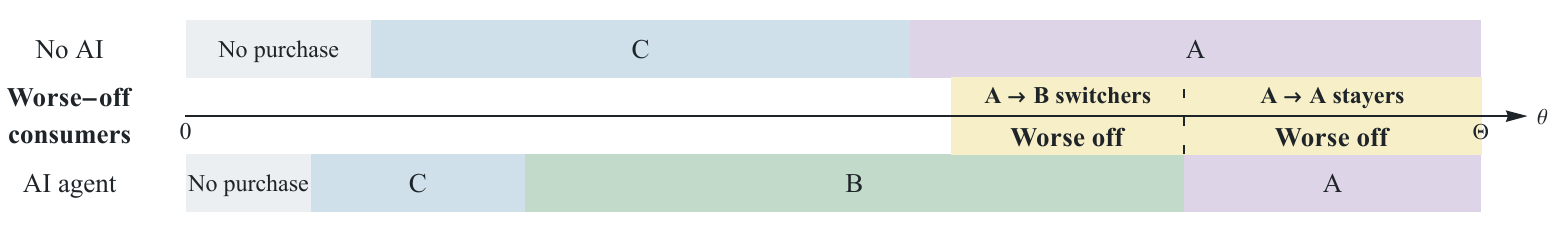}
         \caption{Individual consumer purchase and welfare changes under \(S_P=\{A,C\}\)}
         \label{fig: agent_CS_c}
       \end{subfigure}

       \begin{subfigure}{0.78\linewidth}
         \vspace*{0.7em}
         \centering
         \includegraphics[width=\linewidth]{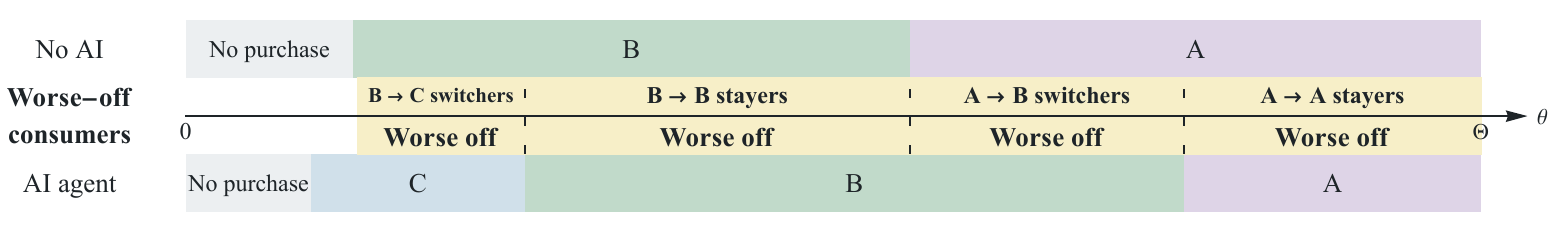}
         \caption{Individual consumer purchase and welfare changes under \(S_P=\{A,B\}\)}
         \label{fig: agent_CS_d}
       \end{subfigure}
       \end{minipage}%
     }
	   {Effects of the AI Agent on Consumer Surplus\label{fig: agent_CS}}
	   {\scriptsize \(\Theta=2\); \(q_C=0.5\); \(Q=2\) in Figure~\ref{fig: agent_CS}(a) and \(Q=0.5\) in Figure~\ref{fig: agent_CS}(b). Figures~\ref{fig: agent_CS}(c) and (d) adopt the same format as Figure~\ref{fig: assis_CS} and show how the AI agent changes individual purchases and welfare relative to the no-AI benchmark. Switchers and stayers are defined as before.}
    \end{figure}
Proposition~\ref{prop: agent_CS} places consumer-surplus gains on opposite sides of the quality-position thresholds.
Under \(S_P=\{A,C\}\), the condition \(\lambda>\underline\lambda_{CS}^g\) requires the newly accessible \(B\) to be sufficiently high in quality.
For fixed \(q_C\) and \(Q\), a higher \(\lambda\) improves this additional option and brings it closer to \(A\).
Under \(S_P=\{A,B\}\), the condition is instead \(\lambda<\overline\lambda_{CS}^g\): a low \(\lambda\) places the hosted product \(B\) close in quality to the newly accessible \(C\).
The quality of \(C\) is fixed in this comparison; its strength is relative to the hosted alternative.
In each assortment, the threshold combines the gains from access and undistorted selection with the gains or losses caused by firms' price responses.

At the individual level, when both hosted prices fall, every consumer is weakly better off under the agent.
Each consumer can retain the original purchase at a lower price, and true-utility-maximizing selection delivers an outcome at least as good; consumers who previously did not buy can still choose the outside option.
This guarantee contrasts with the assistant, under which a newly curated comparison can induce a worse purchase despite lower hosted prices.

When \(A\)'s price rises while the other hosted firm's price falls, former buyers of the lower-quality hosted product remain weakly better off because their original product is still available at a lower price.
Consumers who continue buying \(A\) lose, while former \(A\) buyers who switch to \(B\) can lose when the quality reduction exceeds the price saving.
These consumers choose the best option at the new prices, but that option can yield less true utility than their original purchase.
Figure~\ref{fig: agent_CS}(c) illustrates this case under \(S_P=\{A,C\}\); the corresponding case under \(S_P=\{A,B\}\) is in Appendix~\ref{sec:proof}.

When both hosted prices rise under \(S_P=\{A,B\}\), consumers who keep buying \(A\) or \(B\) pay more, and former \(A\) buyers who move to \(B\) also lose.
Some consumers who switch from \(B\) to \(C\) lose when their quality reduction outweighs the price saving, as illustrated in Figure~\ref{fig: agent_CS}(d).
Aggregate consumer surplus falls in this case: the access and selection gains do not offset the losses.
Under the assistant, by contrast, the boundary niche equilibrium can raise aggregate consumer surplus even when both hosted prices rise.

Stronger salience distortion makes delegation less attractive to consumers in the equilibrium comparison.
As Section~\ref{sec:benchmark} shows, a lower \(\delta\) reduces both no-AI prices proportionally while leaving participation and product choices unchanged.
The same consumers buy the same products for less, so no-AI consumer surplus rises.
The agent's prices and allocation remain fixed as \(\delta\) changes, and its consumer-surplus advantage consequently shrinks.
This explains the threshold movements in Proposition~\ref{prop: agent_CS}: stronger distortion (lower \(\delta\)) weakly narrows the consumer-gain region under either assortment, as illustrated in Figure~\ref{fig: agent_CS}(a) and (b).

We next ask whether delegation can raise both platform profit and aggregate consumer surplus.
Combining Propositions~\ref{prop: agent_price and profits} and~\ref{prop: agent_CS} yields the following result; Appendix~\ref{sec: technical details} gives the corresponding parameter conditions.
\begin{proposition}\label{prop: agent_joint}
Compared with no AI, the AI agent can generate a strict platform--consumer win--win under \(S_P=\{A,C\}\), but not under \(S_P=\{A,B\}\). Specifically:
\begin{enumerate}[\upshape(i)]
\item Under \(S_P=\{A,C\}\), it may benefit both parties, hurt both parties, or benefit one party while hurting the other.
\item Under \(S_P=\{A,B\}\), it may hurt both parties or benefit one party while hurting the other.
\end{enumerate}
\end{proposition}

\begin{figure}[h]
	   \FIGURE
	   {%
       \begin{minipage}{\linewidth}
       \centering
       \begin{subfigure}[t]{\regionplotwidth}
         \centering
         \includegraphics[width=\linewidth]{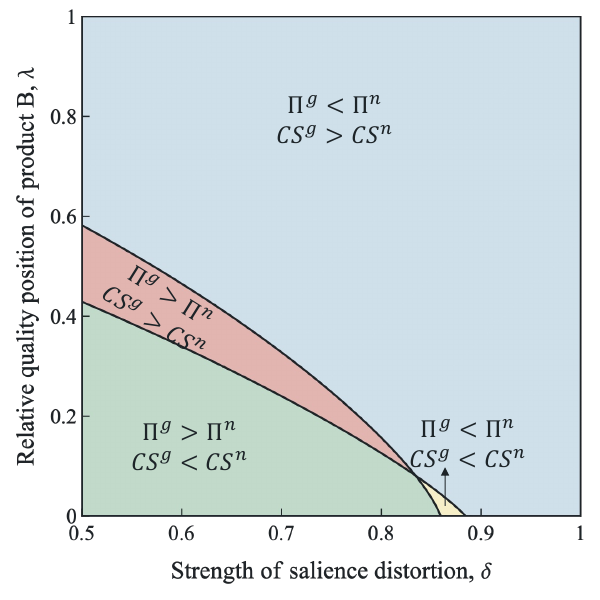}
         \caption{\(S_P=\{A,C\}\)}
         \label{fig: agent_joint_a}
       \end{subfigure}%
      \hspace{0.030\textwidth}%
       \begin{subfigure}[t]{\regionplotwidth}
         \centering
         \includegraphics[width=\linewidth]{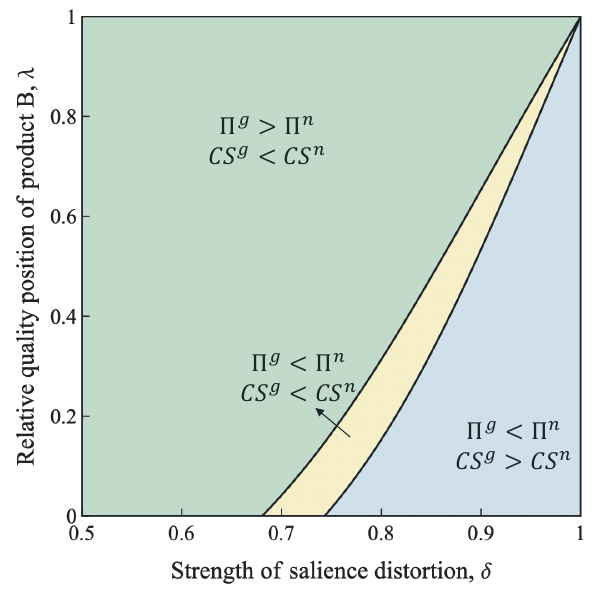}
         \caption{\(S_P=\{A,B\}\)}
         \label{fig: agent_joint_b}
       \end{subfigure}
       \end{minipage}%
     }
	   {Joint Effects of the AI Agent on the Platform and Consumers\label{fig: agent_joint}}
	   {\scriptsize \(\Theta=2\); \(q_C=0.5\); \(Q=2\) in Figure~\ref{fig: agent_joint}(a) and \(Q=0.5\) in Figure~\ref{fig: agent_joint}(b).}
    \end{figure}

Under \(S_P=\{A,C\}\), a strict win--win arises exactly when \(Q>\underline Q^g\) and \(\underline\lambda_{CS}^g<\lambda<\overline\lambda_P^g\).
The off-platform product \(B\) must be high enough in quality to deliver a net consumer gain, yet sufficiently separated from \(A\) for the premium firm's additional profit to offset \(C\)'s loss.
The two thresholds define the overlap between these requirements, illustrated in Figure~\ref{fig: agent_joint}(a).
This is an aggregate gain for consumers; it can coexist with losses for some types.

Under \(S_P=\{A,B\}\), aggregate consumer surplus increases when \(\lambda<\overline\lambda_{CS}^g\), whereas platform profit increases when \(\lambda>\underline\lambda_P^g\).
The proof establishes \(\overline\lambda_{CS}^g\leq\underline\lambda_P^g\), so the platform and consumers cannot both gain.
When \(C\) is a sufficiently close substitute for the hosted pair, consumers gain while the platform loses; when \(C\) is sufficiently distant, the platform gains while consumers lose.
Between the thresholds, both can lose, as Figure~\ref{fig: agent_joint}(b) illustrates.

The feasible win--win therefore depends on the pairing of mode and assortment: it can arise under delegation with \(\{A,C\}\) and under assistance with \(\{A,B\}\).
The assistant's gain comes from niche specialization and the resulting reallocation of purchases; the agent's gain combines expanded access with undistorted valuation of the premium product.
The next section lets the platform choose the pairing rather than taking the assortment as given.

\section{Platform Design and Cross-Scenario Outcomes}\label{sec:platform design}
The preceding sections hold the platform's assortment fixed.
We now let the platform choose among all three two-product assortments and, when AI is available, between assistance and delegation.
AI availability remains exogenous, as in Section~\ref{sec:timing_equil}; the platform chooses its preferred design within each shopping scenario.
We compare designs only where all relevant pricing subgames have the characterized pure-strategy equilibria.
Using the assistant regions in Lemma~\ref{lem: assis_equil}, the common equilibrium-existence domain is \(\mathcal R_s\equiv\mathcal R_1\cup\mathcal R_{2a}\cup\mathcal R_{2b}\), together with the maintained existence conditions for the no-AI and agent subgames.
The preferred mode is not determined in \(\mathcal R_{\varnothing}\).

Without AI, hosting determines which products consumers discover and therefore affects price competition.
Within either AI mode, all products remain accessible, so the assortment determines which sales generate commissions without changing prices or consumer purchases.
The platform thus chooses both the AI mode and the pair of sellers whose revenues it shares.
Let \((\Pi^*,CS^*)\) and \((\Pi^{n*},CS^{n*})\) denote platform profit and aggregate consumer surplus under the platform's preferred AI and no-AI designs, respectively.

\begin{proposition}\label{prop: platform design}
On the common equilibrium-existence domain \(\mathcal R_s\), where all compared pricing subgames admit the characterized pure-strategy equilibria, the platform's preferred designs and their implications are as follows:
\begin{enumerate}[\upshape(i)]
\item In the no-AI scenario, the platform prefers \(S_P=\{A,C\}\).
In the AI scenario, the platform weakly prefers
\(S_P=\{A,B\}\) with the AI agent.

\item Relative to the no-AI optimum, the outcome under AI increases platform
profit iff \(Q>\underline Q_P^*\) and
\(\underline{\lambda}_P^*<\lambda<\overline{\lambda}_P^*\), and increases
consumer surplus iff \(\lambda>\underline{\lambda}_{CS}^{g}\).
Otherwise, each of these outcomes weakly decreases.

\item Jointly, the outcome under AI may benefit both the platform and
consumers, hurt both, or benefit either party while hurting the other.
\end{enumerate}
The thresholds are defined in Appendix~\ref{sec: technical details}.
\end{proposition}

Without AI, hosting \(\{A,C\}\) pairs the products with the widest quality gap and reduces competition between the hosted products.
The resulting platform profit is higher than under either alternative assortment.
With AI, all three products compete regardless of hosting, and \(C\) generates less revenue than either \(A\) or \(B\) in each characterized equilibrium.
Hosting \(\{A,B\}\) therefore captures the two larger profit streams without changing the pricing game within a mode.

With AI, the mode comparison then asks how much profit the platform earns from these two hosted firms.
In the market-sharing assistant equilibrium, platform profit is \(\delta\) times that under the agent, so delegation removes the reduction in revenue associated with price salience.
The niche equilibria allow the hosted firms to protect margins, but the platform still earns weakly less than under the agent, as established in the proof.
Although the niche assistant equilibria raise platform profit relative to no AI with \(\{A,B\}\), they do not outperform the agent.
This mode ranking holds for the available hosting arrangements and the fixed commission rate.

\begin{figure}[h]
	   \FIGURE
	   {%
       \begin{minipage}{\linewidth}
       \centering
       \begin{subfigure}[t]{\regionplotwidth}
         \centering
         \includegraphics[width=\linewidth]{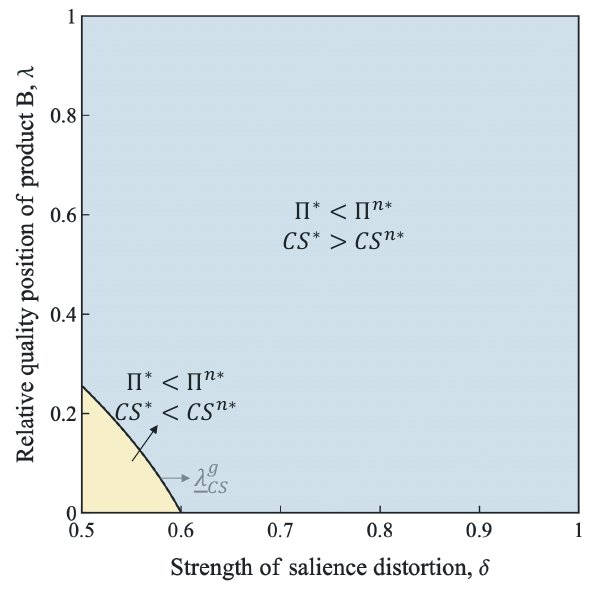}
         \caption{\(Q\leq\underline Q_P^*\)}
         \label{fig: final_joint_a}
       \end{subfigure}%
       \hspace{0.030\textwidth}%
       \begin{subfigure}[t]{\regionplotwidth}
         \centering
         \includegraphics[width=\linewidth]{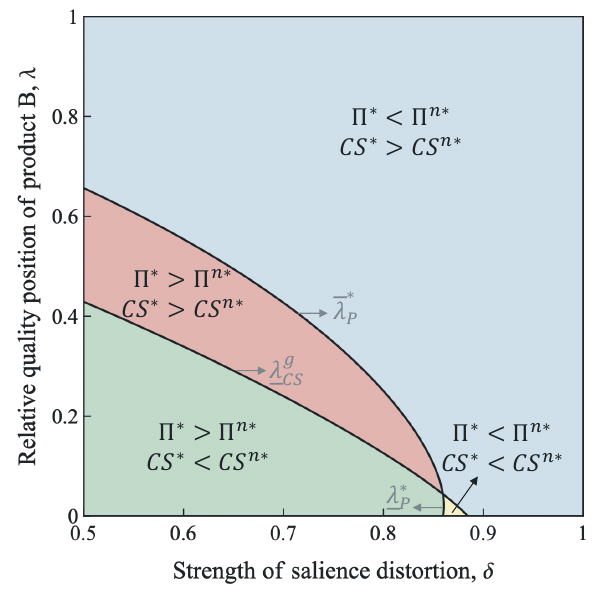}
         \caption{\(Q>\underline Q_P^*\)}
         \label{fig: final_joint_b}
       \end{subfigure}
       \end{minipage}%
     }
	   {
        Joint Effects of an Agent with Assortment \(\{A,B\}\) Relative to No AI with Assortment \(\{A,C\}\)\label{fig: final_joint}}
	   {\scriptsize \(\Theta=2\); \(q_C=0.5\); \(Q=0.05\) in Figure~\ref{fig: final_joint}(a) and \(Q=2\) in Figure~\ref{fig: final_joint}(b).
The starred quantities in the plots refer to the two named configurations wherever their pricing equilibria exist.
They denote preferred designs only on the common equilibrium-existence domain of Proposition~\ref{prop: platform design}.}
    \end{figure}

The possibility of a joint gain under the preferred AI design is consistent with Proposition~\ref{prop: agent_joint}.
That result rules out a strict win--win for delegation when the no-AI benchmark also hosts \(\{A,B\}\); here the no-AI optimum hosts \(\{A,C\}\).
Moving to the preferred AI design therefore combines delegation's effects on access, choice, and pricing with a shift in commission revenue from \(C\) to \(B\).
 
Figure~\ref{fig: final_joint} compares these two configurations wherever their pricing equilibria are characterized; their interpretation as preferred designs applies on \(\mathcal R_s\).

The platform gains only if \(Q>\underline Q_P^*\) and \(\underline\lambda_P^*<\lambda<\overline\lambda_P^*\). The gain interval reflects two forces: \(B\) contributes little profit as \(\lambda\) approaches zero, while competition between \(A\) and \(B\) compresses their margins as \(\lambda\) approaches one. At a given \(\delta\),  the interval may have a zero lower endpoint or be empty; a sufficiently wide quality span therefore does not guarantee a gain.  Figure~\ref{fig: final_joint}(a) and (b) illustrate these two quality-span cases.

Consumers face the same agent prices and product set regardless of the hosted pair.
Proposition~\ref{prop: agent_CS}(i) therefore applies to the comparison with no AI under \(\{A,C\}\): aggregate consumer surplus increases iff \(\lambda>\underline\lambda_{CS}^g\).
A sufficiently high-quality \(B\) provides an additional option whose benefit outweighs any losses from  price changes.
A strict win--win requires this consumer-gain region to overlap the platform's profit-gain interval.

Holding the two configurations fixed, stronger salience distortion lowers no-AI prices without changing purchases or the agent equilibrium. It lowers no-AI platform profit and raises  consumer surplus, weakly expanding the platform-gain region while  narrowing the consumer-gain region. Stronger distortion therefore does not uniformly favor a joint gain. Overall, the platform's preferred AI design need not increase either platform profit or consumer surplus relative to no AI.

\section{Conclusion}\label{sec:conclusion}
This paper studies how AI's role in product selection changes price competition, platform profit, and consumer welfare.
An assistant curates the products consumers compare, while an agent selects a product directly, subject to the consumer's purchase decision.
Both expand access to off-platform products, but only the assistant leaves the salience-based comparison stage intact.
The resulting price, profit, and welfare effects depend on the original platform assortment, the strength of salience distortion, and firms' pricing responses.

Under the assistant, the middle-quality seller can reach two recommendation groups while charging a single price, creating an incentive to forgo low-valuation demand to protect its premium margin.
Greater visibility can therefore coexist with a retreat from one customer segment.
The \textit{alignment paradox} identifies a separate consequence of assistance: accurate product rankings can still induce harmful switching because the comparison partner changes how consumers weigh quality against price.
Alignment here concerns product rankings rather than optimization of the consumer's induced purchase.
Delegation removes this comparison distortion, but firms' price responses can still leave consumers worse off.
With the assortment held fixed, each AI mode can generate joint gains in platform profit and aggregate consumer surplus, under different configurations.


For \textit{platform operators}, the analysis distinguishes the competitive effects of the AI mode from the revenues generated by the hosted sellers.
Once AI makes all products accessible, changing hosted sellers changes the commission base without changing prices or consumer purchases within a given mode.
On the common equilibrium-existence domain, the platform weakly prefers the specialized-premium assortment with an agent.
This preferred AI design guarantees neither higher platform profit nor higher aggregate consumer surplus relative to the no-AI optimum.

For \textit{retailers and brands}, recommendation frequency should be evaluated together with the alternatives shown alongside a product and the purchases those comparisons induce.
An assistant can expose the middle-quality product to both high- and low-valuation consumers, yet its seller may  forgo low-valuation sales to protect  its premium margin.
Broader exposure therefore need not lead to sales across more customer segments.
Under delegation,   a firm's profit change relative to no AI  depends on its quality position and the rivals consumers could previously consider.
For firms hosted on a specialized-premium platform, a sufficiently weak off-platform alternative can allow higher prices and profits than without AI.
These gains are relative to the restricted-access benchmark; changing the hosted pair alone does not change competition under either AI mode.

For \textit{policymakers and regulators}, the alignment paradox shows that aligning product rankings with consumer preferences is insufficient to ensure beneficial consumer choices.
An assistant can include the consumer's best product yet induce a worse purchase because the other product in the shortlist changes how price and quality are compared.
Recommendation systems should therefore be evaluated on both ranking accuracy and induced choices.
This conclusion concerns our top-two rule, not  an assistant that optimizes shortlists for realized consumer welfare.
Delegation removes the  comparison distortion, but consumer welfare remains sensitive to the resulting price response.


Our analysis rests on several assumptions that future research could relax.
First, the assistant recommends two products in our three-product setting, matching the size of the no-AI comparison set while changing its composition.
Extending the model to markets with more products and larger personalized shortlists could reveal how choice overload interacts with salience \citep{long2025choice}.
Second, alignment under assistance concerns the ranking of individual products by true utility.
The assistant recommends the top two without optimizing the consumer's subsequent salience-based choice.
A choice-aware assistant could instead select the shortlist to maximize the true utility of the induced purchase.
Analyzing that policy would require characterizing its recommendation rule and the resulting pricing game.
Platform revenue or engagement objectives could also change both shortlist composition and sellers' incentives.
Third, we assume AI perfectly observes each consumer's type.
Incomplete or noisy preference signals could produce recommendation or selection errors that interact with salience and firms' pricing responses.
Allowing AI to learn from such signals would extend the analysis to imperfect preference information.
Fourth, AI availability, seller participation, and hosting terms are taken as given.
Allowing platforms to choose whether to offer AI and sellers to decide whether to participate and on what terms would connect the design comparisons to adoption and contracting.

\makeatletter
\renewcommand{\@biblabel}[1]{}
\let\OLDthebibliography\thebibliography
\renewcommand{\thebibliography}[1]{%
  \OLDthebibliography{#1}%
  \setlength{\itemsep}{1pt}%
  \setlength{\parskip}{1pt}%
  \setlength{\baselineskip}{15pt}%
}
\makeatother
\bibliographystyle{ormsv080-nourl}
\bibliography{2026_Shopping_Agents.bib}
\newpage
\vspace{-20pt}
\begin{APPENDICES}
\numberwithin{lemma}{section}
\numberwithin{proposition}{section}
\numberwithin{figure}{section}
\numberwithin{table}{section}
\numberwithin{equation}{section}
\renewcommand{\theHsection}{appendix.\Alph{section}}
\renewcommand{\theHsubsection}{appendix.\Alph{section}.\arabic{subsection}}
\renewcommand{\theHsubsubsection}{appendix.\Alph{section}.\arabic{subsection}.\arabic{subsubsection}}
\newpage
\section{Technical Details}\label{sec: technical details}
This appendix collects notation, parameter regions, equilibrium outcomes, price and profit comparison summaries, and
threshold definitions used in the main text. All derivations are provided in
Appendix~\ref{sec:proof}.

\subsection{Notation}
We summarize the main notation in Table~\ref{tab:notation}.

\begin{table}[h]
\centering
\caption{Notation}
\label{tab:notation}
{\footnotesize
\setlength{\tabcolsep}{5pt}
\renewcommand{\arraystretch}{1.18}
\begin{tabular}{@{}p{0.25\textwidth}p{0.69\textwidth}@{}}
\toprule
Notation & Description \\
\midrule
\(\mathcal I\) & Full product set, \(\mathcal I=\{A,B,C\}\). \\
\(S_P;\ k\) & Platform's hosted assortment, \(|S_P|=2\); off-platform product \(k\notin S_P\). \\
\(q_i,p_i\) & Quality and price of product \(i\in\mathcal I\). \\
\(Q,\lambda\) & Quality span \(Q=q_A-q_C\) and relative quality position of \(B\), \(\lambda=(q_B-q_C)/Q\). \\
\(\theta,\Theta\) & Consumer willingness to pay for quality, \(\theta\sim U[0,\Theta]\); \(\Theta>0\) is the upper bound. \\
\(u_i(\theta),\tilde u_i(\theta\mid T)\) & True and perceived utility from product \(i\), respectively. \\
\(m\in\{n,s,g\}\) & Outcome index: \(n\) for no AI, \(s\) for assistance, and \(g\) for delegation. \\
\(T\) & Two-product comparison set for consumer choice. \\
\(T^n\) & No-AI comparison set, \(T^n=S_P\). \\
\(T_\theta^s\) & Personalized two-product shortlist recommended by the assistant. \\
\(i_\theta^1,i_\theta^2\) & First- and second-ranked products by true utility for a type-\(\theta\) consumer. \\
\(i_\theta^g\) & Product selected by the agent for a type-\(\theta\) consumer, \(i_\theta^g=i_\theta^1\). \\
\(\bar q(T),\bar p(T)\) & Average quality and price in comparison set \(T\). \\
\(\sigma(a,\bar a)\) & Salience of attribute value \(a\) relative to reference \(\bar a\). \\
\(\delta\) & The strength of salience distortion; lower \(\delta\) means stronger distortion. \\
\(p_i^m\) & Equilibrium price of firm \(i\) under \(m\). \\
\(D_i^m\) & Demand for product \(i\) under \(m\). \\
\(\pi_i^m,\Pi^m\) & Firm and platform profits under \(m\). \\
\(CS^m\) & Consumer surplus under \(m\). \\
\(x^m(\theta)\) & Product purchased by type \(\theta\) under \(m\); \(0\) denotes no purchase. \\
\(\mathcal R_r,\mathcal R_{\varnothing}\) & Assistant equilibrium regions for \(r\in\{1,2a,2b\}\) and the region without a pure-strategy pricing equilibrium. \\
\bottomrule
\end{tabular}
}
\end{table}

\subsection{Equilibrium Regions}

Write \(\mathbf p=(p_A,p_B,p_C)\) for the price vector and \(\mathbf p_{-i}\) for the prices of firms other than \(i\).
Let \(\omega\equiv(q_C,Q,\delta,\lambda)\). The full admissible region is
\[
\Omega\equiv\left\{\omega:q_C>0,\ Q>0,\
\frac12\leq\delta\leq1,\ 0<\lambda<1\right\}.
\]
The local Type-1 region is \(\mathcal L_1\equiv\Omega\). Define
\[
\begin{alignedat}{2}
\underline Q^s&\equiv\frac{7+\sqrt{97}}8q_C,\qquad&
\bar\delta^s&\equiv
\sqrt{\frac{Q^2+2Qq_C}{q_C^2}}-\frac Q{q_C},\\
\bar\lambda^s&\equiv
\frac{2Q(1-\delta)-\delta^2q_C}{\delta(2Q+\delta q_C)},\qquad&
\underline\lambda_1^s&\equiv
\frac{2Q(1-\delta)-\delta^2q_C}{(2-\delta)(2Q+\delta q_C)},\\
\underline\lambda_2^s&\equiv
\mathrlap{\frac{2(1-\delta)Q^2-\delta q_C[(\delta+2)Q+2\delta q_C]}
{\delta Q(2Q+\delta q_C)},\qquad
\underline\lambda_{2b}^s\equiv
\frac{\delta q_C(Q+\delta q_C)}{(1-\delta)Q(2Q+\delta q_C)}.}
\end{alignedat}
\]
For \(Q>\underline Q^s\), let \(\delta_0\) be the unique root in
\((1/2,\bar\delta^s)\) of
\[
\delta^3(Qq_C+q_C^2)+\delta^2(2Q^2-2q_C^2)
-\delta(4Q^2+2Qq_C)+2Q^2=0.
\]
The Type-2 local regions are divided into \(\mathcal L_{2a}\) and
\(\mathcal L_{2b}\):
\[
\mathcal L_{2a}=\left\{\omega:q_C>0,\quad
\begin{array}{ll}
\frac{q_C}{4}<Q\leq\underline Q^s,\ \frac12\leq\delta<\bar\delta^s,\
\ \underline\lambda_1^s<\lambda<\bar\lambda^s;\\[0.3em]
\text{or }Q>\underline Q^s,\ \frac12\leq\delta<\delta_0,\
\ \underline\lambda_2^s<\lambda<\bar\lambda^s;\\[0.3em]
\text{or }Q>\underline Q^s,\ \delta_0\leq\delta<\bar\delta^s,\
\ \underline\lambda_1^s<\lambda<\bar\lambda^s
\end{array}\right\},
\]
\[
\mathcal L_{2b}=\left\{\omega:q_C>0,\ Q>\underline Q^s,\
\frac12\leq\delta<\delta_0,\
\underline\lambda_{2b}^s<\lambda\leq\underline\lambda_2^s\right\}.
\]
The local regions identify candidate price profiles satisfying the relevant
pricing conditions and threshold ordering. Global validity additionally requires
that no firm have a profitable unilateral price deviation, whether within or
across demand regions.
For each \(j\in\{1,2a,2b\}\), denote the corresponding Type-\(j\)
candidate price vector by \(\mathbf p^{sj}\). Let \(\mathcal P_i^j(\omega)\) contain
\(p_i^{sj}\), all attained deviation breakpoints, and all interval-specific
stationary points. Let \(\kappa_i(S_P)=1-\tau\) if \(i\in S_P\) and
\(\kappa_i(S_P)=1\) otherwise. Let \(\mathcal W_i^j(\omega)\) contain all unattained
one-sided payoff suprema. Define
\[
M_i^j(\omega)\equiv
\max\left(\{\kappa_i(S_P)yD_i(y,\mathbf p_{-i}^{sj}):y\in\mathcal P_i^j(\omega)\}
\cup\mathcal W_i^j(\omega)\right),\quad
V_i^j(\omega)\equiv M_i^j(\omega)-\kappa_i(S_P)p_i^{sj}D_i(\mathbf p^{sj}).
\]
Thus \(V_i^j=0\) exactly when the candidate admits no profitable unilateral deviation.
The globally valid regions are
\[
\begin{aligned}
\mathcal R_1&\equiv\{\omega\in\mathcal L_1:V_A^1=V_B^1=V_C^1=0\},\\
\mathcal R_{2a}&\equiv\{\omega\in\mathcal L_{2a}:V_A^{2a}=V_B^{2a}=V_C^{2a}=0\},\\
\mathcal R_{2b}&\equiv\{\omega\in\mathcal L_{2b}:V_A^{2b}=V_B^{2b}=V_C^{2b}=0\},\\
\mathcal R_{\varnothing}&\equiv\Omega\setminus
(\mathcal R_1\cup\mathcal R_{2a}\cup\mathcal R_{2b}).
\end{aligned}
\]
The four regions \(\mathcal R_1\), \(\mathcal R_{2a}\),
\(\mathcal R_{2b}\), and \(\mathcal R_{\varnothing}\) form a partition of
\(\Omega\). For each \(j\in\{1,2a,2b\}\) and every
\(\omega\in\mathcal R_j\), there is a unique pure-strategy pricing equilibrium
of Type \(j\). For every \(\omega\in\mathcal R_{\varnothing}\), no
pure-strategy pricing equilibrium exists.

\subsection{Equilibrium Outcomes}

\noindent\textbf{Auxiliary expressions.}
For compactness, let
\[
x\equiv\frac{q_C}{Q},\qquad
D\equiv3x+4\lambda-\lambda^2,\qquad
L\equiv(\lambda+3)x+4\lambda,\qquad
E\equiv6(1+\delta x)+\delta(1-\lambda)(2+\delta x),
\]
\[
K\equiv4-2\delta\lambda+2\delta x-\delta^2\lambda x-\delta^2x^2.
\]
Define
\[
\Xi(u,v)\equiv
\frac{u\left[(12-8\delta)u^2-(1-6\delta)uv-2(1-\delta)v^2\right]}
{2(4u-v)^2},
\]
and the welfare polynomials
{\footnotesize
\[
\begin{aligned}
\Psi_1={}&36q_C^3+q_C^2Q[27+110\lambda-29\lambda^2
-\delta(18+4\lambda-22\lambda^2)]\\
&+q_CQ^2[(72-48\delta)\lambda+(79+14\delta)\lambda^2
-(46-36\delta)\lambda^3+(3-2\delta)\lambda^4]
+Q^3[(48-32\delta)\lambda^2-(4-24\delta)\lambda^3
-(8-8\delta)\lambda^4],\\[0.4em]
\Psi_{2a}={}&
\delta^2[6+\delta(1-\lambda)]^2q_C^3+q_C^2Q\left[
36\delta(2+\delta)+4\delta^2(8-7\delta)(1-\lambda)
-3\delta^4(1-\lambda)^2+\delta^4(3-2\delta)(1-\lambda)^3
\right]\\
&+q_CQ^2\left[
36+72\delta+4\delta(4-11\delta)(1-\lambda)
-4\delta^2(3-\delta)(1-\lambda)^2
+4\delta^3(3-2\delta)(1-\lambda)^3
\right]\\
&+Q^3\left[
36-4(1+4\delta)(1-\lambda)
-4\delta(2-\delta)(1-\lambda)^2
+4\delta^2(3-2\delta)(1-\lambda)^3
\right],\\[0.4em]
\Psi_{2b}={}&\delta^4q_C^5
-\delta^3[4-2\delta^2-3\delta\lambda+2\delta^2\lambda]Qq_C^4
-\delta^2[4+4\delta-4\delta^2-2\delta^3+4\delta^2\lambda
-3\delta^2\lambda^2+2\delta^3\lambda^2]Q^2q_C^3\\
&+\delta[16-12\delta^2+6\delta^3-20\delta\lambda
+8\delta^2\lambda-\delta^3\lambda+2\delta^4\lambda
+8\delta^2\lambda^2-4\delta^3\lambda^2-2\delta^4\lambda^2]Q^3q_C^2\\
&+[16+16\delta-25\delta^2+6\delta^3-16\delta\lambda
+14\delta^2\lambda-12\delta^3\lambda+8\delta^4\lambda
-\delta^2\lambda^2+10\delta^3\lambda^2-8\delta^4\lambda^2]Q^4q_C\\
&+4[3-2\delta+\lambda-3\delta^2\lambda+2\delta^3\lambda
-2\delta\lambda^2+4\delta^2\lambda^2-2\delta^3\lambda^2]Q^5,\\[0.4em]
\Psi_g={}&36q_C^3+q_C^2Q(9+106\lambda-7\lambda^2)
+q_CQ^2(24\lambda+93\lambda^2-10\lambda^3+\lambda^4)
+Q^3(16\lambda^2+20\lambda^3).
\end{aligned}
\]}
Normalize these polynomials as
\[
\begin{aligned}
\widehat\Psi_1(x,\delta,\lambda)&\equiv\frac{\Psi_1}{Q^3},&
\widehat\Psi_{2a}(x,\delta,\lambda)&\equiv\frac{\Psi_{2a}}{Q^3},\\
\widehat\Psi_{2b}(x,\delta,\lambda)&\equiv\frac{\Psi_{2b}}{Q^5},&
\widehat\Psi_g(x,\lambda)&\equiv\frac{\Psi_g}{Q^3}.
\end{aligned}
\]

\noindent\textbf{No-AI benchmark.}
For \(S_P=\{i,j\}\), where \(q_i>q_j\),
\[
p_i^n=\frac{2\delta\Theta q_i(q_i-q_j)}{4q_i-q_j},\qquad
p_j^n=\frac{\delta\Theta q_j(q_i-q_j)}{4q_i-q_j},
\]
\[
\pi_i^n=(1-\tau)\frac{4\delta\Theta q_i^2(q_i-q_j)}{(4q_i-q_j)^2},\qquad
\pi_j^n=(1-\tau)\frac{\delta\Theta q_iq_j(q_i-q_j)}{(4q_i-q_j)^2}.
\]
\[
CS^{n,ij}=\Theta\Xi(q_i,q_j).
\]
For \(S_P=\{i,j\}\), platform profit is
\(\Pi^{n,ij}=\frac{\tau}{1-\tau}(\pi_i^n+\pi_j^n)\).

\noindent\textbf{AI assistant.}
\[
\alpha_1=\frac{\delta\Theta(1-\lambda)Q}{6q_C+2(4\lambda-\lambda^2)Q},\quad
\alpha_{2a}=\frac{\delta\Theta(1-\lambda)Q}{6(Q+\delta q_C)+\delta(1-\lambda)(2Q+\delta q_C)},
\]
\[
\alpha_{2b}=\frac{\delta\Theta(1-\lambda)Q}
{(2-\delta\lambda)Q(2Q+\delta q_C)-\delta^2q_C^2}.
\]
The Type-1 outcomes are
{\footnotesize
\[
\begin{gathered}
p_A^{s1}=\alpha_1[(\lambda+3)q_C+4\lambda Q],\quad
p_B^{s1}=2\alpha_1\lambda(q_C+\lambda Q),\quad
p_C^{s1}=\alpha_1\lambda q_C,\\
\pi_A^{s1}=\kappa_A(S_P)\frac{\alpha_1[(\lambda+3)q_C+4\lambda Q]^2}{2QD},\quad
\pi_B^{s1}=\kappa_B(S_P)\frac{2\alpha_1\lambda(q_C+\lambda Q)^2}{QD},\\
\pi_C^{s1}=\kappa_C(S_P)\frac{\alpha_1\lambda(1-\lambda)q_C(q_C+\lambda Q)}{2QD},
\quad CS^{s1}=\frac{\Theta\Psi_1}{8Q^2D^2}.
\end{gathered}
\]}
The Type-2a outcomes are
{\footnotesize
\[
\begin{gathered}
p_A^{s2a}=4\alpha_{2a}(Q+\delta q_C),\quad
p_B^{s2a}=\alpha_{2a}[2(Q+\delta q_C)-\delta(1-\lambda)(2Q+\delta q_C)],\quad
p_C^{s2a}=2\alpha_{2a}\delta q_C,\\
\pi_A^{s2a}=\kappa_A(S_P)\frac{16\alpha_{2a}(Q+\delta q_C)^2}{QE},\quad
\pi_B^{s2a}=\kappa_B(S_P)\frac{\alpha_{2a}[2(Q+\delta q_C)-\delta(1-\lambda)(2Q+\delta q_C)]^2}{QE},\\
\pi_C^{s2a}=\kappa_C(S_P)\frac{4\alpha_{2a}\delta^2(1-\lambda)q_C(Q+\delta q_C)}{QE},
\quad CS^{s2a}=\frac{\Theta\Psi_{2a}}{2Q^2E^2}.
\end{gathered}
\]}
The Type-2b outcomes are
{\footnotesize
\[
\begin{gathered}
p_A^{s2b}=2\alpha_{2b}Q(Q+\delta q_C),\quad
p_B^{s2b}=\alpha_{2b}\delta(q_C+\lambda Q)(2Q+\delta q_C),\quad
p_C^{s2b}=\alpha_{2b}\delta q_CQ,\\
\pi_A^{s2b}=\kappa_A(S_P)\frac{4\alpha_{2b}(Q+\delta q_C)^2}{K},\quad
\pi_B^{s2b}=\kappa_B(S_P)\frac{\alpha_{2b}\delta(q_C+\lambda Q)(2Q+\delta q_C)
[2(1-\delta)-\delta^2x(1+x)]}{K},\\
\pi_C^{s2b}=\kappa_C(S_P)\frac{\alpha_{2b}\delta^2(1-\lambda)q_C(Q+\delta q_C)}{K},
\quad CS^{s2b}=\frac{\Theta\Psi_{2b}}{2Q^4K^2}.
\end{gathered}
\]}
For \(t\in\{1,2a,2b\}\) and
\(S_P\in\{\{A,C\},\{A,B\},\{B,C\}\}\), platform profit is
\[
\Pi^{st,S_P}=\frac{\tau}{1-\tau}\sum_{i\in S_P}\pi_i^{st}.
\]

\noindent\textbf{AI agent.}
{\footnotesize
\[
\begin{gathered}
p_A^g=\frac{\Theta(1-\lambda)[(\lambda+3)q_C+4\lambda Q]}{2D},\quad
p_B^g=\frac{\Theta(1-\lambda)\lambda(q_C+\lambda Q)}{D},\quad
p_C^g=\frac{\Theta(1-\lambda)\lambda q_C}{2D},\\
\pi_A^g=\kappa_A(S_P)\frac{\Theta(1-\lambda)[(\lambda+3)q_C+4\lambda Q]^2}{4QD^2},\quad
\pi_B^g=\kappa_B(S_P)\frac{\Theta(1-\lambda)\lambda(q_C+\lambda Q)^2}{QD^2},\\
\pi_C^g=\kappa_C(S_P)\frac{\Theta(1-\lambda)^2\lambda q_C(q_C+\lambda Q)}{4QD^2},
\quad CS^g=\frac{\Theta\Psi_g}{8Q^2D^2}.
\end{gathered}
\]}
For \(S_P\in\{\{A,C\},\{A,B\},\{B,C\}\}\), platform profit is
\[
\Pi^{g,S_P}=\frac{\tau}{1-\tau}\sum_{i\in S_P}\pi_i^g.
\]

\subsection{Price and Profit Comparison Summaries}
Tables~\ref{tab:assistant_price_profit_summary} and~\ref{tab:agent_price_profit_summary} summarize the effects of assistance and delegation on prices and profits relative to the no-AI benchmark for the same assortment.

\begin{table}[h]
\centering
\caption{Summary of the AI Assistant's Effects on Prices and Profits}
\label{tab:assistant_price_profit_summary}
{\footnotesize
\setlength{\tabcolsep}{5pt}
\renewcommand{\arraystretch}{1.45}
\resizebox{0.98\textwidth}{!}{%
\begin{tabular}{ccccc}
\toprule
Assortment & Region & Prices & Firm profits & Platform profit \\
\midrule
\(S_P=\{A,C\}\)
& \(\{1,2a,2b\}\)
& \(p_A^{sr}<p_A^n,\; p_C^{sr}<p_C^n\)
& \(\pi_A^{sr}<\pi_A^n,\; \pi_C^{sr}<\pi_C^n\)
& \(\Pi^{sr}<\Pi^n\) \\
\midrule
\multirow[c]{3}{*}[-1.1ex]{\(S_P=\{A,B\}\)}
& \(1\)
& \(p_A^{s1}<p_A^n,\; p_B^{s1}<p_B^n\)
& \(\pi_A^{s1}<\pi_A^n,\; \pi_B^{s1}<\pi_B^n\)
& \(\Pi^{s1}<\Pi^n\) \\
\cmidrule(lr){2-5}
& \(2a\)
& \(p_A^{s2a}>p_A^n,\; p_B^{s2a}>p_B^n\)
& \(\pi_A^{s2a}>\pi_A^n,\;
     \pi_B^{s2a}>\pi_B^n\ \text{iff}\ \lambda<\bar\lambda_B^s\)
& \(\Pi^{s2a}>\Pi^n\) \\
\cmidrule(lr){2-5}
& \(2b\)
& \(p_A^{s2b}>p_A^n,\; p_B^{s2b}>p_B^n\)
& \(\pi_A^{s2b}>\pi_A^n,\;
     \pi_B^{s2b}>\pi_B^n\ \text{iff}\ \lambda>\underline\lambda_B^s\)
& \(\Pi^{s2b}>\Pi^n\) \\
\bottomrule
\end{tabular}}}

\vspace{2pt}
\parbox{0.96\textwidth}{\scriptsize\textit{Note.} If an iff condition fails,
the corresponding comparison is weakly reversed.}
\end{table}

\begin{table}[h]
\centering
\caption{Summary of the AI Agent's Effects on Prices and Profits}
\label{tab:agent_price_profit_summary}
{\footnotesize
\setlength{\tabcolsep}{5pt}
\renewcommand{\arraystretch}{1.35}
\resizebox{0.98\textwidth}{!}{%
\begin{tabular}{cccc}
\toprule
Assortment & Prices & Firm profits & Platform profit \\
\midrule
\(S_P=\{A,C\}\)
& \(p_C^g<p_C^n,\;
     p_A^g>p_A^n\ \text{iff}\ \lambda<\overline\lambda_{A1}^g\)
& \(\pi_C^g<\pi_C^n,\;
     \pi_A^g>\pi_A^n\ \text{iff}\ \lambda<\overline\lambda_{A2}^g\)
& \(\Pi^g>\Pi^n\ \text{iff}\
     Q>\underline Q^g,\ \lambda<\overline\lambda_P^g\) \\
\midrule
\(S_P=\{A,B\}\)
& \(p_i^g>p_i^n\ \text{iff}\ \lambda>\underline\lambda_{i1}^g,
     \; i\in\{A,B\}\)
& \(\pi_i^g>\pi_i^n\ \text{iff}\ \lambda>\underline\lambda_{i2}^g,
     \; i\in\{A,B\}\)
& \(\Pi^g>\Pi^n\ \text{iff}\ \lambda>\underline\lambda_P^g\) \\
\bottomrule
\end{tabular}}}

\vspace{2pt}
\parbox{0.96\textwidth}{\scriptsize\textit{Note.} If an iff condition fails,
the corresponding comparison is weakly reversed.}
\end{table}

\subsection{Threshold Definitions}

\subsubsection{Assistant thresholds under
\texorpdfstring{\(S_P=\{A,B\}\)}{S\_P = \{A,B\}}}
\mbox{}\par\noindent
\(\bar\lambda_B^s\) is, when such a root exists, the unique root in the
Type-2a \(\lambda\)-interval satisfying \(\Phi_{2a}\geq0\) of
\(G_B^{2a}(x,\delta,\lambda)=0\), where
\[
G_B^{2a}(x,\delta,\lambda)\equiv
\left[2(1-\delta(1-\lambda))+\delta x(2-\delta(1-\lambda))\right]^2
(3x+4-\lambda)^2-(1+x)(x+\lambda)E^2,
\]
\[
\Phi_{2a}=\lambda(2+\delta x)
[2+\delta(-4+x+\delta(2+(-4+\delta)x))]
-\delta^2[x^2+\lambda^2(2+\delta x)^2].
\]
Set \(\bar\lambda_B^s\) equal to the lower endpoint when
\(G_B^{2a}\leq0\) for all such \(\lambda\), and to the upper endpoint when
\(G_B^{2a}>0\) for all such \(\lambda\), with the lower endpoint used by
convention if no such \(\lambda\) exists.

\(\underline\lambda_{B}^s\) is, when such a root exists, the unique interior
root in the Type-2b \(\lambda\)-interval of
\(G_B^{2b}(x,\delta,\lambda)=0\), where
\[
G_B^{2b}(x,\delta,\lambda)\equiv
8(1+x)-\delta\bigl[16+2x(8+x)+\lambda^2(2+\delta x)
-2\lambda(2+x)(2+\delta x)+\delta x\{8+x(12+5x)\}\bigr].
\]
Set \(\underline\lambda_{B}^s\) equal to the lower endpoint when
\(G_B^{2b}>0\) throughout that interval and to the upper endpoint when
\(G_B^{2b}\leq0\) throughout that interval.

\(\bar\lambda_{CS}^s\) is, when such a root exists, the unique interior root
in the Type-2b \(\lambda\)-interval of
\[
\frac{\widehat\Psi_{2b}(x,\delta,\lambda)}{2K^2}
-\Xi(1+x,x+\lambda)=0.
\]
Set \(\bar\lambda_{CS}^s\) equal to the lower endpoint when the left-hand
side is negative throughout that interval.

\subsubsection{Agent thresholds under \texorpdfstring{\(S_P=\{A,C\}\)}{S\_P = \{A,C\}}}
\[
\underline Q^g
\equiv
\frac{q_C\bigl[9\delta-6+\sqrt{\delta(\delta+8)}\bigr]}{8(1-\delta)}
\quad\text{for }\delta\in\left[\frac{1}{2},1\right),
\qquad
\underline Q^g\equiv+\infty\quad\text{at }\delta=1.
\]

\[
\overline{\lambda}_{A1}^g
\equiv
\begin{cases}
\displaystyle
1-
\frac{
6\delta(1+x)
}{
3x+4+2\delta+\sqrt{
(3x+4+2\delta)^2
-3\delta\bigl[(x+4)(3x+4)-4\delta(1+x)\bigr]
}
},
&
\displaystyle
\delta<\frac{4+3x}{4(1+x)},
\\[1.2em]
0,
&
\displaystyle
\delta\geq\frac{4+3x}{4(1+x)}.
\end{cases}
\]

\(\overline{\lambda}_{A2}^g\) is the unique interior root, when one exists, of
\[
\frac{(1-\lambda)L^2(3x+4)^2}{16(x+1)^2D^2}=\delta.
\]
Set \(\overline{\lambda}_{A2}^g=0\) when \(\pi_A^g>\pi_A^{n,AC}\) never holds and \(\overline{\lambda}_{A2}^g=1\) when it holds throughout.

\(\overline{\lambda}_P^g\) is the unique interior root, when one exists, of
\[
\frac{(1-\lambda)\bigl[L^2+(1-\lambda)\lambda x(x+\lambda)\bigr](3x+4)^2}
{4(x+1)(5x+4)D^2}
=\delta.
\]
Set \(\overline{\lambda}_P^g=0\) when \(\Pi^g>\Pi^{n,AC}\) never holds and \(\overline{\lambda}_P^g=1\) when it holds throughout.

\(\underline{\lambda}_{CS}^g\) is the unique interior root, when one exists, of
\[
\frac{\widehat{\Psi}_g(x,\lambda)}{8D^2}-\Xi(1+x,x)=0.
\]
Set \(\underline{\lambda}_{CS}^g=1\) when \(CS^g>CS^{n,AC}\) never holds and \(\underline{\lambda}_{CS}^g=0\) when it holds throughout.

\subsubsection{Agent thresholds under \texorpdfstring{\(S_P=\{A,B\}\)}{S\_P = \{A,B\}}}
\[
\underline{\lambda}_{A1}^g
\equiv
\begin{cases}
0,
&
\displaystyle
\delta\leq\frac{4+3x}{4(1+x)},
\\[1.2em]
\displaystyle
1-
\frac{
24(1-\delta)(1+x)
}{
3x+8(1-\delta)+\sqrt{
\bigl[3x+8(1-\delta)\bigr]^2
-48(1-\delta)\bigl[3x-4(1-\delta)(1+x)\bigr]
}
},
&
\displaystyle
\delta>\frac{4+3x}{4(1+x)}.
\end{cases}
\]

\[
\underline{\lambda}_{B1}^g
\equiv
\frac{
6\delta x
}{
3x+4(1-\delta)
+\sqrt{\bigl[3x+4(1-\delta)\bigr]^2-12\delta(1-\delta)x}
}.
\]

\(\underline{\lambda}_{A2}^g\) is the unique interior root, when one exists, of
\[
\frac{L^2(3x+4-\lambda)^2}{16(x+1)^2D^2}=\delta.
\]
Set \(\underline{\lambda}_{A2}^g=1\) when \(\pi_A^g>\pi_A^{n,AB}\) never holds and \(\underline{\lambda}_{A2}^g=0\) when it holds throughout.

\(\underline{\lambda}_{B2}^g\) is the unique interior root, when one exists, of
\[
\frac{\lambda(x+\lambda)(3x+4-\lambda)^2}{(x+1)D^2}=\delta.
\]
Set \(\underline{\lambda}_{B2}^g=1\) when \(\pi_B^g>\pi_B^{n,AB}\) never holds and \(\underline{\lambda}_{B2}^g=0\) when it holds throughout.

\(\underline{\lambda}_P^g\) is the unique interior root, when one exists, of
\[
\frac{\bigl[L^2+4\lambda(x+\lambda)^2\bigr](3x+4-\lambda)^2}
{4(x+1)(5x+4+\lambda)D^2}
=\delta.
\]
Set \(\underline{\lambda}_P^g=1\) when \(\Pi^g>\Pi^{n,AB}\) never holds and \(\underline{\lambda}_P^g=0\) when it holds throughout.

\(\overline{\lambda}_{CS}^g\) is the unique interior root, when one exists, of
\[
\frac{\widehat{\Psi}_g(x,\lambda)}{8D^2}
-\Xi(1+x,x+\lambda)=0.
\]
Set \(\overline{\lambda}_{CS}^g=0\) when \(CS^g>CS^{n,AB}\) never holds and \(\overline{\lambda}_{CS}^g=1\) when it holds throughout.

\subsubsection{Platform-design thresholds}
\(\underline{\lambda}_P^*\) and \(\overline{\lambda}_P^*\) are,
respectively, the lower and upper thresholds defined by the roots of
\(G_P(x,\delta,\lambda)=0\) in \((0,1)\), allowing the two thresholds to
coincide at an interior double root, where
\[
G_P(x,\delta,\lambda)
\equiv
\frac{(1-\lambda)\bigl[L^2+4\lambda(x+\lambda)^2\bigr]}{4D^2}
-\frac{\delta(x+1)(5x+4)}{(3x+4)^2}.
\]
Set \(\underline{\lambda}_P^*=0\) when only the upper interior root exists,
and set both thresholds to zero when \(G_P(x,\delta,\lambda)<0\) for all
\(\lambda\in(0,1)\).
For fixed \(q_C\), let \(\underline Q_P^*\) be the unique \(Q\) satisfying
\(\underline{\lambda}_P^*=\overline{\lambda}_P^*\in(0,1)\) at
\(\delta=1/2\). As shown in Appendix~\ref{sec:proof}, there exists an
admissible pair
\((\delta,\lambda)\) such that
\(\Pi^{g,AB}>\Pi^{n,AC}\) iff \(Q>\underline Q_P^*\).

\subsection{Joint-Effect Conditions}

\subsubsection{AI assistant (Proposition~\ref{prop: assis_joint})}
Under \(S_P=\{A,C\}\), every assistant equilibrium satisfies
\[
\Pi^s<\Pi^{n,AC},\qquad CS^s>CS^{n,AC}.
\]
Under \(S_P=\{A,B\}\), the complete conditions are
\[
\begin{array}{c|c}
\text{condition}&(\Pi^s-\Pi^{n,AB},\ CS^s-CS^{n,AB})\\ \hline
\mathcal R_1&(<0,>0)\\
\mathcal R_{2a}&(>0,<0)\\
\mathcal R_{2b},\ \lambda<\bar\lambda_{CS}^s&(>0,>0)\\
\mathcal R_{2b},\ \lambda\geq\bar\lambda_{CS}^s&(>0,\leq0)
\end{array}
\]

\subsubsection{AI agent (Proposition~\ref{prop: agent_joint})}
Under \(S_P=\{A,C\}\),
\[
\begin{array}{c|c}
\text{condition}&(\Pi^g-\Pi^{n,AC},\ CS^g-CS^{n,AC})\\ \hline
Q>\underline Q^g,\ \lambda<\overline{\lambda}_P^g,\ \lambda>\underline{\lambda}_{CS}^g&(>0,>0)\\
\bigl(Q\leq\underline Q^g\ \text{or}\ \lambda\geq\overline{\lambda}_P^g\bigr),\ \lambda>\underline{\lambda}_{CS}^g&(\leq0,>0)\\
Q>\underline Q^g,\ \lambda<\overline{\lambda}_P^g,\ \lambda\leq\underline{\lambda}_{CS}^g&(>0,\leq0)\\
\bigl(Q\leq\underline Q^g\ \text{or}\ \lambda\geq\overline{\lambda}_P^g\bigr),\ \lambda\leq\underline{\lambda}_{CS}^g&(\leq0,\leq0)
\end{array}
\]
Under \(S_P=\{A,B\}\), using
\(\overline{\lambda}_{CS}^g\leq\underline{\lambda}_P^g\),
\[
\begin{array}{c|c}
\text{condition}&(\Pi^g-\Pi^{n,AB},\ CS^g-CS^{n,AB})\\ \hline
\lambda>\underline{\lambda}_P^g&(>0,<0)\\
\lambda<\overline{\lambda}_{CS}^g&(<0,>0)\\
\overline{\lambda}_{CS}^g\leq\lambda\leq\underline{\lambda}_P^g&(\leq0,\leq0)
\end{array}
\]

\subsubsection{Preferred-design comparisons
(Proposition~\ref{prop: platform design})}
The following preferred-design comparisons apply on the common equilibrium-existence domain of Proposition~\ref{prop: platform design}.
Let \(I_P^*\equiv(\underline{\lambda}_P^*,\overline{\lambda}_P^*)\). Then
\[
\begin{array}{c|c}
\text{condition}&(\Pi^*-\Pi^{n*},\ CS^*-CS^{n*})\\ \hline
Q>\underline Q_P^*,\ \lambda\in I_P^*,\
\lambda>\underline{\lambda}_{CS}^g&(>0,>0)\\
Q>\underline Q_P^*,\ \lambda\in I_P^*,\
\lambda\leq\underline{\lambda}_{CS}^g&(>0,\leq0)\\
\bigl(Q\leq\underline Q_P^*\ \text{or}\ \lambda\notin I_P^*\bigr),\
\lambda>\underline{\lambda}_{CS}^g&(\leq0,>0)\\
\bigl(Q\leq\underline Q_P^*\ \text{or}\ \lambda\notin I_P^*\bigr),\
\lambda\leq\underline{\lambda}_{CS}^g&(\leq0,\leq0)
\end{array}
\]

\newpage
\section{Proofs of Statements}\label{sec:proof}
We present the proofs of the lemmas and propositions in this appendix.
At any price profile \(\mathbf p\) and fixed assortment \(S_P\), actual firm and platform profits are
\[
\pi_i(\mathbf p;S_P)=\kappa_i(S_P)p_iD_i(\mathbf p),\qquad
\Pi(\mathbf p;S_P)=\tau\sum_{i\in S_P}p_iD_i(\mathbf p),
\]
where \(\kappa_i(S_P)=1-\tau\) for a hosted firm and \(1\) otherwise. We use normalized profits in the proofs to simplify the calculations:
\[
\widehat{\pi}_i(\mathbf p;S_P)\equiv
\frac{\pi_i(\mathbf p;S_P)}{\kappa_i(S_P)}=p_iD_i(\mathbf p),\qquad
\widehat{\Pi}(\mathbf p;S_P)\equiv\frac{\Pi(\mathbf p;S_P)}{\tau}
=\sum_{i\in S_P}\widehat{\pi}_i(\mathbf p;S_P).
\]
Since \(\kappa_i(S_P)>0\) is fixed for each firm within an assortment, normalization leaves equilibrium prices and comparisons of that firm's profit across modes unchanged. For any assortment, the hosted firms' combined actual profit is \((1-\tau)\widehat{\Pi}\), and platform profit is \(\tau\widehat{\Pi}\). Because \(\tau\) is fixed across assortments, all three measures give the same assortment ranking. The normalized platform-profit comparisons below retain their direction when stated in actual profits; Appendix~\ref{sec: technical details} presents them in that form.
The finite upper bound ensures compact strategy sets and is chosen to be nonbinding.
Throughout, we take \(p^{\max}\geq \frac{\Theta q_A}{\delta}\) and
restrict attention to prices in
\([0,\frac{\Theta q_A}{\delta}]\). This restriction is without loss
because, for any \(p_i>\frac{\Theta q_A}{\delta}\), the highest possible quality-salient,
price-salient, and true utilities are, respectively,
\(\Theta q_A-\delta p_i<0\),
\(\delta\Theta q_A-p_i<0\), and
\(\Theta q_A-p_i<0\). Such a price therefore generates zero demand in
every region and cannot be a strictly profitable deviation.

\noindent\textit{Proof of Lemma \ref{lem:no AI_equil}:} Consider an on-platform assortment $S_P=\{i,j\}$, where products $i$ and $j$ are characterized by price-quality pairs $(p_i,q_i)$ and $(p_j,q_j)$, respectively, with $q_i>q_j$. We first note that no equilibrium can involve a zero price. If exactly one firm charges zero, its product is purchased by a positive measure of consumers: product $i$ dominates when $p_i=0<p_j$, whereas product $j$ is purchased by sufficiently low types when $p_j=0<p_i$. In either case, a sufficiently small price increase retains positive demand and gives the zero-priced firm strictly positive profit. If both prices are zero, both firms earn zero profit, whereas a deviation by firm $i$ to any $p_i'\in(0,\delta\Theta(q_i-q_j))$ attracts a positive measure of high types and gives firm $i$ strictly positive profit. Thus both prices are strictly positive at any equilibrium. Moreover, no equilibrium can have $p_i\leq p_j$: product $i$ then has both higher quality and a weakly lower price, so product $j$ has zero demand and can profitably deviate to a sufficiently small positive price. Hence it is enough to analyze the remaining profiles with $p_i>p_j>0$ in three subcases.

\noindent\textbf{(i)} $q_i/p_i>q_j/p_j$, implying that quality is salient in assortment $S_P$. Consumers' perceived utilities from purchasing products $i$ and $j$ are
$\tilde{u}_i(\theta\mid S_P)=\theta q_i-\delta p_i$ and
$\tilde{u}_j(\theta\mid S_P)=\theta q_j-\delta p_j$. Consumers purchase product $i$ if it provides the highest nonnegative utility, i.e.,
$\tilde{u}_i(\theta\mid S_P)\geq \tilde{u}_j(\theta\mid S_P)$ and $\tilde{u}_i(\theta\mid S_P)\geq 0$,
which implies $\theta \geq \max\left\{\frac{\delta(p_i-p_j)}{q_i-q_j},\frac{\delta p_i}{q_i}\right\}$. Consumers purchase product $j$ if $\tilde{u}_j(\theta\mid S_P) > \tilde{u}_i(\theta\mid S_P)$ and
$\tilde{u}_j(\theta\mid S_P)\geq 0$, which requires
$\frac{\delta p_j}{q_j} \leq \theta < \frac{\delta(p_i-p_j)}{q_i-q_j}$. However, when $q_i/p_i>q_j/p_j$, we have
$\frac{\delta(p_i-p_j)}{q_i-q_j} < \frac{\delta p_i}{q_i} < \frac{\delta p_j}{q_j}$.
Therefore, the interval $\frac{\delta p_j}{q_j} \leq \theta < \frac{\delta(p_i-p_j)}{q_i-q_j}$ is empty, implying that no consumer purchases product $j$. Hence its normalized profit is $\widehat{\pi}_j=0$. For product $i$, demand is $D_i=1-\frac{\delta p_i}{\Theta q_i}$ if $\frac{\delta p_i}{q_i}\le \Theta$, and zero otherwise. The corresponding normalized profit is $\widehat{\pi}_i=(1-\frac{\delta p_i}{\Theta q_i})p_i$ if $\frac{\delta p_i}{q_i}\leq \Theta$, and zero otherwise.

\noindent\textbf{(ii)} $q_i/p_i=q_j/p_j$, implying that quality and price are equally salient in $S_P$. Consumers' perceived utilities from purchasing products $i$ and $j$ are
$\tilde{u}_i(\theta\mid S_P)=\theta q_i-p_i$ and
$\tilde{u}_j(\theta\mid S_P)=\theta q_j-p_j$. Consumers purchase product $i$ if it provides the highest nonnegative utility, i.e.,
$\tilde{u}_i(\theta\mid S_P)\geq \tilde{u}_j(\theta\mid S_P)$ and $\tilde{u}_i(\theta\mid S_P)\geq 0$,
which implies $\theta \geq \max\left\{\frac{p_i-p_j}{q_i-q_j},p_i/q_i\right\}$. Consumers purchase product $j$ if $\tilde{u}_j(\theta\mid S_P) > \tilde{u}_i(\theta\mid S_P)$ and
$\tilde{u}_j(\theta\mid S_P)\geq 0$, which requires
$p_j/q_j \leq \theta < \frac{p_i-p_j}{q_i-q_j}$. However, when $q_i/p_i=q_j/p_j$, we have
$\frac{p_i-p_j}{q_i-q_j} = p_i/q_i = p_j/q_j$. Therefore, the interval $p_j/q_j \leq \theta < \frac{p_i-p_j}{q_i-q_j}$ is empty, implying that no consumer purchases product $j$. Hence its normalized profit is $\widehat{\pi}_j=0$. For product $i$, demand is $D_i=1-\frac{p_i}{\Theta q_i}$ if $p_i/q_i< \Theta$, and zero otherwise. The corresponding normalized profit is $\widehat{\pi}_i=(1-\frac{p_i}{\Theta q_i})p_i$ if $p_i/q_i<\Theta$, and zero otherwise.

\noindent\textbf{(iii)} $q_i/p_i<q_j/p_j$, implying that price is salient in $S_P$. Consumers' perceived utilities from purchasing products $i$ and $j$ are
$\tilde{u}_i(\theta\mid S_P)=\theta \delta q_i-p_i$ and
$\tilde{u}_j(\theta\mid S_P)=\theta \delta q_j-p_j$. Consumers purchase product $i$ if it provides the highest nonnegative utility, i.e.,
$\tilde{u}_i(\theta\mid S_P)\geq \tilde{u}_j(\theta\mid S_P)$ and $\tilde{u}_i(\theta\mid S_P)\geq 0$,
which implies $\theta \geq \max\left\{\frac{p_i-p_j}{\delta(q_i-q_j)},\frac{p_i}{\delta q_i}\right\}$. Consumers purchase product $j$ if $\tilde{u}_j(\theta\mid S_P) > \tilde{u}_i(\theta\mid S_P)$ and
$\tilde{u}_j(\theta\mid S_P)\geq 0$, which requires
$\frac{p_j}{\delta q_j} \leq \theta < \frac{p_i-p_j}{\delta(q_i-q_j)}$. When $q_i/p_i<q_j/p_j$, we have $\frac{p_j}{\delta q_j}<\frac{p_i}{\delta q_i}<\frac{p_i-p_j}{\delta(q_i-q_j)}$. Thus, the demand for product $i$ is $D_i=1-\frac{p_i-p_j}{\delta \Theta(q_i-q_j)}$ if $\frac{p_i-p_j}{\delta(q_i-q_j)}<\Theta$, otherwise it is zero. The corresponding normalized profit is $\widehat{\pi}_i=[1-\frac{p_i-p_j}{\delta \Theta(q_i-q_j)}]p_i$ if $\frac{p_i-p_j}{\delta(q_i-q_j)}\leq \Theta$, and zero otherwise. For product $j$, its demand is $D_j=1-\frac{p_j}{\delta \Theta q_j}$ if $\frac{p_j}{\delta q_j}<\Theta\leq \frac{p_i-p_j}{\delta(q_i-q_j)}$, $D_j=\frac{p_i-p_j}{\delta \Theta(q_i-q_j)}-\frac{p_j}{\delta \Theta q_j}$ if $\frac{p_i-p_j}{\delta(q_i-q_j)}<\Theta$, otherwise it is zero. The corresponding normalized profit is $\widehat{\pi}_j=(1-\frac{p_j}{\delta \Theta q_j})p_j$ if $\frac{p_j}{\delta q_j}<\Theta\leq \frac{p_i-p_j}{\delta(q_i-q_j)}$, $\widehat{\pi}_j=[\frac{p_i-p_j}{\delta \Theta(q_i-q_j)}-\frac{p_j}{\delta \Theta q_j}]p_j$ if $\frac{p_i-p_j}{\delta(q_i-q_j)}<\Theta$, and zero otherwise.

We next examine firms' pricing incentives. In cases (i) and (ii), product $j$ receives zero demand, so firm $j$ earns zero profit. Because production cost is zero, firm $j$ can profitably deviate by choosing any sufficiently small $p_j'>0$. Indeed, choosing $p_j'<\min\{\frac{p_iq_j}{q_i},\delta\Theta q_j\}$ makes price salient and gives product $j$ a nonempty purchase interval, since $\frac{p_j'}{\delta q_j}<\min\{\Theta,\frac{p_i-p_j'}{\delta(q_i-q_j)}\}$. Hence firm $j$ obtains strictly positive demand and profit, contradicting equilibrium. Therefore, no equilibrium can arise in cases (i) or (ii). Now consider case (iii). If $\Theta\leq \frac{p_i-p_j}{\delta(q_i-q_j)}$, then product $i$ receives zero demand and firm $i$ earns zero profit. Since $p_j>0$, firm $i$ can choose a sufficiently small $p_i'\in(0,\min\{p_j,\delta\Theta q_i\})$. Product $i$ then has both higher quality and a lower price, so it attracts a positive measure of consumers and earns strictly positive profit. Hence, case (iii) can arise in equilibrium only if $\frac{p_i-p_j}{\delta(q_i-q_j)}<\Theta$, under which both firms receive positive demand. In this case, $\widehat{\pi}_i=[1-\frac{p_i-p_j}{\delta\Theta(q_i-q_j)}]p_i$ and $\widehat{\pi}_j=[\frac{p_i-p_j}{\delta\Theta(q_i-q_j)}-\frac{p_j}{\delta\Theta q_j}]p_j$. The first-order conditions are
$\frac{\partial \widehat{\pi}_i}{\partial p_i}=0$ and $\frac{\partial \widehat{\pi}_j}{\partial p_j}=0$, which yield
$p_i=\frac{\delta\Theta(q_i-q_j)+p_j}{2}$ and
$p_j=\frac{p_i q_j}{2q_i}$.
Solving these two equations jointly gives
$p_i^n=\frac{2\delta\Theta q_i(q_i-q_j)}{4q_i-q_j}$ and
$p_j^n=\frac{\delta\Theta(q_i-q_j)q_j}{4q_i-q_j}$.
Moreover,
$\frac{p_j^n}{\delta q_j}<\frac{p_i^n}{\delta q_i}<\frac{p_i^n-p_j^n}{\delta(q_i-q_j)}<\Theta$,
so the candidate equilibrium indeed lies in case (iii). Since $\frac{\partial^2 \widehat{\pi}_i}{\partial p_i^2}<0$ and $\frac{\partial^2 \widehat{\pi}_j}{\partial p_j^2}<0$, the normalized profit functions are strictly concave in own prices within this subcase. It remains to check deviations that change the salience or demand configuration.

First hold $p_j=p_j^n$ fixed and let firm $i$ choose any $y\in[0,p^{\max}]$. On the ordered-price domain $y>p_j^n$, the salience boundary is $(q_i/q_j)p_j^n=p_i^n/2$. If $y>p_i^n/2$ and firm $i$ has positive demand, its price-salient normalized profit satisfies
\[
\widehat{\pi}_i(y,p_j^n)-\widehat{\pi}_i(p_i^n,p_j^n)
=-\frac{(y-p_i^n)^2}{\delta\Theta(q_i-q_j)}\leq0;
\]
if firm $i$ has zero demand, its deviation profit is zero and the same conclusion follows. For $0\leq y\leq p_i^n/2$, firm $i$'s demand is no greater than its quality-salient monopoly demand, so its normalized profit is bounded above by $y(1-\frac{\delta y}{\Theta q_i})$. This bound is increasing over the relevant interval, and at its upper endpoint its difference from firm $i$'s candidate normalized profit is
\[
\frac{p_i^n}{2}\left(1-\frac{\delta p_i^n}{2\Theta q_i}\right)-\widehat{\pi}_i(p_i^n,p_j^n)
=-\frac{\delta\Theta q_i(q_i-q_j)[q_j+\delta^2(q_i-q_j)]}{(4q_i-q_j)^2}<0.
\]
This argument also covers deviations with $y\leq p_j^n$, where product $i$ has both higher quality and a weakly lower price, as well as the salience tie at $y=p_i^n/2$. Hence firm $i$ has no profitable deviation.

Next hold $p_i=p_i^n$ fixed. For $0<y<(q_j/q_i)p_i^n=2p_j^n$, price is salient, both firms have positive demand, and
\[
\widehat{\pi}_j(y,p_i^n)-\widehat{\pi}_j(p_j^n,p_i^n)
=-\frac{q_i(y-p_j^n)^2}{\delta\Theta q_j(q_i-q_j)}\leq0.
\]
At $y=0$, firm $j$ earns zero profit. At $y=2p_j^n$, salience is tied and product $j$ has zero demand. For $y>2p_j^n$, product $j$ either has zero demand in the quality-salient configuration or, if $y\geq p_i^n$, is dominated by product $i$. Thus firm $j$ also has no profitable deviation, and $(p_i^n,p_j^n)$ is a Nash equilibrium.

Finally, the preceding zero-price and zero-demand arguments imply that any pure-strategy equilibrium must be an interior, price-salient profile at which both firms have positive demand. It must therefore satisfy the two first-order conditions above, whose unique solution is $(p_i^n,p_j^n)$. Hence this candidate is the unique pure-strategy equilibrium.

Substituting the equilibrium prices into the normalized profit and consumer-surplus
expressions gives
$\widehat{\pi}_i^n=\frac{4\delta\Theta q_i^2(q_i-q_j)}{(4q_i-q_j)^2}$,
$\widehat{\pi}_j^n=\frac{\delta\Theta q_i(q_i-q_j)q_j}{(4q_i-q_j)^2}$, and
$CS^n=\frac{\Theta q_i[(12-8\delta)q_i^2-(1-6\delta)q_iq_j-2(1-\delta)q_j^2]}{2(4q_i-q_j)^2}$.

We reparametrize the quality structure as $Q\equiv q_A-q_C$, $\lambda\equiv\frac{q_B-q_C}{q_A-q_C}\in(0,1)$. Thus, the equilibrium prices, firm normalized profits, normalized platform profit, and consumer surplus under the three possible two-product assortments are as follows. Define
\[
\Xi(x,y)\equiv
\frac{x\left[(12-8\delta)x^2-(1-6\delta)xy-2(1-\delta)y^2\right]}
{2(4x-y)^2}.
\]

For \(S_P=\{A,C\}\), the equilibrium outcomes are
{\footnotesize
\[
p_A^{n,AC}=\frac{2\delta\Theta(q_C+Q)Q}{3q_C+4Q},\quad
p_C^{n,AC}=\frac{\delta\Theta Qq_C}{3q_C+4Q};
\]
\[
\begin{gathered}
\widehat{\pi}_A^{n,AC}=\frac{4\delta\Theta(q_C+Q)^2Q}{(3q_C+4Q)^2},\quad
\widehat{\pi}_C^{n,AC}=\frac{\delta\Theta(q_C+Q)Qq_C}{(3q_C+4Q)^2},\\
\widehat{\Pi}^{n,AC}=\frac{\delta\Theta Q(q_C+Q)(5q_C+4Q)}{(3q_C+4Q)^2},\quad
CS^{n,AC}=\Theta\Xi(q_C+Q,q_C).
\end{gathered}
\]}

For \(S_P=\{A,B\}\), the equilibrium outcomes are
{\footnotesize
\[
p_A^{n,AB}=\frac{2\delta\Theta(q_C+Q)(1-\lambda)Q}{3q_C+(4-\lambda)Q},\quad
p_B^{n,AB}=\frac{\delta\Theta(1-\lambda)Q(q_C+\lambda Q)}{3q_C+(4-\lambda)Q};
\]
\[
\begin{gathered}
\widehat{\pi}_A^{n,AB}=\frac{4\delta\Theta(q_C+Q)^2(1-\lambda)Q}{[3q_C+(4-\lambda)Q]^2},\quad
\widehat{\pi}_B^{n,AB}=\frac{\delta\Theta(q_C+Q)(1-\lambda)Q(q_C+\lambda Q)}{[3q_C+(4-\lambda)Q]^2},\\
\widehat{\Pi}^{n,AB}=\frac{\delta\Theta(1-\lambda)Q(q_C+Q)[5q_C+(4+\lambda)Q]}{[3q_C+(4-\lambda)Q]^2},\quad
CS^{n,AB}=\Theta\Xi(q_C+Q,q_C+\lambda Q).
\end{gathered}
\]}

For \(S_P=\{B,C\}\), the equilibrium outcomes are
{\footnotesize
\[
p_B^{n,BC}=\frac{2\delta\Theta(q_C+\lambda Q)\lambda Q}{3q_C+4\lambda Q},\quad
p_C^{n,BC}=\frac{\delta\Theta\lambda Qq_C}{3q_C+4\lambda Q};
\]
\[
\begin{gathered}
\widehat{\pi}_B^{n,BC}=\frac{4\delta\Theta(q_C+\lambda Q)^2\lambda Q}{(3q_C+4\lambda Q)^2},\quad
\widehat{\pi}_C^{n,BC}=\frac{\delta\Theta(q_C+\lambda Q)\lambda Qq_C}{(3q_C+4\lambda Q)^2},\\
\widehat{\Pi}^{n,BC}=\frac{\delta\Theta\lambda Q(q_C+\lambda Q)(5q_C+4\lambda Q)}{(3q_C+4\lambda Q)^2},\quad
CS^{n,BC}=\Theta\Xi(q_C+\lambda Q,q_C).
\end{gathered}
\]}
For each \(S_P=\{i,j\}\), the corresponding actual equilibrium profits are
\(\pi_i^{n,ij}=(1-\tau)\widehat{\pi}_i^{n,ij}\),
\(\pi_j^{n,ij}=(1-\tau)\widehat{\pi}_j^{n,ij}\), and
\(\Pi^{n,ij}=\tau(\widehat{\pi}_i^{n,ij}+\widehat{\pi}_j^{n,ij})\).
\(\Halmos\)\vspace{3mm}

\noindent\textit{Proof of Lemma \ref{lem: assis_equil}:} We proceed in
three steps.  First, we characterize the AI assistant's recommendation rule
and consumers' purchase decisions conditional on the recommendation set.
Second, we identify all recommendation--salience configurations that can
support strictly positive demand for all three firms.  Third, we solve the
pricing problems in the remaining configurations, impose every local and
global no-deviation condition, and derive the equilibrium outcomes.
Throughout the proof, the assistant and consumers choose the higher-quality
product in the event of a tie.  Since a consumer-type tie has measure zero,
this convention does not affect demand or profit.  We maintain
\(q_A>q_B>q_C>0\) and \(\Theta>0\).
We restrict attention to \(\delta\in[1/2,1]\).  Under this restriction,
the non-salient dimension retains at least half the weight of the salient
dimension.  It also limits the number of relevant configurations: as Step 2
shows, several additional positive-demand configurations can arise only when
\(\delta<1/2\).
This preserves tractability: we do not characterize the additional equilibrium candidates and potential multiplicity for \(\delta<1/2\).
We use
\(Q\equiv q_A-q_C>0\) and
\(\lambda\equiv\frac{q_B-q_C}{Q}\in(0,1)\), so that
\(q_A=q_C+Q\) and \(q_B=q_C+\lambda Q\), throughout.

\medskip
\noindent\textbf{Step 1: The recommendation and purchase rules.}
Under assistance, the assistant searches the full
product set \(\{A,B,C\}\) and recommends to each consumer the two products that yield
the highest true utilities.  Consumer \(\theta\)'s true utility from product
\(i\) is \(u_i(\theta)=\theta q_i-p_i\).  Depending on the ranking of these
utilities, there are three possible recommendation sets:
\begin{enumerate}[\upshape(i)]
\item It recommends \(\{A,B\}\) if
\(\theta\geq\max\{\frac{p_A-p_C}{q_A-q_C},
\frac{p_B-p_C}{q_B-q_C}\}\).
\item It recommends \(\{A,C\}\) if
\(\frac{p_A-p_B}{q_A-q_B}\leq\theta<\frac{p_B-p_C}{q_B-q_C}\).
\item It recommends \(\{B,C\}\) if
\(\theta<\min\{\frac{p_A-p_B}{q_A-q_B},
\frac{p_A-p_C}{q_A-q_C}\}\).
\end{enumerate}
The three recommendation thresholds are not independent because
\[
\frac{p_A-p_C}{q_A-q_C}
=\frac{q_A-q_B}{q_A-q_C}\frac{p_A-p_B}{q_A-q_B}
+\frac{q_B-q_C}{q_A-q_C}\frac{p_B-p_C}{q_B-q_C}.
\]
Thus, \(\frac{p_A-p_C}{q_A-q_C}\) lies between the other two thresholds.
There are only two strict recommendation orderings.  If
\begin{equation}
\frac{p_A-p_B}{q_A-q_B}>
\frac{p_A-p_C}{q_A-q_C}>
\frac{p_B-p_C}{q_B-q_C},
\label{eq:rec-case1}
\end{equation}
the assistant recommends \(\{B,C\}\) below
\(\frac{p_A-p_C}{q_A-q_C}\) and \(\{A,B\}\) above it.  If all inequalities
in~\eqref{eq:rec-case1} are reversed, there is an intermediate
\(\{A,C\}\) interval.  The equality case has no intermediate interval.

{Within a recommended pair \(\{i,j\}\), where \(q_i>q_j\),
substituting the pair averages into the salience function gives the
pairwise salience values \(S_q(i,j)=\frac{q_i-q_j}{q_i+q_j}\) and
\(S_p(i,j)=\frac{|p_i-p_j|}{p_i+p_j}\) if \(p_i+p_j>0\), with
\(S_p(i,j)=0\) when both prices are zero.  These values are the same
for both products.  Quality is salient when \(S_q(i,j)>S_p(i,j)\);
if \(p_i>p_j>0\), this is equivalent to \(q_i/p_i>q_j/p_j\).}
Consumers then choose \(i\) when
\[
\theta\geq\max\left\{\frac{\delta p_i}{q_i},
\frac{\delta(p_i-p_j)}{q_i-q_j}\right\},
\]
and choose \(j\) when
\(\frac{\delta p_j}{q_j}\leq\theta<\frac{\delta(p_i-p_j)}{q_i-q_j}\).
{Price is salient when \(S_q(i,j)<S_p(i,j)\); if
\(p_i>p_j>0\), this is equivalent to \(q_i/p_i<q_j/p_j\).}
Consumers then choose \(i\) when
\[
\theta\geq\max\left\{\frac{p_i}{\delta q_i},
\frac{p_i-p_j}{\delta(q_i-q_j)}\right\},
\]
and choose \(j\) when
\(\frac{p_j}{\delta q_j}\leq\theta<\frac{p_i-p_j}{\delta(q_i-q_j)}\).
{At \(S_q(i,j)=S_p(i,j)\), perceived utility coincides with true
utility.}  Intersecting
these purchase intervals with the recommendation intervals above gives the
demands in each case.

We next show that every firm must have positive demand in equilibrium.  First,
all equilibrium prices are strictly positive.  Otherwise, the highest-quality
firm \(i\) among those charging zero can raise its price to a sufficiently
small \(\varepsilon>0\).  Positive-priced rivals have negative true and
perceived utility on an interval of types close to zero.  For every
lower-quality zero-priced rival \(j\), product \(i\) is both recommended and
preferred once
\(\theta>\max\{\frac{\varepsilon}{q_i-q_j},
\frac{\varepsilon}{\delta(q_i-q_j)}\}\).  Taking \(\varepsilon\) small makes
these lower bounds lie inside the low-type interval and also guarantees
participation.  Thus the deviating firm is purchased by a positive measure
of consumers and earns positive profit instead of zero.
Second, suppose that all prices are positive but firm \(i\) has zero demand.
Choose an interval \([a,b]\subset(0,\Theta)\) sufficiently close to zero
that both rivals have negative true and perceived utility throughout it.
After firm \(i\) deviates to any sufficiently small
\(\varepsilon<\delta aq_i\), its true and perceived utility is positive on
\([a,b]\), regardless of which attribute is salient.  It is therefore
recommended and purchased on this positive-measure interval and earns
positive profit, again a contradiction.  Hence every pure-strategy
equilibrium has strictly positive prices and demands for all three firms.
{Furthermore, if \(q_i>q_j\) but \(p_i\leq p_j\), both products
have the same salience weights in their pair and
\(\widetilde u_i(\theta)-\widetilde u_j(\theta)
=\alpha\theta(q_i-q_j)+\beta(p_j-p_i)>0\) for almost every type,
where \((\alpha,\beta)\in\{(1,\delta),(\delta,1),(1,1)\}\).
Product \(i\) also weakly dominates \(j\) in true utility, so any
recommendation containing \(j\) must contain \(i\).  Therefore
\(D_j=0\), contradicting positive demand.  Every equilibrium thus has
\(p_A>p_B>p_C>0\); on this ordered domain the quality-to-price
comparisons used in Steps 2--3 are equivalent to the pairwise salience
values.}

\medskip
\noindent\textbf{Step 2: The three possible candidates.}
We now consider the two strict recommendation orderings and their common
boundary.  The case analysis serves two
purposes.  It first identifies the product, if any, that has zero demand in
each recommendation--salience configuration.  It then divides every
all-price-salient configuration into its smooth endpoint cases and checks
the intervening boundaries.  Table~\ref{tab:assistant_case_summary}
summarizes the conclusions; the resulting demand and normalized profit expressions
follow the table.

\begin{table}[!ht]
\centering
\caption{Summary of recommendation--salience configurations}
\label{tab:assistant_case_summary}
{\footnotesize
\setlength{\tabcolsep}{4pt}
\renewcommand{\arraystretch}{1.35}
\begin{tabular}{c p{0.43\textwidth} p{0.36\textwidth}}
\hline
Case & Salience or endpoint condition & Conclusion \\
\hline
1(i) & Quality salient in both recommended pairs & \(D_C=0\) \\
1(ii) & Price salient in \(\{A,B\}\), quality salient in \(\{B,C\}\)
& \(D_C=0\) \\
1(iii-a) & Price salient; \(\theta^{\mathrm{rec}}<z_B<m\)
& The first-order conditions and endpoint ordering require
\(\delta<1/2\) \\
1(iii-b) & Price salient; \(z_B<m<\theta^{\mathrm{rec}}\) & Type 1 \\
1(iii-c) & Price salient; \(z_B<\theta^{\mathrm{rec}}<m\) & Type 2a \\
2(i) & Quality salient in all three pairs & \(D_C=0\) \\
2(ii)--(iii) & Mixed quality and price salience & \(D_B=0\) \\
2(iv-a)--(iv-b) & Price salient; the two positive-demand endpoint cases
& The first-order conditions and endpoint ordering require
\(\delta<1/2\) \\
2(iv-c) & Boundary \(z_B=\delta m\) & No kink maximum for
\(\delta\geq1/2\) \\
3(i) & Equal quality-to-price ratios & \(D_B=0\) \\
3(ii) & Quality salient in both recommended pairs & \(D_C=0\) \\
3(iii) & Price salient in both recommended pairs & Upward derivative jump
if \(\delta<1\); \(D_B=0\) if \(\delta=1\) \\
Boundaries & \(m=\theta^{\mathrm{rec}}\), salience ties, and
\(z_B=\theta^{\mathrm{rec}}\) & Only \(z_B=\theta^{\mathrm{rec}}\)
survives, giving Type 2b \\
\hline
\end{tabular}}
\end{table}

\smallskip
\noindent\textbf{Case 1:}
Suppose~\eqref{eq:rec-case1} holds.  The assistant recommends \(\{B,C\}\)
when \(\theta<\frac{p_A-p_C}{Q}\) and \(\{A,B\}\) when
\(\theta\geq\frac{p_A-p_C}{Q}\).  The pairwise quality-to-price comparisons and
the Case-1 threshold inequalities leave four of the six strict rankings of
\((q_A/p_A,q_B/p_B,q_C/p_C)\):
\[
\frac{q_A}{p_A}>\frac{q_B}{p_B}>\frac{q_C}{p_C},\qquad
\frac{q_B}{p_B}>\frac{q_A}{p_A}>\frac{q_C}{p_C},
\]
\[
\frac{q_B}{p_B}>\frac{q_C}{p_C}>\frac{q_A}{p_A},\qquad
\frac{q_C}{p_C}>\frac{q_B}{p_B}>\frac{q_A}{p_A}.
\]
The second and third rankings are equivalent for demand because both have
\(q_B/p_B\) larger than the other two ratios and \(\{A,C\}\) is never
recommended.  Hence there are three economically distinct salience
configurations, listed next.

\textbf{(i)} If \(q_A/p_A>q_B/p_B>q_C/p_C\), quality is salient in both
pairs.
Within \(\{B,C\}\), firm \(C\)'s purchase interval would be
\[
\frac{\delta p_C}{q_C}\leq\theta<
\frac{\delta(p_B-p_C)}{q_B-q_C}.
\]
But \(q_B/p_B>q_C/p_C\) implies
\(\frac{\delta p_C}{q_C}>\frac{\delta(p_B-p_C)}{q_B-q_C}\), so the interval is empty
and \(D_C=0\).

\textbf{(ii)} If \(q_A/p_A<q_B/p_B\) but
\(q_B/p_B>q_C/p_C\), price is salient in \(\{A,B\}\) and quality is
salient in \(\{B,C\}\).  Here \(C\)'s purchase interval is empty because
\(q_B/p_B>q_C/p_C\) implies
\(\frac{\delta p_C}{q_C}>\frac{\delta(p_B-p_C)}{q_B-q_C}\), so no consumer can lie
between \(C\)'s participation and indifference thresholds.

\textbf{(iii)} Suppose
\begin{equation}
\frac{q_A}{p_A}<\frac{q_B}{p_B}<\frac{q_C}{p_C}.
\label{eq:all-price-salient}
\end{equation}
Price is then salient in both recommendation pairs.  Define the purchase
thresholds
\[
h\equiv\frac{p_A-p_B}{\delta(q_A-q_B)},\qquad
m\equiv\frac{p_B-p_C}{\delta(q_B-q_C)},
\]
the recommendation threshold
\(\theta^{\mathrm{rec}}\equiv\frac{p_A-p_C}{Q}\), and the participation thresholds
\(z_B\equiv\frac{p_B}{\delta q_B}\) and
\(z_C\equiv\frac{p_C}{\delta q_C}\).  The \(A\)--\(B\) and \(B\)--\(C\) recommendation thresholds
are \(\delta h\) and \(\delta m\), so Case 1 is equivalently
\(\delta h>\theta^{\mathrm{rec}}>\delta m\).
Moreover, \(q_B/p_B<q_C/p_C\) is equivalent to \(z_B<m\), and
\(q_A/p_A<q_B/p_B\) is equivalent to \(z_B<h\).  Hence the
all-price-salient part of Case 1 can be divided by comparing only
\(\theta^{\mathrm{rec}}\) with \(z_B\) and \(m\).  The strict trichotomy is
\[
\theta^{\mathrm{rec}}<z_B<m,\qquad
z_B<\theta^{\mathrm{rec}}<m,\qquad
z_B<m<\theta^{\mathrm{rec}}.
\]
The equalities \(\theta^{\mathrm{rec}}=z_B\) and
\(\theta^{\mathrm{rec}}=m\) are the intervening faces.  Thus no other
positive-demand endpoint ordering is omitted.

Firm \(A\) is purchased above \(h\).  In the low recommendation group,
firm \(C\) is purchased from \(z_C\) up to
\(\min\{m,\theta^{\mathrm{rec}}\}\), and firm \(B\) is purchased above
\(m\) when \(m<\theta^{\mathrm{rec}}\).  In the high recommendation
group, \(B\) is purchased from
\(\max\{z_B,\theta^{\mathrm{rec}}\}\) up to \(h\).  Thus, when the
relevant endpoints lie in \((0,\Theta)\),
\begin{equation}
\begin{aligned}
D_A&=\frac{\Theta-h}{\Theta},\\
D_B&=\frac{[\theta^{\mathrm{rec}}-\max\{m,z_B\}]_+
+[h-\max\{\theta^{\mathrm{rec}},z_B\}]_+}{\Theta},\\
D_C&=\frac{\min\{m,\theta^{\mathrm{rec}}\}-z_C}{\Theta},
\end{aligned}
\label{eq:case1-demand}
\end{equation}
where \([v]_+\equiv\max\{v,0\}\).

\textbf{(iii-a)} If \(\theta^{\mathrm{rec}}<z_B<m\), the low recommendation group contains only buyers of
\(C\), while the participation threshold \(z_B\) is the lower endpoint of
\(B\)'s high-group demand.  The complete ordering and demands are
\[
0<z_C<\theta^{\mathrm{rec}}<z_B<m<h<\Theta,\qquad
\delta h>\theta^{\mathrm{rec}}>\delta m,
\]
and
\[
D_A=\frac{\Theta-h}{\Theta},\qquad
D_B=\frac{h-z_B}{\Theta},\qquad
D_C=\frac{\theta^{\mathrm{rec}}-z_C}{\Theta}.
\]
The corresponding normalized profits are
\[
\widehat{\pi}_A=\frac{p_A(\Theta-h)}{\Theta},\qquad
\widehat{\pi}_B=\frac{p_B(h-z_B)}{\Theta},\qquad
\widehat{\pi}_C=\frac{p_C(\theta^{\mathrm{rec}}-z_C)}{\Theta}.
\]
Their first-order conditions are
\[
1-\frac{2p_A-p_B}{\delta\Theta(q_A-q_B)}=0,
\qquad
\frac{p_A-2p_B}{\delta\Theta(q_A-q_B)}
-\frac{2p_B}{\delta\Theta q_B}=0,
\qquad
\frac{p_A-2p_C}{\Theta Q}
-\frac{2p_C}{\delta\Theta q_C}=0.
\]
Solving gives
\[
p_A=\frac{2\delta\Theta q_A(q_A-q_B)}{4q_A-q_B},
\qquad
p_B=\frac{\delta\Theta q_B(q_A-q_B)}{4q_A-q_B},
\qquad
p_C=\frac{\delta^2\Theta q_Aq_C(q_A-q_B)}
{(4q_A-q_B)[q_A-(1-\delta)q_C]}.
\]
Substitution into \(\theta^{\mathrm{rec}}<z_B\), together with the remaining displayed
ordering, yields \(\delta<1/2\).  Thus Case~1(iii-a) is impossible on the
maintained domain.

\textbf{(iii-b)} If \(z_B<m<\theta^{\mathrm{rec}}\), the two parts of \(B\)'s demand join at the recommendation
threshold, and
\(D_A=\frac{\Theta-h}{\Theta}\),
\(D_B=\frac{h-m}{\Theta}\), and
\(D_C=\frac{m-z_C}{\Theta}\).  This is the
market-sharing configuration, Type 1.  Its complete endpoint ordering is
\begin{equation}
0<z_C<z_B<m<\theta^{\mathrm{rec}}<h<\Theta,
\qquad \delta h\geq\theta^{\mathrm{rec}}\geq\delta m.
\label{eq:type1-order}
\end{equation}

\textbf{(iii-c)} If \(z_B<\theta^{\mathrm{rec}}<m\), firm \(B\) is never purchased in the low group and
\(D_A=\frac{\Theta-h}{\Theta}\),
\(D_B=\frac{h-\theta^{\mathrm{rec}}}{\Theta}\), and
\(D_C=\frac{\theta^{\mathrm{rec}}-z_C}{\Theta}\).  This is the interior-niche
configuration, Type 2a, with complete ordering
\begin{equation}
0<z_C<z_B<\theta^{\mathrm{rec}}<m<h<\Theta,
\qquad \delta h\geq\theta^{\mathrm{rec}}\geq\delta m.
\label{eq:type2a-order}
\end{equation}
The equalities \(m=\theta^{\mathrm{rec}}\) and
\(z_B=\theta^{\mathrm{rec}}\) are not smooth.  They are checked
separately below.

\smallskip
\noindent\textbf{Case 2:}
Now reverse all inequalities in~\eqref{eq:rec-case1}.  The assistant
recommends \(\{B,C\}\) below \(\frac{p_A-p_B}{q_A-q_B}\), \(\{A,C\}\)
between \(\frac{p_A-p_B}{q_A-q_B}\) and
\(\frac{p_B-p_C}{q_B-q_C}\), and \(\{A,B\}\) above the latter threshold.
Of the six strict rankings of the three quality-to-price ratios, exactly four
are compatible with this recommendation ordering:
\[
\frac{q_A}{p_A}>\frac{q_B}{p_B}>\frac{q_C}{p_C},\qquad
\frac{q_A}{p_A}>\frac{q_C}{p_C}>\frac{q_B}{p_B},
\]
\[
\frac{q_C}{p_C}>\frac{q_A}{p_A}>\frac{q_B}{p_B},\qquad
\frac{q_C}{p_C}>\frac{q_B}{p_B}>\frac{q_A}{p_A}.
\]
They correspond to subcases (i)--(iv), respectively; the other two rankings
are inconsistent with the Case-2 threshold inequalities.

\textbf{(i)} If \(q_A/p_A>q_B/p_B>q_C/p_C\), quality is salient in every
pair.
In particular, within \(\{B,C\}\), product \(C\) would be purchased only
if
\(\frac{\delta p_C}{q_C}\leq\theta<\frac{\delta(p_B-p_C)}{q_B-q_C}\).
But \(q_B/p_B>q_C/p_C\) reverses these two endpoints, so \(D_C=0\).

\textbf{(ii)} If \(q_A/p_A>q_B/p_B\),
\(q_B/p_B<q_C/p_C\), and \(q_A/p_A>q_C/p_C\), quality is salient in
\(\{A,B\}\) and \(\{A,C\}\), whereas price is salient in
\(\{B,C\}\).  Firm \(B\) is not chosen from \(\{A,B\}\).  It is also
not chosen from \(\{B,C\}\), because that pair is recommended only below
\(\frac{p_A-p_B}{q_A-q_B}\), whereas choosing \(B\) requires
\[
\theta\geq\frac{p_B-p_C}{\delta(q_B-q_C)}>
\frac{p_A-p_B}{q_A-q_B}.
\]
Thus \(D_B=0\).

\textbf{(iii)} The third mixed ordering,
\(q_A/p_A>q_B/p_B\), \(q_B/p_B<q_C/p_C\), and
\(q_A/p_A<q_C/p_C\), changes the salience of \(\{A,C\}\), which does not
contain \(B\).  In \(\{A,B\}\), quality salience again makes \(B\)'s
purchase interval empty.  In \(\{B,C\}\), choosing \(B\) requires
\(\theta\geq\frac{p_B-p_C}{\delta(q_B-q_C)}\), strictly above the
recommendation threshold \(\frac{p_A-p_B}{q_A-q_B}\).  Hence \(D_B=0\).

\textbf{(iv)} It remains to consider~\eqref{eq:all-price-salient}.  Case 2 requires
\(\delta h<\theta^{\mathrm{rec}}<\delta m\).  In the low \(\{B,C\}\) group, the
\(B\)--\(C\) purchase threshold is \(m>\delta m\), so only \(C\) can be
purchased.  In the high \(\{A,B\}\) group, \(B\) is chosen below \(h\)
and \(A\) above it.  Therefore positive demand for \(B\) requires
\(h>\max\{z_B,\delta m\}\), and in particular \(h>\delta m\).
Since \(\theta^{\mathrm{rec}}>\delta h\), this implies
\[
\frac{\theta^{\mathrm{rec}}}{\delta}>h>\delta m.
\]
The \(A\)--\(C\) purchase threshold \(\theta^{\mathrm{rec}}/\delta\)
consequently lies above the
entire middle recommendation interval.  Hence every participating consumer
in that interval chooses \(C\); the apparent alternative ordering
\(\theta^{\mathrm{rec}}/\delta<\delta m\) is incompatible with positive
demand for \(B\).
The demands reduce to
\begin{equation}
D_A=\frac{\Theta-h}{\Theta},\qquad
D_B=\frac{h-\max\{z_B,\delta m\}}{\Theta},\qquad
D_C=\frac{\delta m-z_C}{\Theta}.
\label{eq:case2-demand}
\end{equation}
There are exactly two smooth positive-demand cases.

\textbf{(iv-a)} If \(z_B<\delta m\), the complete relevant ordering is
\[
0<z_C<z_B<\delta m<h<\Theta,\qquad
\delta h<\theta^{\mathrm{rec}}<\delta m,
\]
and
\((D_A,D_B,D_C)=\frac{(\Theta-h,h-\delta m,\delta m-z_C)}{\Theta}\).
The three
first-order conditions are
\[
1-\frac{2p_A-p_B}{\delta\Theta(q_A-q_B)}=0,\qquad
\frac{p_A-2p_B}{\delta\Theta(q_A-q_B)}
 -\frac{2p_B-p_C}{\Theta(q_B-q_C)}=0,\qquad
\frac{p_B-2p_C}{\Theta(q_B-q_C)}
-\frac{2p_C}{\delta\Theta q_C}=0.
\]

\textbf{(iv-b)} If \(\delta m<z_B\), the ordering is
\[
0<z_C<\delta m<z_B<h<\Theta,\qquad
\delta h<\theta^{\mathrm{rec}}<\delta m,
\]
and
\((D_A,D_B,D_C)=\frac{(\Theta-h,h-z_B,\delta m-z_C)}{\Theta}\).
The first-order
conditions for \(A\) and \(C\) are unchanged, while that of \(B\) is
\[
\frac{p_A-2p_B}{\delta\Theta(q_A-q_B)}
-\frac{2p_B}{\delta\Theta q_B}=0.
\]
All three profit functions are strictly concave in own price within either
case.  Solving each displayed linear system and substituting the solution
into its complete ordering gives, respectively,
\[
\begin{aligned}
&q_A>q_B>q_C>0, \delta>0, z_B<\delta m,
\ \delta h<\theta^{\mathrm{rec}}<\delta m, D_A,D_B,D_C>0
\quad\Longrightarrow\quad\delta<\frac12,\\
&q_A>q_B>q_C>0, \delta>0, \delta m<z_B,
\ \delta h<\theta^{\mathrm{rec}}<\delta m, D_A,D_B,D_C>0
\quad\Longrightarrow\quad\delta<\frac12.
\end{aligned}
\]

\textbf{(iv-c)} At the boundary \(z_B=\delta m\), firm \(A\) must still
satisfy its displayed smooth first-order condition.  Hence
\(2p_A-p_B=\delta\Theta(q_A-q_B)\).  The boundary equation and
\(p_C>0\) imply \(q_C-(1-\delta)q_B>0\), so increasing \(p_B\) switches
the lower endpoint of \(B\)'s demand from \(z_B\) to \(\delta m\).  A kink
maximum therefore requires the derivative on the \(z_B\) side to be
nonnegative.  After substituting \(A\)'s first-order condition, the Case-2
ordering \(\delta h<z_B\) and this derivative condition require, respectively,
\[
\frac{p_B}{q_A-q_B}>
\frac{\delta\Theta}
{1+2(q_A-q_B)/(\delta q_B)},
\qquad
\frac{p_B}{q_A-q_B}\leq
\frac{\delta\Theta}
{3+4(q_A-q_B)/q_B}.
\]
These inequalities can hold together only if
\[
\frac{q_A-q_B}{q_B}\left(\frac1\delta-2\right)>1,
\]
which is impossible for \(\delta\geq1/2\).  Thus this boundary cannot be a
kink maximum.  Together with (iv-a)--(iv-b), Case 2 has no positive-demand
candidate when \(\delta\geq1/2\).

\smallskip
\noindent\textbf{Case 3:}
Finally, suppose all three true recommendation thresholds are equal to
\(t\).  The assistant recommends \(\{B,C\}\) below \(t\) and
\(\{A,B\}\) above it.  Their equality implies that prices have the
form \(p_i=tq_i+c\) for a common intercept \(c\).  Therefore
\(q_i/p_i=\frac{q_i}{tq_i+c}\) is increasing in quality when \(c>0\), decreasing
when \(c<0\), and constant when \(c=0\).  Consequently the three subcases
below---\(q_A/p_A>q_B/p_B>q_C/p_C\),
\(q_C/p_C>q_B/p_B>q_A/p_A\), and equality of all three ratios---are
exhaustive.  A tie of only two ratios is impossible on this recommendation
face.

\textbf{(i)} If the three quality-to-price ratios are equal,
perceived and true utilities coincide.  Firm \(B\) loses to \(C\) below
\(t\) and to \(A\) above \(t\), so \(D_B=0\).

\textbf{(ii)} If
\(q_A/p_A>q_B/p_B>q_C/p_C\), quality is salient in both pairs.  In the
low \(\{B,C\}\) group, \(C\)'s participation threshold is
\(\frac{\delta p_C}{q_C}\), while its upper purchase threshold is
\(\frac{\delta(p_B-p_C)}{q_B-q_C}\).  The displayed quality-to-price ordering
makes the participation threshold strictly larger than the upper purchase
threshold, so \(D_C=0\).

\textbf{(iii)} Under~\eqref{eq:all-price-salient}, the common threshold is
\(t=\theta^{\mathrm{rec}}\).  Equality of the three true thresholds gives
\(\delta h=\theta^{\mathrm{rec}}=\delta m\).  Consumers below
\(\theta^{\mathrm{rec}}\) are recommended \(\{B,C\}\), but the purchase
threshold \(m=\theta^{\mathrm{rec}}/\delta\) lies above this interval, so
they purchase \(C\) whenever they participate.  Consumers above
\(\theta^{\mathrm{rec}}\) are recommended \(\{A,B\}\): they purchase
\(B\) below \(h=\theta^{\mathrm{rec}}/\delta\) and \(A\) above it.  Thus all
three firms can have positive demand only when \(\delta<1\) and the
participation thresholds lie
inside these intervals.  On that part of the face,
\[
D_A=\frac{\Theta-h}{\Theta},\qquad
D_B=\frac{h-\max\{z_B,\theta^{\mathrm{rec}}\}}{\Theta},\qquad
D_C=\frac{\theta^{\mathrm{rec}}-z_C}{\Theta}.
\]
If \(\delta=1\), then
\(h=\theta^{\mathrm{rec}}=m\), and the displayed formula gives
\(D_B=0\), so Step 1 excludes the profile.  Suppose therefore that
\(\delta<1\).  This boundary separates Cases 1 and 2.  Lowering \(p_C\)
enters Case 2, while raising \(p_C\) enters Case 1, and direct
differentiation gives
\[
\left.\frac{\partial\widehat{\pi}_C}{\partial p_C}\right|_{\mathrm{Case\,1}}
-\left.\frac{\partial\widehat{\pi}_C}{\partial p_C}\right|_{\mathrm{Case\,2}}
=\frac{p_C(q_A-q_B)}
{\Theta(q_A-q_C)(q_B-q_C)}>0.
\]
Thus firm \(C\)'s derivative jumps upward.  A kink maximum requires the left
derivative to be nonnegative and the right derivative to be nonpositive, so
the boundary is excluded, independently of \(z_B\).

We must also check the remaining faces that separate the strict configurations.  At
\(m=\theta^{\mathrm{rec}}\), firm \(B\)'s right derivative minus its left derivative is
\(\frac{p_B}{\delta\Theta(q_B-q_C)}>0\).  The derivative jumps upward, so
this face cannot maximize \(B\)'s profit.  It remains to check partial
salience ties.  In Case 1, the only
compatible partial ties are
\[
\frac{q_A}{p_A}=\frac{q_B}{p_B}>\frac{q_C}{p_C},\qquad
\frac{q_B}{p_B}>\frac{q_A}{p_A}=\frac{q_C}{p_C},\qquad
\frac{q_B}{p_B}=\frac{q_C}{p_C}>\frac{q_A}{p_A}.
\]
In the first two, quality is salient in \(\{B,C\}\), and \(C\)'s
participation threshold lies weakly above its pairwise indifference threshold.
In the third, salience is tied in \(\{B,C\}\), so those two thresholds are
equal.  Because \(\{A,C\}\) is not recommended in Case 1, all three
partial-tie profiles have \(D_C=0\).

In Case 2, the only compatible partial ties are
\[
\frac{q_C}{p_C}>\frac{q_A}{p_A}=\frac{q_B}{p_B},\qquad
\frac{q_A}{p_A}=\frac{q_C}{p_C}>\frac{q_B}{p_B},\qquad
\frac{q_A}{p_A}>\frac{q_B}{p_B}=\frac{q_C}{p_C}.
\]
For each ordering, \(B\) cannot be purchased from \(\{A,B\}\): its
participation threshold is weakly above the pairwise indifference threshold.
Nor can it be purchased from \(\{B,C\}\), because that pair is
recommended below the \(A\)--\(B\) recommendation threshold, which is
strictly below the \(B\)--\(C\) recommendation threshold.  Hence all
three Case-2 partial ties have \(D_B=0\).  Finally, if all three ratios
tie, prices are proportional to qualities, all three recommendation
thresholds tie, and Case 3(i) gives \(D_B=0\).  These partial ties,
together with the six strict rankings enumerated above, exhaust all weak
orderings of the three quality-to-price ratios.

One face remains.  It is
\begin{equation}
z_B=\theta^{\mathrm{rec}},
\label{eq:type2b-face}
\end{equation}
with
\[
0<z_C<z_B=\theta^{\mathrm{rec}}<m<h<\Theta,\qquad
\delta h\geq\theta^{\mathrm{rec}}\geq\delta m.
\]
To the
left of this face, \(B\)'s lower demand boundary is
\(\theta^{\mathrm{rec}}\), as in Type 2a;
to the right, it is the participation threshold \(z_B\).  Hence \(B\)'s profit
has two different slopes and can attain a maximum at the kink even though
neither smooth first-order condition is zero.  This is the Type-2b
boundary-niche candidate.  Type 2b does not introduce a new market pattern:
its recommendation and purchase allocation coincides with the limiting
Type-2a allocation, but \(p_B\) is determined by a different condition.

We have now exhausted both strict recommendation orderings and their common
face.  Step 1 excludes every endpoint face on which a price or a demand is
zero.  Case~1(iii-a) and both positive-demand Case-2 cases require
\(\delta<1/2\), while the intervening Case-2 kink has
a strictly negative left derivative.  The face
\(m=\theta^{\mathrm{rec}}\) has an upward derivative jump.  The common
face where the recommendation thresholds coincide has an upward derivative
jump when \(\delta<1\) and gives \(D_B=0\) when \(\delta=1\). Every salience-tie
face gives zero demand to \(B\) or \(C\).  The only unexcluded
face is \(z_B=\theta^{\mathrm{rec}}\), which gives Type 2b.  Therefore Types 1, 2a, and 2b
exhaust the pure-strategy candidates for \(\delta\geq1/2\).

\medskip
\noindent\textbf{Step 3: Candidate prices, equilibrium regions, and outcomes.}
For each of the three configurations retained in Step 2, we first solve the
firms' pricing conditions and impose the complete threshold ordering of that
configuration.  This gives a candidate price vector and its local
admissibility region.  We then compare each firm's profit at that vector with
its profit from every unilateral price.  The second comparison is essential:
a candidate that is optimal while its threshold ordering is fixed need not be a
best response after a deviation changes the recommendation or salience
ordering.

\smallskip
\noindent\textit{Type 1 (market sharing).}
Using~\eqref{eq:type1-order}, normalized profits are
\[
\begin{aligned}
\widehat{\pi}_A&=p_A\left[1-\frac{p_A-p_B}
{\delta\Theta(q_A-q_B)}\right],\\
\widehat{\pi}_B&=p_B\left[\frac{p_A-p_B}{\delta\Theta(q_A-q_B)}
-\frac{p_B-p_C}{\delta\Theta(q_B-q_C)}\right],\\
\widehat{\pi}_C&=p_C\left[\frac{p_B-p_C}{\delta\Theta(q_B-q_C)}
-\frac{p_C}{\delta\Theta q_C}\right].
\end{aligned}
\]
Each normalized profit function is strictly concave in the firm's own price on this branch.
The three first-order conditions are
\[
\begin{aligned}
1-\frac{2p_A-p_B}{\delta\Theta(q_A-q_B)}&=0,\qquad
\frac{p_A-2p_B}{q_A-q_B}-\frac{2p_B-p_C}{q_B-q_C}&=0,\qquad
\frac{p_B-2p_C}{q_B-q_C}-\frac{2p_C}{q_C}&=0.
\end{aligned}
\]
Solving them gives
\[
\begin{aligned}
p_A^{s1}&=\alpha_1[(\lambda+3)q_C+4\lambda Q],&
p_B^{s1}&=2\alpha_1\lambda(q_C+\lambda Q),&
p_C^{s1}&=\alpha_1\lambda q_C,
\end{aligned}
\]
where
\[
\alpha_1\equiv
\frac{\delta\Theta(1-\lambda)Q}
{6q_C+2(4\lambda-\lambda^2)Q}.
\]
Substituting these prices into every inequality
in~\eqref{eq:type1-order} shows that the complete ordering holds for every
admissible \(q_C,Q,\delta,\lambda\).  Hence
\[
\mathcal L_1=
\left\{(q_C,Q,\delta,\lambda):q_C>0,\ Q>0,\quad
\frac12\leq\delta\leq1,\ 0<\lambda<1\right\}.
\]

\smallskip
\noindent\textit{Type 2a (interior niche).}
Using~\eqref{eq:type2a-order}, normalized profits are
\[
\begin{aligned}
\widehat{\pi}_A&=p_A\left[1-\frac{p_A-p_B}
{\delta\Theta(q_A-q_B)}\right],\\
\widehat{\pi}_B&=p_B\left[\frac{p_A-p_B}{\delta\Theta(q_A-q_B)}
-\frac{p_A-p_C}{\Theta Q}\right],\\
\widehat{\pi}_C&=p_C\left[\frac{p_A-p_C}{\Theta Q}
-\frac{p_C}{\delta\Theta q_C}\right].
\end{aligned}
\]
The first-order conditions are
\[
\begin{aligned}
1-\frac{2p_A-p_B}{\delta\Theta(q_A-q_B)}&=0,\qquad
\frac{p_A-2p_B}{\delta(q_A-q_B)}-
\frac{p_A-p_C}{Q}&=0,\qquad
\frac{p_A-2p_C}{Q}-\frac{2p_C}{\delta q_C}&=0.
\end{aligned}
\]
Solving them gives
\[
\begin{aligned}
p_A^{s2a}&=4\alpha_{2a}(Q+\delta q_C),\qquad
p_C^{s2a}=2\alpha_{2a}\delta q_C,\\
p_B^{s2a}&=\alpha_{2a}
\left[2(Q+\delta q_C)-\delta(1-\lambda)(2Q+\delta q_C)\right],
\end{aligned}
\]
where
\[
\alpha_{2a}\equiv
\frac{\delta\Theta(1-\lambda)Q}
{6(Q+\delta q_C)+\delta(1-\lambda)(2Q+\delta q_C)}.
\]
To state the result of substituting these prices into every inequality
in~\eqref{eq:type2a-order}, define
\[
\begin{alignedat}{2}
\underline{Q}^{\,s}&\equiv\frac{7+\sqrt{97}}8q_C,\qquad&
\bar{\delta}^{\,s}&\equiv
\sqrt{\frac{Q^2+2Qq_C}{q_C^2}}-\frac Q{q_C},\\
\bar{\lambda}^{\,s}&\equiv\frac{2Q(1-\delta)-\delta^2q_C}
{\delta(2Q+\delta q_C)},\qquad&
\underline{\lambda}_{1}^{\,s}&\equiv
\frac{2Q(1-\delta)-\delta^2q_C}
{(2-\delta)(2Q+\delta q_C)},\\
\underline{\lambda}_{2}^{\,s}&\equiv
\mathrlap{\frac{2(1-\delta)Q^2-\delta q_C[(\delta+2)Q+2\delta q_C]}
{\delta Q(2Q+\delta q_C)}.}
\end{alignedat}
\]
When \(Q>\underline{Q}^{\,s}\), let \(\delta_0\) be the unique root in
\((1/2,\bar{\delta}^{\,s})\) of
\[
\delta^3(Qq_C+q_C^2)+\delta^2(2Q^2-2q_C^2)
-\delta(4Q^2+2Qq_C)+2Q^2=0.
\]
Substitution into the complete ordering then gives
{\footnotesize
\[
\mathcal L_{2a}=
\left\{(q_C,Q,\delta,\lambda):q_C>0,\quad
\begin{array}{ll}
\displaystyle
\frac{q_C}{4}<Q\leq \underline{Q}^{\,s},\quad
\frac12\leq\delta<\bar{\delta}^{\,s},\quad
\underline{\lambda}_{1}^{\,s}<\lambda<\bar{\lambda}^{\,s};\\[0.7em]
\text{or}\quad
\displaystyle Q>\underline{Q}^{\,s},\quad
\frac12\leq\delta<\delta_0,\quad
\underline{\lambda}_{2}^{\,s}<\lambda<\bar{\lambda}^{\,s};\\[0.7em]
\text{or}\quad
\displaystyle Q>\underline{Q}^{\,s},\quad
\delta_0\leq\delta<\bar{\delta}^{\,s},\quad
\underline{\lambda}_{1}^{\,s}<\lambda<\bar{\lambda}^{\,s}
\end{array}\right\}.
\]
}
Thus \(\bar{\lambda}^{\,s}\) is always the upper bound.  The lower bound is
\(\underline{\lambda}_{1}^{\,s}\) in the first and third cases and
\(\underline{\lambda}_{2}^{\,s}\) in the second case; the equation for
\(\delta_0\) is where these two lower bounds exchange order.

\smallskip
\noindent\textit{Type 2b (boundary niche).}
At~\eqref{eq:type2b-face}, firms \(A\) and \(C\) retain the Type-2a
first-order conditions, while \(B\)'s price is determined by
\[
\frac{p_B}{\delta\Theta q_B}
=\frac{p_A-p_C}{\Theta Q}.
\]
Solving these three equations gives
\[
\begin{aligned}
p_A^{s2b}&=2\alpha_{2b}Q(Q+\delta q_C),\qquad
p_B^{s2b}=\alpha_{2b}\delta(q_C+\lambda Q)(2Q+\delta q_C),\\
p_C^{s2b}&=\alpha_{2b}\delta q_CQ,
\end{aligned}
\]
where
\[
\alpha_{2b}\equiv
\frac{\delta\Theta(1-\lambda)Q}
{(2-\delta\lambda)Q(2Q+\delta q_C)-\delta^2q_C^2}.
\]
The kink is a local maximum for \(B\) exactly when its left derivative is
nonnegative and its right derivative is nonpositive.  Define
\[
\underline{\lambda}_{2b}^{\,s}\equiv
\frac{\delta q_C(Q+\delta q_C)}
{(1-\delta)Q(2Q+\delta q_C)}.
\]
Substituting \(\mathbf p^{s2b}\) into these two one-sided conditions and every
remaining inequality in~\eqref{eq:type2b-face} gives
\[
\mathcal L_{2b}=
\left\{(q_C,Q,\delta,\lambda):q_C>0,\ Q>\underline{Q}^{\,s},\quad
\frac12\leq\delta<\delta_0,\
\underline{\lambda}_{2b}^{\,s}<\lambda\leq\underline{\lambda}_{2}^{\,s}\right\}.
\]
At \(\lambda=\underline{\lambda}_{2}^{\,s}\), direct simplification gives
\(\mathbf p^{s2a}=\mathbf p^{s2b}\).  We assign the equality to Type 2b and use the strict
condition \(\underline{\lambda}_{2}^{\,s}<\lambda\) in the relevant Type-2a region.

\smallskip
\noindent\textit{Global best responses and equilibrium regions.}
The regions \(\mathcal L_1,\mathcal L_{2a}\), and \(\mathcal L_{2b}\)
establish only that the candidate price profiles satisfy the pricing conditions and
the threshold ordering of the corresponding configuration.  They do not rule
out a deviation that crosses a participation, recommendation, salience, or
purchase boundary.  We now impose the global conditions.

Let
\[
\omega\equiv(q_C,Q,\delta,\lambda),\qquad
\Omega\equiv\left\{(q_C,Q,\delta,\lambda):q_C>0,\ Q>0,\
\frac12\leq\delta\leq1,\ 0<\lambda<1\right\}.
\]
Fix \(j\in\{1,2a,2b\}\) and \(i\in\{A,B,C\}\), hold
\(\mathbf p_{-i}^{sj}\) fixed, and write the deviation price as \(y\).
As shown above, it is enough to consider
\(0\leq y\leq\frac{\Theta q_A}{\delta}\).  Let
\(D_i(y,\mathbf p_{-i}^{sj})\) denote firm \(i\)'s demand obtained from the
recommendation and purchase rules in Step 1, and write
\(D_i(\mathbf p^{sj})\equiv D_i(p_i^{sj},\mathbf p_{-i}^{sj})\).

{For each rival \(k\ne i\), the quality and price salience
values can be equal at \(y=p_k^{sj}q_i/q_k\) or
\(y=p_k^{sj}q_k/q_i\).  Retain the feasible such prices, remove
duplicates, and sort them; there are at most four.  Together with the
price-domain endpoints, they partition the domain into intervals on
which every pair's salience ranking is fixed.  On each such interval,
the recommendation, participation, and purchase endpoints, together
with \(0\) and \(\Theta\), are affine in \(y\).  Add all prices within
the interval at which two endpoints coincide; the resulting
subintervals have a fixed active-endpoint ordering and
\(D_i(y,\mathbf p_{-i}^{sj})=a+by\).  When \(b\ne0\), also add the
profit stationary point \(-a/(2b)\) if it lies in the subinterval.}

Let \(\mathcal P_i^j(\omega)\) contain the price-domain endpoints, the
candidate price \(p_i^{sj}\), {all feasible salience-tie prices}, all endpoint-crossing
prices, and all such stationary points.  At every breakpoint, its payoff is
evaluated under the maintained tie rule and included through the corresponding
price in \(\mathcal P_i^j(\omega)\).  Let \(\mathcal W_i^j(\omega)\) contain
each relevant left- or right-hand payoff limit not attained at that breakpoint;
such a limit enters only as a payoff value, not as a feasible price.  Define
\[
\begin{aligned}
\widehat V_i^j(\omega)
&\equiv\max\Bigl(
\bigl\{\widehat{\pi}_i(y,\mathbf p_{-i}^{sj};S_P)
:y\in\mathcal P_i^j(\omega)\bigr\}\\
&\qquad\cup\bigl\{w/\kappa_i(S_P)
:w\in\mathcal W_i^j(\omega)\bigr\}\Bigr)
-\widehat{\pi}_i(\mathbf p^{sj};S_P).
\end{aligned}
\]
The corresponding actual payoff gap is \(V_i^j(\omega)=\kappa_i(S_P)\widehat V_i^j(\omega)\).
Because \(p_i^{sj}\in\mathcal P_i^j(\omega)\), \(V_i^j(\omega)\geq0\).
The finite set includes the maximum of every quadratic piece and every
possible boundary payoff or one-sided supremum.  Hence firm \(i\) has no
profitable unilateral deviation iff \(V_i^j(\omega)=0\).

The globally valid regions are therefore
\begin{equation}
\begin{aligned}
\mathcal R_1&\equiv\left\{\omega\in\mathcal L_1:
V_A^1(\omega)=V_B^1(\omega)=V_C^1(\omega)=0\right\},\\
\mathcal R_{2a}&\equiv\left\{\omega\in\mathcal L_{2a}:
V_A^{2a}(\omega)=V_B^{2a}(\omega)=V_C^{2a}(\omega)=0\right\},\\
\mathcal R_{2b}&\equiv\left\{\omega\in\mathcal L_{2b}:
V_A^{2b}(\omega)=V_B^{2b}(\omega)=V_C^{2b}(\omega)=0\right\}.
\end{aligned}
\label{eq:lemma2-global-regions}
\end{equation}
We retain \(V_i^j=0\) in this finite-maximum form.  A full characterization
in terms of the model parameters would require many additional cases.  We
evaluate a particular deviation explicitly only when it is needed for a
subsequent comparison.

These conditions are necessary and sufficient.  If
\(\omega\in\mathcal R_j\), then membership in \(\mathcal L_j\) gives the
Type-\(j\) candidate and its threshold ordering, while \(V_i^j=0\) rules out a
profitable unilateral price deviation by any firm.  Hence \(\mathbf p^{sj}\) is a
Nash equilibrium.  Conversely, if a Type-\(j\) candidate is a Nash
equilibrium, its defining conditions imply \(\omega\in\mathcal L_j\), and the
absence of a profitable unilateral price deviation implies
\(V_A^j=V_B^j=V_C^j=0\).  Therefore \(\omega\in\mathcal R_j\).

It remains to rule out multiple distinct equilibria.  We compare firm
\(B\)'s profits from the two relevant cross-configuration deviations.
For \(j\in\{1,2a,2b\}\), define
\[
s_B^j\equiv\frac{p_C^{sj}q_B}{q_C},\qquad
b_{12}^j\equiv p_C^{sj}+\delta\lambda(p_A^{sj}-p_C^{sj}),\qquad
b_{2b}^j\equiv\frac{\delta q_B(p_A^{sj}-p_C^{sj})}{Q},
\]
and
\[
\begin{aligned}
\widehat{\pi}_B^1(y;p_A,p_C)&\equiv
y\left[\frac{p_A-y}{\delta\Theta(q_A-q_B)}
-\frac{y-p_C}{\delta\Theta(q_B-q_C)}\right],\\
\widehat{\pi}_B^2(y;p_A,p_C)&\equiv
y\left[\frac{p_A-y}{\delta\Theta(q_A-q_B)}
-\frac{p_A-p_C}{\Theta Q}\right].
\end{aligned}
\]
The first quadratic has stationary point
\(v_1(p_A,p_C)=\frac{(q_B-q_C)p_A+(q_A-q_B)p_C}{2Q}\), and the second has
stationary point
\(v_2(p_A,p_C)=\frac{p_A-\delta(1-\lambda)(p_A-p_C)}{2}\).
For the Type-1 candidate, set
\[
t_1\equiv p_C^{s1}+\lambda(p_A^{s1}-p_C^{s1})=2p_B^{s1}.
\]
Firm \(B\)'s deviation normalized profit equals
\(\widehat{\pi}_B^2(y;p_A^{s1},p_C^{s1})\) for
\(y\in[b_{12}^1,\min\{b_{2b}^1,t_1\}]\), with the appropriate
one-sided interpretation at an open endpoint.
Direct substitution gives
\[
\begin{aligned}
b_{2b}^1-b_{12}^1
&=\frac{\delta\Theta(1-\lambda)q_C
[\lambda(4\delta-1)Q+3\delta q_C]}
{2[3q_C+(4\lambda-\lambda^2)Q]}>0,\\
b_{12}^{2a}-s_B^{2a}
&=2\alpha_{2a}\delta\lambda(Q+\delta q_C)>0,\\
b_{12}^{2b}-s_B^{2b}
&=\alpha_{2b}\delta\lambda Q(Q+\delta q_C)>0.
\end{aligned}
\]
The three intervals below are strictly ordered.  On the
Type-2a and Type-2b local domains, the intervals \(I_{21}^j\) satisfy
the ordered-price and price-salience conditions for the Type-1 deviation,
and their upper purchase cutoff at \(s_B^j\) is below \(\Theta\).  If an
endpoint belongs to an open branch, its value below denotes the corresponding
one-sided payoff limit; the payoff assigned at the breakpoint by the tie rule
is included separately in \(\mathcal P_i^j\).  Write
\(\operatorname{clip}_{[a,b]}(v)\equiv\min\{b,\max\{a,v\}\}\), and define
\begin{equation}
\begin{aligned}
I_{12}&\equiv[b_{12}^1,b_{2b}^1],\qquad
I_{21}^j\equiv[s_B^j,b_{12}^j],\quad j\in\{2a,2b\},\\
y_{12}&\equiv\operatorname{clip}_{I_{12}}
\bigl(v_2(p_A^{s1},p_C^{s1})\bigr),\\
y_{21}^j&\equiv\operatorname{clip}_{I_{21}^j}
\bigl(v_1(p_A^{sj},p_C^{sj})\bigr),\qquad j\in\{2a,2b\},\\
\widehat{M}_{B,\times}^{1}&\equiv
\max_{y\in I_{12}}\widehat{\pi}_B^2(y;p_A^{s1},p_C^{s1})
=\widehat{\pi}_B^2(y_{12};p_A^{s1},p_C^{s1}),\\
\widehat{M}_{B,\times}^{j}&\equiv
\max_{y\in I_{21}^j}\widehat{\pi}_B^1(y;p_A^{sj},p_C^{sj})
=\widehat{\pi}_B^1(y_{21}^j;p_A^{sj},p_C^{sj}),\quad j\in\{2a,2b\},\\
F_{12}&\equiv\widehat{\pi}_B^{s1}-\widehat{M}_{B,\times}^{1},\\
F_{21}^j&\equiv\widehat{\pi}_B^{sj}-\widehat{M}_{B,\times}^{j}.
\end{aligned}
\label{eq:lemma2-B-cross-deviation}
\end{equation}
Both quadratic objectives are strictly concave, so their
clipped stationary points attain the stated maxima.  A Type-1 equilibrium
with \(y_{12}\leq t_1\) requires \(F_{12}\geq0\), and a Type-\(j\)
equilibrium, \(j\in\{2a,2b\}\), requires \(F_{21}^j\geq0\).

For the remaining comparisons, maintain the shared notation
\[
x\equiv\frac{q_C}{Q}>0,\quad
Z\equiv\delta\Theta Q>0,\quad
D\equiv3x+4\lambda-\lambda^2>0,
\]
\[
E\equiv6(1+\delta x)+\delta(1-\lambda)(2+\delta x)>0,
\qquad
K\equiv4-2\delta\lambda+2\delta x-\delta^2\lambda x-\delta^2x^2.
\]
For the Type-1 candidate, define
\[
\begin{aligned}
H&\equiv[1-2\lambda-\delta(1-\lambda)](3x+4\lambda)-\lambda x,\\
J&\equiv(1-\delta)(1-\lambda)(3x+4\lambda)-(x+\lambda).
\end{aligned}
\]
Direct substitution gives
\[
v_2(p_A^{s1},p_C^{s1})-t_1=\frac{Z(1-\lambda)H}{4D},
\qquad J=H+(4\lambda-1)(x+\lambda).
\]
When \(t_1\leq b_{2b}^1\), firm \(B\)'s deviation normalized profit at
\(y=t_1\) satisfies
\begin{equation}
\widehat{\pi}_B(t_1,\mathbf p_{-B}^{s1})-\widehat{\pi}_B^{s1}
=\frac{p_B^{s1}}{D}J.
\label{eq:lemma2-B-recommendation-face-gain}
\end{equation}
At this price the recommendation tie affects only a measure-zero type.
On \(\mathcal L_{2b}\), \(Q>\underline{Q}^{\,s}>q_C\), so \(0<x<1\), and
\(K=4-2\delta\lambda+\delta x(2-\delta\lambda)-\delta^2x^2
>2-x^2>0\).
For Type 2a, the stationary point in
\eqref{eq:lemma2-B-cross-deviation} is below the upper endpoint.  If it is
weakly below the lower endpoint, that endpoint already yields a profitable
deviation.  Hence it is interior whenever the candidate can be an
equilibrium.  The Type-2b local inequalities always place its stationary
point in the interior.  Substitution therefore gives the two necessary
cross-configuration incentive conditions
\begin{equation}
\begin{aligned}
\frac{\widehat{\pi}_B^{s2a}-\widehat{M}_{B,\times}^{2a}}{Z}
&=\frac{(1-\lambda)^2\Phi_{2a}}{\lambda E^2}\geq0,
&&\text{so }\Phi_{2a}\geq0,\\
\frac{\widehat{\pi}_B^{s2b}-\widehat{M}_{B,\times}^{2b}}{Z}
&=-\frac{(1-\lambda)\Phi_{2b}}{4\lambda K^2}\geq0,
&&\text{so }\Phi_{2b}\leq0.
\end{aligned}
\label{eq:lemma2-B-cross-IC}
\end{equation}
Here,
\[
\begin{aligned}
\Phi_{2a}={}&\lambda(2+\delta x)
[2+\delta(-4+x+\delta(2+(-4+\delta)x))]
-\delta^2[x^2+\lambda^2(2+\delta x)^2],\\
\Phi_{2b}={}&\delta^2x^2+2\delta\lambda x(2+\delta x)[-3+2\delta(2+\delta x(1+x))]
+\lambda^2(2+\delta x)[2+\delta(-8+x+4\delta(2+\delta x(1+x)))] .
\end{aligned}
\]
For the Type-1 candidate, the values of
\(\widehat{\pi}_B^{s1}-\widehat{\pi}_B^2(y;p_A^{s1},p_C^{s1})\) at
\(y=b_{12}^1,v_2(p_A^{s1},p_C^{s1}),b_{2b}^1\), respectively, are
\[
\frac{Z(1-\lambda)\lambda
[\lambda(4\delta-2)+(3\delta-1)x]^2}{4D^2},\qquad
\frac{Z(1-\lambda)^2\Phi_I}{16D^2},\qquad
\frac{Z(1-\lambda)(\lambda+x)\Phi_R}{4D^2},
\]
where
\[
\begin{aligned}
\Phi_I&\equiv\lambda\{16[-1+\delta(2+\delta(-1+\lambda))]\lambda-24x
+8\delta[6+3\delta(-1+\lambda)+\lambda]x+(1+3\delta)^2x^2\}
-9(1-\delta)^2x^2,\\
\Phi_R&\equiv4\lambda^2(1-4\delta+4\delta^2)
+4\lambda x(1-6\delta+6\delta^2-\delta\lambda+4\delta^2\lambda)
+3\delta x^2(-3+3\delta-\lambda+8\delta\lambda+3\delta x).
\end{aligned}
\]
All denominators in these comparisons are strictly positive.

Consider first Type 2a.  Under \(\mathcal L_{2a}\),
\(b_{12}^1<v_2(p_A^{s1},p_C^{s1})<b_{2b}^1\).  If
\(v_2(p_A^{s1},p_C^{s1})\leq t_1\), this stationary price lies on the
actual Type-2 deviation branch.  Type 1 then requires \(\Phi_I\geq0\),
whereas the valid reverse deviation from the Type-2a candidate requires
\(\Phi_{2a}\geq0\) by~\eqref{eq:lemma2-B-cross-IC}.  These conditions
are incompatible.  At \(\delta=1\),
\(\Phi_{2a}=-(2\lambda+x+\lambda x)^2<0\).  For \(\delta<1\), the cubic
\(\Phi_I\) has positive leading coefficient, negative discriminant, and
\(\Phi_I(0)<0\), so it has a unique positive root \(\rho_I\) and is
negative below it.  The quadratic \(\Phi_{2a}\) is strictly concave and
negative at both \(0\) and \(1\); if it is nonnegative between its roots
\(\rho_-\leq\rho_+\), comparison of the roots yields
\(\rho_+<\rho_I\).
Thus \(\Phi_{2a}\geq0\) implies \(\Phi_I<0\).

If instead \(v_2(p_A^{s1},p_C^{s1})>t_1\), then
\(t_1<b_{2b}^1\), so the payoff from deviating to \(t_1\) is
given by~\eqref{eq:lemma2-B-recommendation-face-gain}.  We first establish that
\(\mathcal L_{2a}\) and \(\Phi_{2a}\geq0\) imply \(\lambda>1/4\).
The Type-2a recommendation ordering gives
\[
R\equiv(2-\delta)(2+\delta x)\lambda
-[2(1-\delta)-\delta^2x]>0.
\]
For fixed \((\delta,\lambda)\), \(\Phi_{2a}\) is strictly decreasing in
\(x\geq0\).  Indeed, with \(s=1-\delta\) and
\(a=4\delta-\delta^2-1>s^2\), it can be written as
\[
\Phi_{2a}=\lambda(2+\delta x)(2s^2-\delta ax)
-\delta^2[x^2+\lambda^2(2+\delta x)^2].
\]
Suppose \(\lambda\leq1/4\), and set
\(u=s-\lambda(2-\delta)\) and
\(w=\delta+\lambda(2-\delta)>0\).  If \(u\leq0\), then
\(\delta\geq2/3\) and
\(\lambda\geq s/(2-\delta)\geq s^2/\delta^2\).  Hence
\(\Phi_{2a}(0)=4\lambda(s^2-\delta^2\lambda)\leq0\), and strict
decrease gives \(\Phi_{2a}(x)<0\) for \(x>0\).  If \(u>0\), then
\(R>0\) gives \(x>x_0\equiv2u/(\delta w)>0\), while direct
substitution gives
\[
\Phi_{2a}(x_0)
=-\frac{4s^2(1-\lambda)(1-4\lambda)}{w^2}\leq0.
\]
Strict decrease again gives \(\Phi_{2a}(x)<0\), a contradiction.
Therefore \(\lambda>1/4\).  Since the present case has \(H>0\),
\(J=H+(4\lambda-1)(x+\lambda)>0\), and the deviation to \(t_1\)
is strictly profitable for Type 1.  Thus
\(\mathcal R_1\cap\mathcal R_{2a}=\varnothing\).

Consider next Type 2b.  Its reverse deviation gives
\(\Phi_{2b}\leq0\).  Define
\[
\begin{aligned}
G_L&\equiv4\lambda(-1+\delta+\delta\lambda)
+[-3+\lambda+3\delta(1+\lambda)]x,\\
G_R&\equiv[3+\lambda-\delta(3+11\lambda)]x
-4\lambda(-1+\delta+\delta\lambda)-6\delta x^2.
\end{aligned}
\]
On \(\mathcal L_{2b}\), the following implications hold:
\[
\begin{array}{c|c|c}
\text{location of }v_2&\text{condition if }F_{12}\geq0
&\text{resulting sign}\\ \hline
v_2\leq b_{12}^1&G_L\geq0&\Phi_{2b}>0\\
b_{12}^1<v_2<b_{2b}^1&G_L<0,\ G_R<0,\ \Phi_I\geq0&\Phi_{2b}>0\\
v_2\geq b_{2b}^1&G_R\geq0,\ \Phi_R\geq0&\Phi_{2b}>0.
\end{array}
\]
If \(v_2(p_A^{s1},p_C^{s1})\leq t_1\), the clipped price \(y_{12}\)
also lies at or below \(t_1\), so its algebraic payoff is an actual
deviation payoff.  Type 1 requires \(F_{12}\geq0\), and the applicable
row contradicts \(\Phi_{2b}\leq0\).  Suppose instead that
\(v_2(p_A^{s1},p_C^{s1})>t_1\).  If \(b_{2b}^1\leq t_1\), then
\(v_2>b_{2b}^1\), and \(y=b_{2b}^1\) is an actual deviation on the
Type-2 branch.  The right-hand row again yields a contradiction.

It remains to consider \(v_2>t_1\) and \(b_{2b}^1>t_1\).  The
deviation to \(t_1\) now has the actual gain in
\eqref{eq:lemma2-B-recommendation-face-gain}, and \(H>0\).  If
\(\lambda\geq1/4\), then \(J>0\) immediately.  If
\(0<\lambda<1/4\), it suffices to establish the implication
\[
x>0,\quad \tfrac12\leq\delta\leq1,\quad
0<\lambda<\tfrac14,\quad\Phi_{2b}\leq0
\quad\Longrightarrow\quad J>0.
\]
To prove this, write
\(\Phi_{2b}=c_0+c_1x+c_2x^2+c_3x^3+c_4x^4\), where
\[
\begin{aligned}
c_0&=4(2\delta-1)^2\lambda^2,\\
c_1&=4\delta\lambda[-3+4\delta+\lambda-2\delta\lambda
+4\delta^2\lambda],\\
c_2&=\delta^2[1-6\lambda+16\delta\lambda+\lambda^2
+8\delta\lambda^2+4\delta^2\lambda^2],\\
c_3&=4\delta^3\lambda(2+\delta+\delta\lambda),
\qquad c_4=4\delta^4\lambda.
\end{aligned}
\]
Here \(c_2,c_3,c_4>0\).  Put
\(q(x)=c_0+c_1x+c_2x^2\),
\(a_J=3(1-\delta)(1-\lambda)-1\), and
\(b_J=4(1-\delta)(1-\lambda)-1\), so that
\(J=a_Jx+b_J\lambda\).  For \(\delta\geq2/3\), either
\(c_1\geq0\), or \(c_1<0\) and the quadratic discriminant
\(c_1^2-4c_0c_2<0\); hence \(q(x)>0\).  For
\(\delta<2/3\), we have \(b_J>0\).  If \(a_J\geq0\), then \(J>0\).
If \(a_J<0\) and \(J\leq0\), then
\(x\geq x_J\equiv-b_J\lambda/a_J>0\).  For
\(1/2\leq\delta<2/3\), \(0<\lambda<1/4\), and \(a_J<0\),
evaluation at \(x_J=-b_J\lambda/a_J\) gives
\(q(x_J)>0\) and \(q'(x_J)>0\).  Since \(c_2>0\),
\(q(x)>0\) for every \(x\geq x_J\), and thus
\(\Phi_{2b}\geq q(x)>0\), a contradiction.  Consequently
\(J>0\), making the deviation to \(t_1\) strictly profitable.
Therefore \(\mathcal R_1\cap\mathcal R_{2b}=\varnothing\).

Finally, when \(Q>\underline{Q}^{\,s}\) and
\(1/2\leq\delta<\delta_0\), Type 2a requires
\(\underline{\lambda}_{2}^{\,s}<\lambda\), whereas Type 2b requires
\(\lambda\leq\underline{\lambda}_{2}^{\,s}\).  Componentwise substitution gives
\[
\left.(p_A^{s2a}-p_A^{s2b},\ p_B^{s2a}-p_B^{s2b},\
p_C^{s2a}-p_C^{s2b})\right|_{\lambda=\underline{\lambda}_{2}^{\,s}}=(0,0,0).
\]
The two descriptions therefore meet at the same price vector.  The equality
is assigned to Type 2b and excluded from Type 2a, so
\(\mathcal R_{2a}\cap\mathcal R_{2b}=\varnothing\).  The three global
regions are pairwise disjoint.

Define
\[
\mathcal R_{\varnothing}\equiv
\Omega\setminus(\mathcal R_1\cup\mathcal R_{2a}\cup\mathcal R_{2b}).
\]
For \(\omega\in\mathcal R_{\varnothing}\), none of the three candidates
passes all global deviation checks.  Step 2 excludes every additional
positive-demand and endpoint candidate in Cases 1--3.  Every remaining tie
gives zero demand to \(B\) or \(C\) and is excluded by Step 1.  Hence no
pure-strategy pricing equilibrium exists on \(\mathcal R_{\varnothing}\).
Mixed-strategy pricing equilibria are not characterized here.

Substituting the equilibrium prices into the normalized profit and true-utility
expressions gives the following prices, normalized profits, and consumer surplus.

\medskip
\noindent\normalsize\textit{Type-1 Equilibrium:}
{\footnotesize
\begin{equation}
\begin{gathered}
p_A^{s1}=\alpha_1[(\lambda+3)q_C+4\lambda Q],\quad
p_B^{s1}=2\alpha_1\lambda(q_C+\lambda Q),\quad
p_C^{s1}=\alpha_1\lambda q_C;\\
\widehat{\pi}_A^{s1}=\frac{\alpha_1[(\lambda+3)q_C+4\lambda Q]^2}
{2QD},\quad
\widehat{\pi}_B^{s1}=\frac{2\alpha_1\lambda(q_C+\lambda Q)^2}
{QD},\\
\widehat{\pi}_C^{s1}=\frac{\alpha_1\lambda(1-\lambda)q_C(q_C+\lambda Q)}
{2QD},\quad
CS^{s1}=\frac{\Theta\Psi_1}{8Q^2D^2}.
\end{gathered}
\label{eq:lemma2-outcomes-1}
\end{equation}}

\noindent\normalsize\textit{Type-2a Equilibrium:}
{\footnotesize
\begin{equation}
\begin{gathered}
p_A^{s2a}=4\alpha_{2a}(Q+\delta q_C),\quad
p_B^{s2a}=\alpha_{2a}[2(Q+\delta q_C)
-\delta(1-\lambda)(2Q+\delta q_C)],\quad
p_C^{s2a}=2\alpha_{2a}\delta q_C;\\
\widehat{\pi}_A^{s2a}=\frac{16\alpha_{2a}(Q+\delta q_C)^2}
{QE},\quad
\widehat{\pi}_B^{s2a}=\frac{\alpha_{2a}
[2(Q+\delta q_C)-\delta(1-\lambda)(2Q+\delta q_C)]^2}
{QE},\\
\widehat{\pi}_C^{s2a}=\frac{4\alpha_{2a}\delta^2(1-\lambda)q_C(Q+\delta q_C)}
{QE},\quad
CS^{s2a}=\frac{\Theta\Psi_{2a}}{2Q^2E^2}.
\end{gathered}
\label{eq:lemma2-outcomes-2a}
\end{equation}}

\noindent\normalsize\textit{Type-2b Equilibrium:}
{\footnotesize
\begin{equation}
\begin{gathered}
p_A^{s2b}=2\alpha_{2b}Q(Q+\delta q_C),\quad
p_B^{s2b}=\alpha_{2b}\delta(q_C+\lambda Q)(2Q+\delta q_C),\quad
p_C^{s2b}=\alpha_{2b}\delta q_CQ;\\
\widehat{\pi}_A^{s2b}=\frac{4\alpha_{2b}(Q+\delta q_C)^2}{K},\quad
\widehat{\pi}_B^{s2b}=\frac{\alpha_{2b}\delta(q_C+\lambda Q)(2Q+\delta q_C)
[2(1-\delta)-\delta^2x(1+x)]}{K},\\
\widehat{\pi}_C^{s2b}=\frac{\alpha_{2b}\delta^2(1-\lambda)q_C
(Q+\delta q_C)}{K},\quad
CS^{s2b}=\frac{\Theta\Psi_{2b}}{2Q^4K^2},
\end{gathered}
\label{eq:lemma2-outcomes-2b}
\end{equation}}
For any \(j\in\{1,2a,2b\}\), normalized platform profit is
\(\widehat{\Pi}^{sj}=\widehat{\pi}_A^{sj}+\widehat{\pi}_C^{sj}\) under \(S_P=\{A,C\}\) and
\(\widehat{\Pi}^{sj}=\widehat{\pi}_A^{sj}+\widehat{\pi}_B^{sj}\) under \(S_P=\{A,B\}\).
The actual equilibrium profits are \(\pi_i^{sj}=\kappa_i(S_P)\widehat{\pi}_i^{sj}\) for each firm and \(\Pi^{sj,S_P}=\tau\widehat{\Pi}^{sj,S_P}\).

\medskip
\noindent\normalsize where \(\Psi_1,\Psi_{2a}\), and \(\Psi_{2b}\) are
{\footnotesize
\begin{equation}
\begin{aligned}
\Psi_1\equiv{}&36q_C^3
+q_C^2Q[27+110\lambda-29\lambda^2
-\delta(18+4\lambda-22\lambda^2)]\\
&+q_CQ^2[(72-48\delta)\lambda+(79+14\delta)\lambda^2
-(46-36\delta)\lambda^3+(3-2\delta)\lambda^4]
+Q^3[(48-32\delta)\lambda^2-(4-24\delta)\lambda^3
-(8-8\delta)\lambda^4],\\[0.3em]
\Psi_{2a}\equiv{}&\delta^2[6+\delta(1-\lambda)]^2q_C^3
+q_C^2Q[36\delta(2+\delta)+4\delta^2(8-7\delta)(1-\lambda)
-3\delta^4(1-\lambda)^2+\delta^4(3-2\delta)(1-\lambda)^3]\\
&+q_CQ^2[36+72\delta+4\delta(4-11\delta)(1-\lambda)
-4\delta^2(3-\delta)(1-\lambda)^2
+4\delta^3(3-2\delta)(1-\lambda)^3]\\
&+Q^3[36-4(1+4\delta)(1-\lambda)
-4\delta(2-\delta)(1-\lambda)^2
+4\delta^2(3-2\delta)(1-\lambda)^3],\\[0.3em]
\Psi_{2b}\equiv{}&\delta^4q_C^5
-\delta^3[4-2\delta^2-3\delta\lambda+2\delta^2\lambda]Qq_C^4
-\delta^2[4+4\delta-4\delta^2-2\delta^3+4\delta^2\lambda
-3\delta^2\lambda^2+2\delta^3\lambda^2]Q^2q_C^3\\
&+\delta[16-12\delta^2+6\delta^3-20\delta\lambda
+8\delta^2\lambda-\delta^3\lambda+2\delta^4\lambda
+8\delta^2\lambda^2-4\delta^3\lambda^2-2\delta^4\lambda^2]Q^3q_C^2\\
&+[16+16\delta-25\delta^2+6\delta^3-16\delta\lambda
+14\delta^2\lambda-12\delta^3\lambda+8\delta^4\lambda
-\delta^2\lambda^2+10\delta^3\lambda^2-8\delta^4\lambda^2]Q^4q_C\\
&+4[3-2\delta+\lambda-3\delta^2\lambda+2\delta^3\lambda
-2\delta\lambda^2+4\delta^2\lambda^2-2\delta^3\lambda^2]Q^5.
\end{aligned}
\label{eq:lemma2-cs}
\end{equation}}
Equations~\eqref{eq:lemma2-outcomes-1}--\eqref{eq:lemma2-cs} complete the
equilibrium characterization. \(\Halmos\)\vspace{3mm}

\noindent\textit{Proof of Proposition \ref{prop: assis_price and profits}:}
Throughout, comparisons involving the assistant are conditional on the
existence of a pure-strategy equilibrium, that is, on
\(\omega\in\mathcal R_1\cup\mathcal R_{2a}\cup\mathcal R_{2b}\).
We compare firm prices, firm profits, and platform profit with and without the
AI assistant.  Maintain the notation \(x,Z,D,E,K\) from the
proof of Lemma~\ref{lem: assis_equil}.

We first consider \(S_P=\{A,C\}\).

\noindent\textbf{Region \(\mathcal{R}_1\).} Direct simplification gives
\[
\begin{aligned}
p_A^{s1}-p_A^{n,AC}
&=-\frac{Z[3(1+\lambda)^2x^2
+12\lambda(1+\lambda)x+12\lambda^2]}{2D(3x+4)}<0,\\
p_C^{s1}-p_C^{n,AC}
&=-\frac{Z x[3(\lambda^2-\lambda+2)x
+2\lambda(\lambda+2)]}{2D(3x+4)}<0,\\
\widehat{\pi}_A^{s1}-\widehat{\pi}_A^{n,AC}
&=-\frac{Z H_A^{sAC}(x,\lambda)}{4D^2(3x+4)^2}<0,\\
\widehat{\pi}_C^{s1}-\widehat{\pi}_C^{n,AC}
&=-\frac{Z xH_C^{sAC}(x,\lambda)}{4D^2(3x+4)^2}<0,
\end{aligned}
\]
where
\[
\begin{aligned}
H_A^{sAC}(x,\lambda)={}&16\lambda^3(8+\lambda)
+32x\lambda^2(9+8\lambda+\lambda^2)
+16x^2\lambda(15+24\lambda+14\lambda^2+\lambda^3)\\
&+24x^3(3+10\lambda+7\lambda^2+4\lambda^3)
+9x^4(7+3\lambda+5\lambda^2+\lambda^3)>0,\\
H_C^{sAC}(x,\lambda)={}&12\lambda^2(4-\lambda^2)
+x(80\lambda+48\lambda^2-20\lambda^4)\\
&+x^2(36+72\lambda+15\lambda^2-6\lambda^3-9\lambda^4)
+9x^3(4-\lambda+2\lambda^2-\lambda^3)>0
\end{aligned}
\]
for \(x>0\) and \(0<\lambda<1\).  Hence both on-platform firms earn
lower profits under the assistant, and
\(\widehat{\Pi}^{s1}=\widehat{\pi}_A^{s1}+\widehat{\pi}_C^{s1}
<\widehat{\pi}_A^{n,AC}+\widehat{\pi}_C^{n,AC}=\widehat{\Pi}^{n,AC}\).

\noindent\textbf{Region \(\mathcal{R}_{2a}\).} To display
all four comparisons in the same form, divide by
\(Z>0\):
\[
\begin{aligned}
\frac{p_A^{s2a}-p_A^{n,AC}}{Z}
&=\frac{2[2(1-\lambda)(1+\delta x)(3x+4)-(1+x)E]}
{E(3x+4)},\\
\frac{p_C^{s2a}-p_C^{n,AC}}{Z}
&=\frac{x[2\delta(1-\lambda)(3x+4)-E]}
{E(3x+4)},\\
\frac{\widehat{\pi}_A^{s2a}-\widehat{\pi}_A^{n,AC}}{Z}
&=\frac{4[4(1-\lambda)(1+\delta x)^2(3x+4)^2
-(1+x)^2E^2]}{E^2(3x+4)^2},\\
\frac{\widehat{\pi}_C^{s2a}-\widehat{\pi}_C^{n,AC}}{Z}
&=\frac{x[4\delta^2(1-\lambda)^2(1+\delta x)(3x+4)^2
-(1+x)E^2]}{E^2(3x+4)^2}.
\end{aligned}
\]
Every denominator is positive. In the first and third cases defining
\(\mathcal L_{2a}\), \(\lambda>\underline{\lambda}_{1}^{\,s}\); in the
second, \(\lambda>\underline{\lambda}_{2}^{\,s}\geq
\underline{\lambda}_{1}^{\,s}\). The first three bracketed numerators are
nonpositive at \(\lambda=\underline{\lambda}_{1}^{\,s}\) and strictly
decreasing in \(\lambda\), so they are strictly negative throughout
\(\mathcal L_{2a}\). The fourth is also strictly negative: the second numerator gives
\(2\delta(1-\lambda)(3x+4)<E\), which implies
\(4\delta^2(1-\lambda)^2(1+\delta x)(3x+4)^2
<(1+\delta x)E^2\leq(1+x)E^2\). Thus both on-platform prices and
profits are lower throughout \(\mathcal L_{2a}\). Since
\(\mathcal R_{2a}\subseteq\mathcal L_{2a}\), the same conclusion holds at
every Type-2a equilibrium.  Since normalized platform profit is the sum of these two
firms' normalized profits, it follows immediately that \(\widehat{\Pi}^{s2a}<\widehat{\Pi}^{n,AC}\).

\noindent\textbf{Region \(\mathcal{R}_{2b}\).}
Using \(K>0\) on \(\mathcal L_{2b}\), the corresponding comparisons are
\[
\begin{aligned}
\frac{p_A^{s2b}-p_A^{n,AC}}{Z}
&=\frac{2[(1-\lambda)(1+\delta x)(3x+4)-(1+x)K]}
{K(3x+4)},\\
\frac{p_C^{s2b}-p_C^{n,AC}}{Z}
&=\frac{x[\delta(1-\lambda)(3x+4)-K]}{K(3x+4)},\\
\frac{\widehat{\pi}_A^{s2b}-\widehat{\pi}_A^{n,AC}}{Z}
&=\frac{4[(1-\lambda)(1+\delta x)^2(3x+4)^2
-(1+x)^2K^2]}{K^2(3x+4)^2},\\
\frac{\widehat{\pi}_C^{s2b}-\widehat{\pi}_C^{n,AC}}{Z}
&=\frac{x[\delta^2(1-\lambda)^2(1+\delta x)(3x+4)^2
-(1+x)K^2]}{K^2(3x+4)^2}.
\end{aligned}
\]
By definition, every point in \(\mathcal L_{2b}\) satisfies
\(\lambda>\underline{\lambda}_{2b}^{\,s}\). The first three bracketed
numerators are nonpositive at
\(\lambda=\underline{\lambda}_{2b}^{\,s}\) and strictly decreasing in
\(\lambda\), so they are strictly negative throughout
\(\mathcal L_{2b}\). The fourth is also strictly negative: the second
numerator gives \(\delta(1-\lambda)(3x+4)<K\), which implies
\(\delta^2(1-\lambda)^2(1+\delta x)(3x+4)^2
<(1+\delta x)K^2\leq(1+x)K^2\). Thus both on-platform prices and
profits are lower throughout \(\mathcal L_{2b}\). Since
\(\mathcal R_{2b}\subseteq\mathcal L_{2b}\), the same conclusion holds at
every Type-2b equilibrium, and summing the two normalized profit differences gives
\(\widehat{\Pi}^{s2b}<\widehat{\Pi}^{n,AC}\).  Combining the three regions,
the assistant lowers the prices and profits of both on-platform firms, and
therefore platform profit, under \(S_P=\{A,C\}\).

We next consider \(S_P=\{A,B\}\).

\noindent\textbf{Region \(\mathcal{R}_1\).} Again, direct simplification gives
\[
\begin{aligned}
p_A^{s1}-p_A^{n,AB}
&=-\frac{3Z(1-\lambda)^2x(x+\lambda)}
{2D(3x+4-\lambda)}<0,\\
p_B^{s1}-p_B^{n,AB}
&=-\frac{3Z(1-\lambda)^2x(x+\lambda)}
{D(3x+4-\lambda)}<0,\\
\widehat{\pi}_A^{s1}-\widehat{\pi}_A^{n,AB}
&=-\frac{Z(1-\lambda)^2x(x+\lambda)H_A^{sAB}(x,\lambda)}
{4D^2(3x+4-\lambda)^2}<0,\\
\widehat{\pi}_B^{s1}-\widehat{\pi}_B^{n,AB}
&=-\frac{Z(1-\lambda)^2x(x+\lambda)H_B^{sAB}(x,\lambda)}
{D^2(3x+4-\lambda)^2}<0,
\end{aligned}
\]
where
\[
\begin{aligned}
H_A^{sAB}(x,\lambda)
&=63x^2+72x+24\lambda(4-\lambda)
+x\lambda(87-15\lambda)+9x^2\lambda>0,\\
H_B^{sAB}(x,\lambda)
&=9x^2+9x+9x\lambda+\lambda(8+2\lambda-\lambda^2)>0.
\end{aligned}
\]
Both inequalities hold for \(x>0\) and \(0<\lambda<1\).
Hence both on-platform firms earn lower profits under the assistant, and
\(\widehat{\Pi}^{s1}<\widehat{\Pi}^{n,AB}\).

\noindent\textbf{Region \(\mathcal{R}_{2a}\).} The global-deviation
condition is important for this comparison.  By
\eqref{eq:lemma2-B-cross-IC}, every Type-2a equilibrium satisfies
\(\Phi_{2a}\geq0\).  This condition
implies
\[
\Gamma\equiv2(1-\delta)-\delta^2x(1+x)>0.
\]
To see the implication directly, \(\Phi_{2a}\geq0\) first requires \(\delta<1\)
and
\(x<\frac{2(1-\delta)^2}{\delta(4\delta-\delta^2-1)}\).  Since \(\Gamma\)
is decreasing in \(x\), substitution of this upper bound gives
\[
\Gamma>
\frac{2(\delta-1)(1+\delta-7\delta^2+\delta^3)}
{(1-4\delta+\delta^2)^2}>0
\qquad\text{for }\frac12\leq\delta<1.
\]

The two price comparisons now share the same positive factor:
\[
\frac{p_A^{s2a}-p_A^{n,AB}}{Z}
=\frac{2(1-\lambda)^2\Gamma}{(3x+4-\lambda)E}>0,
\qquad
p_B^{s2a}-p_B^{n,AB}=2(p_A^{s2a}-p_A^{n,AB})>0.
\]
Firm \(A\)'s normalized profit difference can be decomposed as
\[
\frac{\widehat{\pi}_A^{s2a}-\widehat{\pi}_A^{n,AB}}{Z}
=\frac{4(1-\lambda)^2\Gamma
\left[2(1+\delta x)(3x+4-\lambda)+(1+x)E\right]}
{E^2(3x+4-\lambda)^2}>0.
\]

The normalized platform profit difference is
\[
\frac{\widehat{\Pi}^{s2a}-\widehat{\Pi}^{n,AB}}{Z}
=\frac{(1-\lambda)G_P^{2a}(x,\delta,\lambda)}
{E^2(3x+4-\lambda)^2},
\]
where
\[
\begin{aligned}
G_P^{2a}(x,\delta,\lambda)
\equiv{}&\left\{16(1+\delta x)^2+
\left[2(1-\delta(1-\lambda))
+\delta x(2-\delta(1-\lambda))\right]^2\right\}
(3x+4-\lambda)^2\\
&-(1+x)(5x+4+\lambda)E^2.
\end{aligned}
\]
Every \(\lambda\in\mathcal L_{2a}\) satisfies
\(\lambda>\underline{\lambda}_{1}^{\,s}\). If
\(13(1+\delta x)\leq8\delta(1+x)(2+\delta x)\), the unique maximizer of
\(\Phi_{2a}\) with respect to \(\lambda\) is no greater than
\(\underline{\lambda}_{1}^{\,s}\). Because
\(\Phi_{2a}(\underline{\lambda}_{1}^{\,s})<0\), the condition
\(\Phi_{2a}\geq0\) can therefore hold only if
\(13(1+\delta x)>8\delta(1+x)(2+\delta x)\).
Differentiating \(G_P^{2a}\) and collecting in powers of
\(1-\lambda\) gives
\[
\begin{aligned}
\frac{\partial G_P^{2a}}{\partial\lambda}
={}&-12(1+x)(1+\delta x)
[13(1+\delta x)-8\delta(1+x)(2+\delta x)]\\
&-8(1-\lambda)\{5-6\delta+4\delta^2
+\delta x(4-\delta+4\delta^2)+\delta^3x^2(5+\delta)
+\delta^3x^3(1+2\delta)+\delta^4x^4\}\\
&-\delta(2+\delta x)(1-\lambda)^2
\{3[14\delta-4+10\delta x+7\delta^2x(1+x)]
+4\delta(2+\delta x)(1-\lambda)\}<0.
\end{aligned}
\]
Since \(G_P^{2a}(x,\delta,1)=0\) and \(\lambda<1\), it follows that
\(G_P^{2a}>0\). Therefore, every Type-2a Nash equilibrium raises both
on-platform prices and increases firm \(A\)'s profit and platform profit
relative to the no-AI \(S_P=\{A,B\}\) benchmark.

Firm \(B\)'s profit need not rise.  Its normalized profit difference is
\[
\frac{\widehat{\pi}_B^{s2a}-\widehat{\pi}_B^{n,AB}}{Z}
=\frac{(1-\lambda)G_B^{2a}(x,\delta,\lambda)}
{E^2(3x+4-\lambda)^2},
\]
where
\[
G_B^{2a}(x,\delta,\lambda)
\equiv
\left[2(1-\delta(1-\lambda))
+\delta x(2-\delta(1-\lambda))\right]^2
(3x+4-\lambda)^2-(1+x)(x+\lambda)E^2.
\]
Direct differentiation gives
\[
\begin{aligned}
\frac{\partial G_B^{2a}}{\partial\lambda}
={}&12(1+x)(1+\delta x)
[-5+8\delta+3\delta x+4\delta^2x(1+x)]\\
&-8(1-\lambda)[2\delta-1+\delta x+\delta^2x(1+x)]
[4\delta-1+3\delta x+2\delta^2x(1+x)]\\
&-\delta(2+\delta x)(1-\lambda)^2
\{3[14\delta-4+10\delta x+7\delta^2x(1+x)]
+4\delta(2+\delta x)(1-\lambda)\}<0,
\end{aligned}
\]
where the inequality follows from \(\Phi_{2a}\geq0\). For fixed
\((x,\delta)\), \(\mathcal L_{2a}\cap\{\Phi_{2a}\geq0\}\) restricts
\(\lambda\) to an interval because \(\mathcal L_{2a}\) restricts
\(\lambda\) to an interval and strict concavity of \(\Phi_{2a}\) implies
that \(\{\lambda:\Phi_{2a}\geq0\}\) is also an interval. Hence
\(G_B^{2a}\) is strictly decreasing and has at most one root on this
interval. Let \(\bar\lambda_B^s\) denote this unique root, when such a root
exists. If no such root exists, set \(\bar\lambda_B^s\) equal to the
upper endpoint of the Type-2a \(\lambda\)-interval when the relevant interval
is nonempty and \(G_B^{2a}>0\) throughout it, and to the lower endpoint
otherwise. Since
\(\mathcal R_{2a}\subseteq
\mathcal L_{2a}\cap\{\Phi_{2a}\geq0\}\), every Type-2a equilibrium lies
in this interval. It follows that every Type-2a equilibrium satisfies
\[
\widehat{\pi}_B^{s2a}>\widehat{\pi}_B^{n,AB}
\quad\text{iff}\quad
\lambda<\bar\lambda_B^s.
\]
Both signs occur at Type-2a equilibria.  In particular,
\[
\renewcommand{\arraystretch}{1.35}
\begin{array}{c|c}
(x,\delta,\lambda)&
(\widehat{\pi}_B^{s2a}-\widehat{\pi}_B^{n,AB})/Z\\ \hline
(\tfrac14,\tfrac12,\tfrac7{10})&
\tfrac{1047187}{1554180615}>0\\
(\tfrac25,\tfrac12,\tfrac12)&
-\tfrac{3029}{106142450}<0.
\end{array}
\]
At both parameter vectors, the conditions defining \(\mathcal L_{2a}\) hold
and \(V_A^{2a}=V_B^{2a}=V_C^{2a}=0\); hence both belong to
\(\mathcal R_{2a}\).

\noindent\textbf{Region \(\mathcal{R}_{2b}\).} Define
\[
F\equiv(2\delta-1)+\delta x+\delta^2x(1+x)>0.
\]
Direct simplification gives
\[
p_A^{s2b}-p_A^{n,AB}
=\frac{2Z(1-\lambda)(x+\lambda)F}
{(3x+4-\lambda)K}>0,
\qquad
p_B^{s2b}-p_B^{n,AB}=2(p_A^{s2b}-p_A^{n,AB})>0.
\]
The normalized profit comparison for firm \(A\) has the factorization
\[
\frac{\widehat{\pi}_A^{s2b}-\widehat{\pi}_A^{n,AB}}{Z}
=\frac{4(1-\lambda)(x+\lambda)F
[(1+\delta x)(3x+4-\lambda)+(1+x)K]}
{K^2(3x+4-\lambda)^2}>0.
\]
The normalized platform profit comparison has the factorization
\[
\begin{aligned}
\frac{\widehat{\Pi}^{s2b}-\widehat{\Pi}^{n,AB}}{Z}
={}&\frac{2(1-\lambda)(x+\lambda)F}{K^2(3x+4-\lambda)^2}
\Bigl\{24-16\delta-2\lambda(1-2\delta+\delta\lambda)\\
&\quad+x[22-4\delta-8\delta^2-2\delta(1-\delta)\lambda
-\delta^2\lambda^2]+(8\delta-14\delta^2)x^2-7\delta^2x^3\Bigr\}.
\end{aligned}
\]
On \(\mathcal L_{2b}\), \(Q>\underline Q^s\) implies
\(0<x<1/2\), while \(1/2\leq\delta<1\) and
\(0<\lambda<1\). Under these parameter bounds, the expression in braces
is at least \((1+x)^2(8-7x)>0\). Since all remaining factors are
positive, \(\widehat{\Pi}^{s2b}>\widehat{\Pi}^{n,AB}\). Thus Type 2b also raises both
on-platform prices, firm \(A\)'s profit, and platform profit.

Firm \(B\)'s normalized profit difference is
\[
\frac{\widehat{\pi}_B^{s2b}-\widehat{\pi}_B^{n,AB}}{Z}
=\frac{2(1-\lambda)(x+\lambda)F G_B^{2b}(x,\delta,\lambda)}
{K^2(3x+4-\lambda)^2},
\]
where
\[
\begin{aligned}
G_B^{2b}(x,\delta,\lambda)
\equiv{}&8(1+x)-\delta\bigl[16+2x(8+x)
+\lambda^2(2+\delta x)-2\lambda(2+x)(2+\delta x)
+\delta x\{8+x(12+5x)\}\bigr].
\end{aligned}
\]
Since all remaining factors are positive, firm \(B\)'s profit increases
iff \(G_B^{2b}>0\). Moreover,
\(\partial G_B^{2b}/\partial\lambda
=2\delta(2+\delta x)(2+x-\lambda)>0\). Thus \(G_B^{2b}\) is strictly
increasing and has at most one root in the Type-2b \(\lambda\)-interval.
Let \(\underline\lambda_{B}^s\) denote the unique interior root in that
interval, when such a root exists; set it equal to the lower endpoint when
\(G_B^{2b}>0\) throughout that interval and to the upper endpoint when
\(G_B^{2b}\leq0\) throughout that interval.
Then
\[
\widehat{\pi}_B^{s2b}>\widehat{\pi}_B^{n,AB}
\quad\text{iff}\quad
\lambda>\underline\lambda_B^s.
\]
Both signs also occur at Type-2b equilibria.  In particular,
\[
\renewcommand{\arraystretch}{1.35}
\begin{array}{c|c}
(x,\delta,\lambda)&
(\widehat{\pi}_B^{s2b}-\widehat{\pi}_B^{n,AB})/Z\\ \hline
(\tfrac14,\tfrac12,\tfrac15)&
\tfrac{568}{9018009}>0\\
(\tfrac14,\tfrac12,\tfrac3{20})&
-\tfrac{13481}{449760032}<0.
\end{array}
\]
At both parameter vectors, the conditions defining \(\mathcal L_{2b}\) hold
and \(V_A^{2b}=V_B^{2b}=V_C^{2b}=0\); hence both belong to
\(\mathcal R_{2b}\).
\(\Halmos\)\vspace{3mm}

\noindent\textit{Proof of Proposition \ref{prop: assis_CS}:}
Throughout, comparisons involving the assistant are conditional on the
existence of a pure-strategy equilibrium, that is, on
\(\omega\in\mathcal R_1\cup\mathcal R_{2a}\cup\mathcal R_{2b}\).
We compare true consumer surplus with and without the AI assistant.
Maintain \(x,D,E,K\) from the proof of Lemma~\ref{lem: assis_equil} and
define
\[
\widehat\Psi_1\equiv\frac{\Psi_1}{Q^3},\qquad
\widehat\Psi_{2a}\equiv\frac{\Psi_{2a}}{Q^3},\qquad
\widehat\Psi_{2b}\equiv\frac{\Psi_{2b}}{Q^5}
\]
after substituting \(q_C=xQ\). The powers of \(Q\) differ because
\(\Psi_1\) and \(\Psi_{2a}\) are homogeneous of degree three, whereas
\(\Psi_{2b}\) is homogeneous of degree five. Since
\(\Xi(q_C+Q,q_C)=Q\Xi(1+x,x)\) and
\(\Xi(q_C+Q,q_C+\lambda Q)=Q\Xi(1+x,x+\lambda)\), every comparison below
has the common positive factor \(\Theta Q\).

We first consider \(S_P=\{A,C\}\).

\noindent\textbf{Region \(\mathcal R_1\).} Direct substitution gives
\[
CS^{s1}-CS^{n,AC}=\Theta Q\left[
\frac{\widehat\Psi_1}{8D^2}-\Xi(1+x,x)\right].
\]
The Type-1 assistant outcome is the same under both platform assortments.
Moreover, direct comparison gives
\(\Xi(1+x,x+\lambda)>\Xi(1+x,x)\), and hence
\(CS^{n,AB}>CS^{n,AC}\). The Type-1 comparison under \(S_P=\{A,B\}\),
established below, therefore implies \(CS^{s1}>CS^{n,AC}\).

\noindent\textbf{Region \(\mathcal R_{2a}\).} Here
\[
CS^{s2a}-CS^{n,AC}=\Theta Q\left[
\frac{\widehat\Psi_{2a}}{2E^2}-\Xi(1+x,x)\right].
\]
The Type-2a individual-utility comparison below shows that every consumer is
weakly better off. Moreover, a positive mass of former \(A\) buyers remain
with \(A\) and strictly benefit from its lower price. Hence
\(CS^{s2a}>CS^{n,AC}\) throughout \(\mathcal R_{2a}\).

\noindent\textbf{Region \(\mathcal R_{2b}\).} The consumer-surplus
comparison for the Type-2b equilibrium is
\[
CS^{s2b}-CS^{n,AC}=\Theta Q\left[
\frac{\widehat\Psi_{2b}}{2K^2}-\Xi(1+x,x)\right].
\]
The Type-2b individual-utility comparison below shows that every consumer is
weakly better off. Moreover, a positive mass of former \(A\) buyers remain
with \(A\) and strictly benefit from its lower price. Hence
\(CS^{s2b}>CS^{n,AC}\) throughout \(\mathcal R_{2b}\). Combining the three cases,
the assistant strictly raises consumer surplus under \(S_P=\{A,C\}\).

We next examine individual utilities under \(S_P=\{A,C\}\).  In this
discussion, write \(\mathbf p^s=\mathbf p^{sj}\) for the equilibrium of Type
\(j\in\{1,2a,2b\}\) under consideration.  Consumers who continue to buy
\(A\) or \(C\) are strictly better off because both on-platform prices are
lower; consumers who continue not to purchase obtain the same utility.  If a
former nonbuyer now purchases \(C\), then
\(\theta\geq\frac{p_C^s}{\delta q_C}\geq p_C^s/q_C\), so this consumer is weakly
better off. If a consumer switches from \(C\) to \(i\in\{A,B\}\), then
\[
\theta\geq\frac{p_i^s-p_C^s}{\delta(q_i-q_C)}
\geq\frac{p_i^s-p_C^s}{q_i-q_C}
>\frac{p_i^s-p_C^{n,AC}}{q_i-q_C},
\]
where the last inequality follows from \(p_C^s<p_C^{n,AC}\).  Hence this
consumer is also strictly better off in true utility.

Consumers who previously purchased \(A\) cannot switch to \(C\), because
\[
\min\left\{
\frac{p_B^s-p_C^s}{\delta(q_B-q_C)},
\frac{p_A^s-p_C^s}{q_A-q_C}
\right\}
<\frac{p_A^{n,AC}-p_C^{n,AC}}{\delta(q_A-q_C)}.
\]
For Type 2a, the conditions defining \(\mathcal L_{2a}\), together with
firm \(B\)'s cross-deviation condition \(\Phi_{2a}\geq0\), imply
\[
\frac{p_A^{s2a}-p_B^{s2a}}{\delta(q_A-q_B)}
\leq
\frac{p_A^{n,AC}-p_C^{n,AC}}{\delta(q_A-q_C)}.
\]
Likewise, \(\mathcal L_{2b}\) and \(\Phi_{2b}\leq0\) imply the same
inequality with \(\mathbf p^{s2a}\) replaced by \(\mathbf p^{s2b}\). Since
\(\mathcal R_{2a}\subseteq\mathcal L_{2a}\cap\{\Phi_{2a}\geq0\}\) and
\(\mathcal R_{2b}\subseteq\mathcal L_{2b}\cap\{\Phi_{2b}\leq0\}\), no former
\(A\) buyer switches to \(B\) in either niche equilibrium.

In Type 1, \(A\)-to-\(B\) switchers may exist. Such a consumer is worse off
precisely when
\(\theta>\frac{p_A^{n,AC}-p_B^{s1}}{q_A-q_B}\). Combining this condition with
the purchase thresholds gives the worse-off interval
\[
\theta\in\left[
\max\left\{
\frac{p_A^{n,AC}-p_C^{n,AC}}{\delta(q_A-q_C)},
\frac{p_A^{n,AC}-p_B^{s1}}{q_A-q_B}
\right\},
\frac{p_A^{s1}-p_B^{s1}}{\delta(q_A-q_B)}
\right].
\]
For example, at
\((x,\delta,\lambda)=(1,27/32,1/40)\), this interval has
length \((120385/14440608)\Theta>0\). At this parameter vector,
the conditions defining \(\mathcal L_1\) hold and
\(V_A^1=V_B^1=V_C^1=0\); hence it belongs to \(\mathcal R_1\).
Figure~\ref{fig: assis_CS}(b) illustrates these consumers.

We next consider \(S_P=\{A,B\}\).

\noindent\textbf{Region \(\mathcal R_1\).} Direct substitution gives
\[
\begin{aligned}
CS^{s1}-CS^{n,AB}
={}&\frac{\Theta Q(1-\lambda)^2x(x+\lambda)}
{8(3x+4-\lambda)^2D^2}\Bigl\{\lambda[224\delta-16-(40+16\delta)\lambda
+(11-10\delta)\lambda^2]\\
&\quad+x[216\delta-36+(198\delta-45)\lambda
-(18\delta+9)\lambda^2]+(198\delta-45)x^2
\Bigr\}.
\end{aligned}
\]
For \(x>0\), \(\delta\geq1/2\), and \(0<\lambda<1\), the
factor preceding the braces and every term inside them are strictly positive.
Hence \(CS^{s1}>CS^{n,AB}\).

\noindent\textbf{Region \(\mathcal R_{2a}\).} The consumer-surplus
difference is
\[
CS^{s2a}-CS^{n,AB}=\Theta QF_{2a}^{AB},\quad\text{where}\quad
F_{2a}^{AB}(x,\delta,\lambda)
\equiv\frac{\widehat\Psi_{2a}}{2E^2}
-\Xi(1+x,x+\lambda).
\]
Every Type-2a equilibrium lies in
\(\mathcal L_{2a}\cap\{\Phi_{2a}\geq0\}\), which is an interval in
\(\lambda\) for fixed \((x,\delta)\). The denominator of \(F_{2a}^{AB}\)
is positive, and its numerator contains the positive factor
\((1-\lambda)^2\). After cancelling this factor, direct differentiation
under the defining inequalities of the interval shows that the remaining
cubic is strictly decreasing in \(\lambda\). Its value at the left endpoint
is strictly negative; depending on the binding constraint, this endpoint is
\(\underline\lambda_1^s\), \(\underline\lambda_2^s\), or the smaller root
of \(\Phi_{2a}=0\). Hence \(F_{2a}^{AB}<0\), and therefore
\(CS^{s2a}<CS^{n,AB}\), at every Type-2a equilibrium.

\noindent\textbf{Region \(\mathcal R_{2b}\).} The consumer-surplus
difference is
\[
CS^{s2b}-CS^{n,AB}=\Theta QF_{2b}^{AB},\quad\text{where}\quad
F_{2b}^{AB}(x,\delta,\lambda)
\equiv\frac{\widehat\Psi_{2b}}{2K^2}
-\Xi(1+x,x+\lambda).
\]
The denominator of \(F_{2b}^{AB}\) is positive, and its numerator contains
the positive factor \(1-\lambda\). After cancelling this factor, direct
differentiation under the inequalities defining \(\mathcal L_{2b}\) shows
that the remaining cubic is strictly decreasing in \(\lambda\). Its value
at the upper endpoint \(\underline\lambda_2^s\) is strictly negative. Hence
\(F_{2b}^{AB}=0\) has at most one root in the Type-2b \(\lambda\)-interval.
If such a root exists, then \(F_{2b}^{AB}>0\) below it and
\(F_{2b}^{AB}<0\) above it; otherwise, \(F_{2b}^{AB}<0\) throughout the
interval.
Since \(\mathcal R_{2b}\subseteq\mathcal L_{2b}\), these conclusions continue
to hold after restricting attention to global Type-2b equilibria. Let \(\bar\lambda_{CS}^{s}\) denote the unique interior root in that interval,
when such a root exists, and set it equal to the lower endpoint when
\(F_{2b}^{AB}<0\) throughout that interval. Then
\[
CS^{s2b}>CS^{n,AB}
\quad\text{iff}\quad
\lambda<\bar\lambda_{CS}^{s},
\qquad
CS^{s2b}<CS^{n,AB}
\quad\text{iff}\quad
\lambda>\bar\lambda_{CS}^{s}.
\]
When the root exists, \(CS^{s2b}=CS^{n,AB}\) at
\(\lambda=\bar\lambda_{CS}^{s}\).

Both signs occur at Type-2b equilibria.  In particular,
\[
\renewcommand{\arraystretch}{1.35}
\begin{array}{c|c}
(x,\delta,\lambda)&(CS^{s2b}-CS^{n,AB})/(\Theta Q)\\ \hline
(\tfrac15,\tfrac12,\tfrac3{25})&
\tfrac{16841}{72257920}>0\\
(\tfrac14,\tfrac12,\tfrac3{20})&
-\tfrac{135745}{312876544}<0.
\end{array}
\]
At both parameter vectors, the conditions defining \(\mathcal L_{2b}\) hold
and \(V_A^{2b}=V_B^{2b}=V_C^{2b}=0\); hence both belong to
\(\mathcal R_{2b}\).

We finally examine individual utilities under \(S_P=\{A,B\}\).  In
\(\mathcal R_1\), both on-platform prices fall.  Consumers whose choices do
not change are therefore weakly better off.  Consumers who previously bought
\(A\) cannot switch to \(C\), because
\[
\min\left\{
\frac{p_B^{s1}-p_C^{s1}}{\delta(q_B-q_C)},
\frac{p_A^{s1}-p_C^{s1}}{q_A-q_C}
\right\}
<\frac{p_A^{n,AB}-p_B^{n,AB}}{\delta(q_A-q_B)}.
\]
Among those who switch from \(A\) to \(B\), the worse-off consumers are
\[
\theta\in\left[
\max\left\{
\frac{p_A^{n,AB}-p_B^{n,AB}}{\delta(q_A-q_B)},
\frac{p_A^{n,AB}-p_B^{s1}}{q_A-q_B}
\right\},
\frac{p_A^{s1}-p_B^{s1}}{\delta(q_A-q_B)}
\right].
\]
Moreover,
\(\frac{p_A^{n,AB}-p_B^{n,AB}}{\delta(q_A-q_B)}
<\frac{p_A^{s1}-p_B^{s1}}{\delta(q_A-q_B)}\), so a consumer who previously
bought \(B\) cannot switch to \(A\).  Among those who switch from \(B\) to
\(C\), the worse-off consumers are
\[
\theta\in\left[
\max\left\{
\frac{p_B^{n,AB}}{\delta q_B},
\frac{p_B^{n,AB}-p_C^{s1}}{q_B-q_C}
\right\},
\min\left\{
\frac{p_B^{s1}-p_C^{s1}}{\delta(q_B-q_C)},
\frac{p_A^{s1}-p_C^{s1}}{q_A-q_C}
\right\}
\right].
\]
The two displayed groups can coexist. At
\((x,\delta,\lambda)=(1,1/2,3/5)\), the
\(A\)-to-\(B\) and \(B\)-to-\(C\) worse-off intervals have lengths
\((5/168)\Theta\) and \((1/42)\Theta\), respectively.
At this parameter vector, the conditions defining \(\mathcal L_1\) hold and
\(V_A^1=V_B^1=V_C^1=0\); hence it belongs to \(\mathcal R_1\).
Figure~\ref{fig: assistant_CS_appendix} illustrates these groups.

\begin{figure}[h!]
  \FIGURE
  {\includegraphics[width=0.78\linewidth]{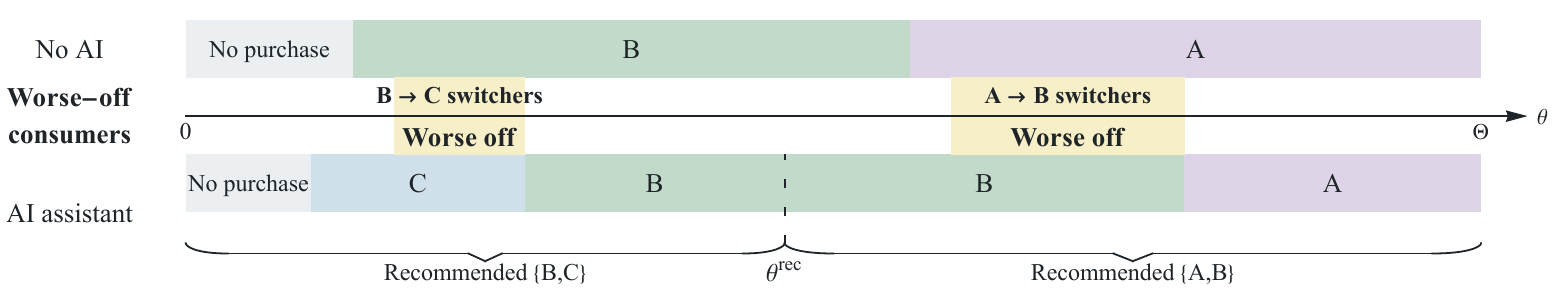}}
  {Individual Consumer Purchase and Welfare Changes under the AI Shopping
  Assistant with \(S_P=\{A,B\}\)\label{fig: assistant_CS_appendix}}
  {\scriptsize Figure follows the convention of
  Figure~\ref{fig: assis_CS}: outcomes above the axis show the no-AI
  benchmark, while those below show the assistant's recommended pairs and
  induced purchases. Switchers and stayers are defined as before.}
\end{figure}

It remains to consider Types 2a and 2b, under which both on-platform prices
rise at every Nash equilibrium of either type. For \(j\in\{2a,2b\}\), the relevant
thresholds satisfy
\[
\frac{p_B^{n,AB}}{\delta q_B}
<\frac{p_A^{sj}-p_C^{sj}}{q_A-q_C}
<\frac{p_A^{sj}-p_B^{sj}}{\delta(q_A-q_B)}
<\frac{p_A^{n,AB}-p_B^{n,AB}}{\delta(q_A-q_B)}.
\]
Consumers in
\[
\theta\in\left[
\frac{p_A^{sj}-p_C^{sj}}{q_A-q_C},
\frac{p_A^{sj}-p_B^{sj}}{\delta(q_A-q_B)}
\right]
\quad\text{and}\quad
\theta\in\left[
\frac{p_A^{n,AB}-p_B^{n,AB}}{\delta(q_A-q_B)},\Theta
\right]
\]
continue to buy \(B\) and \(A\), respectively, and are strictly worse off
because their prices are higher.  Consumers in
\[
\theta\in\left[
\frac{p_B^{n,AB}}{\delta q_B},
\frac{p_A^{sj}-p_C^{sj}}{q_A-q_C}
\right]
\]
switch from \(B\) to \(C\); direct comparison of their true utilities shows
that they are also strictly worse off.

Two other groups benefit. Under
\(\mathcal L_{2a}\cap\{\Phi_{2a}\geq0\}\) for \(j=2a\), and under
\(\mathcal L_{2b}\cap\{\Phi_{2b}\leq0\}\) for \(j=2b\), substitution of the
corresponding prices gives
\[
\frac{p_C^{sj}}{\delta q_C}
<\frac{p_B^{n,AB}}{\delta q_B},
\qquad
\frac{p_A^{sj}-p_B^{n,AB}}{q_A-q_B}
<\frac{p_A^{sj}-p_B^{sj}}{\delta(q_A-q_B)}
<\frac{p_A^{n,AB}-p_B^{n,AB}}{\delta(q_A-q_B)}.
\]
Hence consumers in
\[
\theta\in\left[
\frac{p_A^{sj}-p_B^{sj}}{\delta(q_A-q_B)},
\frac{p_A^{n,AB}-p_B^{n,AB}}{\delta(q_A-q_B)}
\right]
\]
switch from \(B\) to \(A\) and are strictly better off in true utility.
Consumers in
\[
\theta\in\left[
\frac{p_C^{sj}}{\delta q_C},
\frac{p_B^{n,AB}}{\delta q_B}
\right]
\]
previously did not purchase and now buy \(C\).
Figure~\ref{fig: assis_CS}(c) illustrates these groups.
\(\Halmos\)\vspace{3mm}

\noindent\textit{Proof of Proposition \ref{prop: assis_joint}:}
Combine Propositions~\ref{prop: assis_price and profits}
and~\ref{prop: assis_CS}.  Under \(S_P=\{A,C\}\), each of the three
assistant equilibrium types lowers platform profit and raises consumer
surplus.  Thus the assistant benefits consumers but hurts the platform.

Under \(S_P=\{A,B\}\), the joint effect depends on the equilibrium type.  In
\(\mathcal R_1\), platform profit falls and consumer surplus rises.  In
\(\mathcal R_{2a}\), platform profit rises and consumer surplus falls.  In
\(\mathcal R_{2b}\), platform profit rises, whereas Proposition~\ref{prop:
assis_CS} gives
\[
CS^{s2b}>CS^{n,AB}
\quad\text{iff}\quad
\lambda<\bar\lambda_{CS}^{s},
\qquad
CS^{s2b}<CS^{n,AB}
\quad\text{iff}\quad
\lambda>\bar\lambda_{CS}^{s}.
\]
Consequently, Type 2b produces a win--win outcome below the threshold and
benefits only the platform above it; consumer surplus is unchanged at an
interior threshold.  The win--win case is nonempty.  For example, at
\((x,\delta,\lambda)=(1/5,1/2,3/25)\), the Type-2b
profile is a global Nash equilibrium: as verified above,
\(\mathcal L_{2b}\) holds and
\(V_A^{2b}=V_B^{2b}=V_C^{2b}=0\). Moreover,
\[
\frac{\widehat{\Pi}^{s2b,AB}-\widehat{\Pi}^{n,AB}}{\Theta Q}
=\frac{10516}{3951605}>0,
\qquad
\frac{CS^{s2b}-CS^{n,AB}}{\Theta Q}
=\frac{16841}{72257920}>0.
\]
\(\Halmos\)\vspace{3mm}

\noindent\textit{Proof of Lemma \ref{lem: agent_equil}:}
The AI agent selects the product that maximizes
\(u_i(\theta)=\theta q_i-p_i\) from \(\{A,B,C\}\). The consumer purchases
the selected product if its true utility is nonnegative and otherwise does not
purchase. No equilibrium can
contain a zero price: the highest-quality zero-priced firm can raise its
price slightly and retain positive demand.  Nor can a positive-priced firm
receive zero demand.  By charging a sufficiently small positive price,
\(A\) can undercut both lower-quality products, \(C\) can attract sufficiently
low types, and \(B\) can undercut \(C\) while becoming affordable before
\(A\).  Each deviation yields positive demand and profit.  Hence every firm
has positive demand and \(p_A>p_B>p_C>0\).

By single crossing, consumers purchase \(C\), \(B\), and \(A\), respectively,
as \(\theta\) increases, with thresholds
\[
\frac{p_C}{q_C},\qquad
\frac{p_B-p_C}{q_B-q_C},\qquad
\frac{p_A-p_B}{q_A-q_B}.
\]
Thus
\[
\begin{aligned}
D_A&=1-\frac{p_A-p_B}{\Theta(q_A-q_B)},\\
D_B&=\frac{p_A-p_B}{\Theta(q_A-q_B)}
-\frac{p_B-p_C}{\Theta(q_B-q_C)},\\
D_C&=\frac{p_B-p_C}{\Theta(q_B-q_C)}
-\frac{p_C}{\Theta q_C}.
\end{aligned}
\]
The first-order conditions reduce to
\[
p_A=\frac{\Theta(q_A-q_B)+p_B}{2},\qquad
p_B=\frac{(q_B-q_C)p_A+(q_A-q_B)p_C}{2(q_A-q_C)},\qquad
p_C=\frac{q_Cp_B}{2q_B}.
\]
Solving them and substituting \(q_A=q_C+Q\) and
\(q_B=q_C+\lambda Q\) gives the prices stated in the lemma.  Direct
substitution gives
\[
p_A^g>p_B^g>p_C^g>0,\qquad
0<\frac{p_C^g}{q_C}<\frac{p_B^g-p_C^g}{q_B-q_C}
<\frac{p_A^g-p_B^g}{q_A-q_B}<\Theta,
\]
confirming that all three firms receive positive demand.

It remains to verify global best responses.  Holding \(p_B=p_B^g\), firm
\(A\)'s deviation normalized profit at price \(y\) is bounded above by
\(y\{1-\frac{y-p_B^g}{\Theta(q_A-q_B)}\}\) whenever demand is positive.
The difference between this bound and \(\widehat{\pi}_A^g\) is
\[
-\frac{(y-p_A^g)^2}{\Theta(q_A-q_B)}\leq0.
\]
A deviation giving \(A\) zero demand is also unprofitable.  Similarly,
firm \(C\)'s deviation normalized profit, whenever its purchase interval is nonempty,
is bounded above by
\(y\{\frac{p_B^g-y}{\Theta(q_B-q_C)}-\frac{y}{\Theta q_C}\}\).  The
difference from \(\widehat{\pi}_C^g\) is
\[
-\frac{1}{\Theta}
\left(\frac{1}{q_B-q_C}+\frac{1}{q_C}\right)(y-p_C^g)^2\leq0.
\]
If the interval is empty, firm \(C\)'s deviation profit is zero.

For firm \(B\), the lower endpoint of demand is the larger of \(y/q_B\)
and \(\frac{y-p_C^g}{q_B-q_C}\).  The two coincide at
\(y=\frac{q_Bp_C^g}{q_C}=p_B^g/2\).  For \(y\geq p_B^g/2\), the normalized deviation profit
difference is
\[
-\frac{1}{\Theta}
\left(\frac{1}{q_A-q_B}+\frac{1}{q_B-q_C}\right)(y-p_B^g)^2\leq0.
\]
For \(0\leq y\leq p_B^g/2\), deviation normalized profit is
\(y\{\frac{p_A^g-y}{\Theta(q_A-q_B)}-\frac{y}{\Theta q_B}\}\).  This
concave quadratic is increasing up to \(p_B^g/2\), where it equals
\((3/4)\widehat{\pi}_B^g<\widehat{\pi}_B^g\).  Thus the stated prices form a Nash equilibrium.
Conversely, every pure-strategy equilibrium has three positive demands and
satisfies the three first-order conditions above, whose unique solution is
the stated profile.

Maintaining \(D\) from Lemma~\ref{lem: assis_equil}, substitution into the
demand and true-utility expressions gives the following outcomes.

\footnotesize{
\[
p_A^{g}=\frac{\Theta(1-\lambda)[(\lambda+3)q_C+4\lambda Q]}{2D},\quad
p_B^{g}=\frac{\Theta(1-\lambda)\lambda(q_C+\lambda Q)}{D},\quad
p_C^{g}=\frac{\Theta(1-\lambda)\lambda q_C}{2D};
\]
\[
\begin{aligned}
\widehat{\pi}_A^{g}
&=
\frac{\Theta(1-\lambda)[(\lambda+3)q_C+4\lambda Q]^2}
{4QD^2},&
\widehat{\pi}_B^{g}
&=
\frac{\Theta(1-\lambda)\lambda(q_C+\lambda Q)^2}
{QD^2},\\
\widehat{\pi}_C^{g}
&=
\frac{\Theta(1-\lambda)^2\lambda q_C(q_C+\lambda Q)}
{4QD^2},&
CS^{g}
&=
\frac{\Theta\Psi_g}{8Q^2D^2}.
\end{aligned}
\]}
\normalsize
where
{\footnotesize
\[
\Psi_g
=
36q_C^3
+q_C^2Q(9+106\lambda-7\lambda^2)
+q_CQ^2(24\lambda+93\lambda^2-10\lambda^3+\lambda^4)
+Q^3(16\lambda^2+20\lambda^3).
\]}
For either hosted assortment, the actual equilibrium profits are
\(\pi_i^g=\kappa_i(S_P)\widehat{\pi}_i^g\) for each firm and
\(\Pi^{g,S_P}=\tau\sum_{i\in S_P}\widehat{\pi}_i^g\).
\(\Halmos\)\vspace{3mm}

\noindent\textit{Proof of Proposition \ref{prop: agent_price and profits}:}
We compare firm prices, firm profits, and platform profit under the AI shopping
agent with those under the no-AI benchmark.  Maintain \(x\) and \(D\) from
Lemma~\ref{lem: assis_equil}, and let
\(L\equiv(\lambda+3)x+4\lambda\).

\noindent\textbf{Assortment \(S_P=\{A,C\}\).}
Direct simplification gives
\[
\frac{p_C^g}{p_C^{n,AC}}
=\frac{(1-\lambda)\lambda(3x+4)}{2\delta D}<1,
\qquad
\frac{\widehat{\pi}_C^g}{\widehat{\pi}_C^{n,AC}}
=\frac{(1-\lambda)^2\lambda(x+\lambda)(3x+4)^2}
{4\delta(x+1)D^2}<1.
\]
The first inequality follows from \(\delta\geq1/2\) and
\(D-(1-\lambda)\lambda(3x+4)
=3x(1-\lambda)+2\lambda^2>0\); the second follows directly on the same
parameter domain.  Hence \(p_C^g<p_C^{n,AC}\) and
\(\widehat{\pi}_C^g<\widehat{\pi}_C^{n,AC}\).

For firm \(A\)'s price,
\[
\frac{p_A^g}{p_A^{n,AC}}
=\frac{1}{\delta}\frac{(1-\lambda)L(3x+4)}{4(x+1)D}.
\]
The ratio multiplying \(1/\delta\) is weakly decreasing in \(\lambda\) and
equals \(\frac{4+3x}{4(1+x)}\) at \(\lambda=0\).  Accordingly, define
\[
\overline{\lambda}_{A1}^g
\equiv
\begin{cases}
\displaystyle
1-
\frac{
6\delta(1+x)
}{
3x+4+2\delta+\sqrt{
(3x+4+2\delta)^2
-3\delta\bigl[(x+4)(3x+4)-4\delta(1+x)\bigr]
}
},
&
\displaystyle
\delta<\frac{4+3x}{4(1+x)},
\\[1.2em]
0,
&
\displaystyle
\delta\geq\frac{4+3x}{4(1+x)}.
\end{cases}
\]
Then \(p_A^g>p_A^{n,AC}\) iff
\(\lambda<\overline{\lambda}_{A1}^{g}\).  Since a larger \(\delta\) raises
the right-hand side of the defining comparison,
\(\overline{\lambda}_{A1}^{g}\) weakly decreases in \(\delta\).

For firm \(A\)'s profit,
\[
\frac{\widehat{\pi}_A^g}{\widehat{\pi}_A^{n,AC}}
=\frac{1}{\delta}
\frac{(1-\lambda)L^2(3x+4)^2}{16(x+1)^2D^2}.
\]
Let \(\overline{\lambda}_{A2}^{g}\) denote the unique interior root of
\[
\frac{(1-\lambda)L^2(3x+4)^2}{16(x+1)^2D^2}=\delta
\]
when one exists; set it equal to zero when
\(\widehat{\pi}_A^g>\widehat{\pi}_A^{n,AC}\) never holds and to one when the inequality holds
throughout.  Since the left-hand side is weakly decreasing in \(\lambda\),
\(\widehat{\pi}_A^g>\widehat{\pi}_A^{n,AC}\) iff
\(\lambda<\overline{\lambda}_{A2}^{g}\).  The threshold weakly decreases in
\(\delta\).

For normalized platform profit, \(\widehat{\Pi}^g=\widehat{\pi}_A^g+\widehat{\pi}_C^g\), and
\[
\frac{\widehat{\Pi}^g}{\widehat{\Pi}^{n,AC}}
=\frac{1}{\delta}
\frac{(1-\lambda)[L^2+(1-\lambda)\lambda x(x+\lambda)](3x+4)^2}
{4(x+1)(5x+4)D^2}.
\]
Let \(\overline{\lambda}_P^g\) denote the unique interior root of
\[
\frac{(1-\lambda)[L^2+(1-\lambda)\lambda x(x+\lambda)](3x+4)^2}
{4(x+1)(5x+4)D^2}=\delta
\]
when one exists, with the same endpoint convention.  The left-hand side is
weakly decreasing in \(\lambda\).  At \(\lambda=0\), it exceeds \(\delta\)
iff \(Q>\underline Q^g\), where
\[
\underline Q^g
\equiv
\frac{q_C[9\delta-6+\sqrt{\delta(\delta+8)}]}{8(1-\delta)}
\quad\text{for }\delta\in[\tfrac12,1),
\qquad
\underline Q^g\equiv+\infty\quad\text{at }\delta=1.
\]
Hence, if \(Q\leq\underline Q^g\), then
\(\widehat{\Pi}^g\leq\widehat{\Pi}^{n,AC}\) for every \(\lambda\).  If
\(Q>\underline Q^g\), then \(\widehat{\Pi}^g>\widehat{\Pi}^{n,AC}\) iff
\(\lambda<\overline{\lambda}_P^g\).  This threshold weakly decreases in
\(\delta\).

\noindent\textbf{Assortment \(S_P=\{A,B\}\).}
For firm \(A\)'s price,
\[
\frac{p_A^g}{p_A^{n,AB}}
=\frac{1}{\delta}\frac{L(3x+4-\lambda)}{4(x+1)D}.
\]
The ratio multiplying \(1/\delta\) is weakly increasing in \(\lambda\) and
equals \(\frac{4+3x}{4(1+x)}\) at \(\lambda=0\).  Accordingly, define
\[
\underline{\lambda}_{A1}^g
\equiv
\begin{cases}
0,
&
\displaystyle
\delta\leq\frac{4+3x}{4(1+x)},
\\[1.2em]
\displaystyle
1-
\frac{
24(1-\delta)(1+x)
}{
3x+8(1-\delta)+\sqrt{
\bigl[3x+8(1-\delta)\bigr]^2
-48(1-\delta)\bigl[3x-4(1-\delta)(1+x)\bigr]
}
},
&
\displaystyle
\delta>\frac{4+3x}{4(1+x)}.
\end{cases}
\]
Then \(p_A^g>p_A^{n,AB}\) iff
\(\lambda>\underline{\lambda}_{A1}^{g}\).  The threshold weakly increases in
\(\delta\).

For firm \(B\)'s price,
\[
\frac{p_B^g}{p_B^{n,AB}}
=\frac{1}{\delta}\frac{\lambda(3x+4-\lambda)}{D}.
\]
The ratio multiplying \(1/\delta\) is weakly increasing in \(\lambda\).
Accordingly, define
\[
\underline{\lambda}_{B1}^g
\equiv
\frac{
6\delta x
}{
3x+4(1-\delta)
+\sqrt{\bigl[3x+4(1-\delta)\bigr]^2-12\delta(1-\delta)x}
}.
\]
Then \(p_B^g>p_B^{n,AB}\) iff
\(\lambda>\underline{\lambda}_{B1}^{g}\), and the threshold weakly increases
in \(\delta\).

For firm profits,
\[
\frac{\widehat{\pi}_A^g}{\widehat{\pi}_A^{n,AB}}
=\frac{1}{\delta}
\frac{L^2(3x+4-\lambda)^2}{16(x+1)^2D^2},
\qquad
\frac{\widehat{\pi}_B^g}{\widehat{\pi}_B^{n,AB}}
=\frac{1}{\delta}
\frac{\lambda(x+\lambda)(3x+4-\lambda)^2}{(x+1)D^2}.
\]
Let \(\underline{\lambda}_{A2}^{g}\) and
\(\underline{\lambda}_{B2}^{g}\) denote the respective unique interior
roots of
\[
\frac{L^2(3x+4-\lambda)^2}{16(x+1)^2D^2}=\delta,
\qquad
\frac{\lambda(x+\lambda)(3x+4-\lambda)^2}{(x+1)D^2}=\delta,
\]
when they exist.  Set either threshold equal to one when the corresponding
profit increase never occurs and to zero when it occurs throughout.  Both
left-hand sides are weakly increasing in \(\lambda\).  Therefore,
\(\widehat{\pi}_A^g>\widehat{\pi}_A^{n,AB}\) iff
\(\lambda>\underline{\lambda}_{A2}^{g}\), and
\(\widehat{\pi}_B^g>\widehat{\pi}_B^{n,AB}\) iff
\(\lambda>\underline{\lambda}_{B2}^{g}\).  Both thresholds weakly increase
in \(\delta\).

Finally, \(\widehat{\Pi}^g=\widehat{\pi}_A^g+\widehat{\pi}_B^g\), and
\[
\frac{\widehat{\Pi}^g}{\widehat{\Pi}^{n,AB}}
=\frac{1}{\delta}
\frac{[L^2+4\lambda(x+\lambda)^2](3x+4-\lambda)^2}
{4(x+1)(5x+4+\lambda)D^2}.
\]
Let \(\underline{\lambda}_P^g\) denote the unique interior root of
\[
\frac{[L^2+4\lambda(x+\lambda)^2](3x+4-\lambda)^2}
{4(x+1)(5x+4+\lambda)D^2}=\delta
\]
when one exists; set it equal to one when the platform-profit increase never
occurs and to zero when it occurs throughout.  Since the left-hand side is
weakly increasing in \(\lambda\), \(\widehat{\Pi}^g>\widehat{\Pi}^{n,AB}\) iff
\(\lambda>\underline{\lambda}_P^g\).  The threshold weakly increases in
\(\delta\). \(\Halmos\)\vspace{3mm}

\noindent\textit{Proof of Proposition \ref{prop: agent_CS}:}
We compare consumer surplus under the AI agent with that under the
no-AI benchmark, maintaining \(x,D\) from Lemma~\ref{lem: assis_equil} and
\(\Xi\) from Lemma~\ref{lem:no AI_equil}.  Substituting \(q_C=xQ\), define
only the normalized agent polynomial
\[
\widehat{\Psi}_g(x,\lambda)\equiv\frac{\Psi_g}{Q^3}.
\]
The homogeneity of \(\Xi\) gives
\(\Xi(q_C+Q,q_C)=Q\Xi(1+x,x)\) and
\(\Xi(q_C+Q,q_C+\lambda Q)=Q\Xi(1+x,x+\lambda)\).

\noindent\textbf{Assortment \(S_P=\{A,C\}\).}
Direct simplification gives
\[
CS^g-CS^{n,AC}
=\Theta Q F_{AC}(x,\delta,\lambda),
\qquad
F_{AC}(x,\delta,\lambda)
\equiv
\frac{\widehat\Psi_g(x,\lambda)}{8D^2}-\Xi(1+x,x).
\]
Moreover,
\[
\frac{\partial F_{AC}}{\partial\delta}
=\frac{(x+1)(5x+4)}{(3x+4)^2}>0
\]
and
\[
\frac{\partial F_{AC}}{\partial\lambda}
=\frac{(x+\lambda)
[-\lambda^3x+10\lambda^3+63\lambda^2x+56\lambda^2
+51\lambda x^2+70\lambda x+15x^2]}{4D^3}>0.
\]
The last inequality follows from \(0<\lambda<1\) and \(x>0\).  Hence
\(F_{AC}\) is increasing in \(\lambda\).  Let
\(\underline{\lambda}_{CS}^g\) denote its unique interior root when one
exists; set it equal to one when \(CS^g>CS^{n,AC}\) never holds and to zero
when it holds throughout.  Thus \(CS^g>CS^{n,AC}\) iff
\(\lambda>\underline{\lambda}_{CS}^g\).  Since \(F_{AC}\) is increasing in
\(\delta\), this threshold weakly decreases in \(\delta\).

We next examine individual utilities.  When the agent lowers both
on-platform prices, every consumer is weakly better off: without salience
distortion, each consumer can make the same purchase as in the no-AI
benchmark at a weakly lower price.  If the agent instead raises \(A\)'s
price and lowers \(C\)'s price, consumers who previously purchased \(C\)
remain weakly better off.  Consumers who continue to purchase \(A\) are
worse off precisely on
\[
\theta\in
\left[
\max\left\{
\frac{p_A^{n,AC}-p_C^{n,AC}}{\delta(q_A-q_C)},
\frac{p_A^g-p_B^g}{q_A-q_B}
\right\},\Theta
\right].
\]
Consumers who previously purchased \(A\) cannot switch to \(C\), because
\(\frac{p_B^g-p_C^g}{q_B-q_C}
<\frac{p_A^{n,AC}-p_C^{n,AC}}{\delta(q_A-q_C)}\).  It therefore remains to
consider those who switch from \(A\) to \(B\).  Such a consumer is worse off
when \(\theta>\frac{p_A^{n,AC}-p_B^g}{q_A-q_B}\).  Combining the purchase and
welfare conditions gives the interval
\[
\theta\in
\left[
\max\left\{
\frac{p_A^{n,AC}-p_C^{n,AC}}{\delta(q_A-q_C)},
\frac{p_A^{n,AC}-p_B^g}{q_A-q_B}
\right\},
\frac{p_A^g-p_B^g}{q_A-q_B}
\right].
\]
Figure~\ref{fig: agent_CS}(c) illustrates these groups.

\noindent\textbf{Assortment \(S_P=\{A,B\}\).}
Here,
\[
CS^g-CS^{n,AB}
=\Theta Q F_{AB}(x,\delta,\lambda),
\qquad
F_{AB}(x,\delta,\lambda)
\equiv
\frac{\widehat\Psi_g(x,\lambda)}{8D^2}-\Xi(1+x,x+\lambda).
\]
Differentiating with respect to \(\delta\) gives
\(\frac{\partial F_{AB}}{\partial\delta}
=\frac{(1-\lambda)(1+x)(4+5x+\lambda)}{(4+3x-\lambda)^2}>0\).
Solving \(F_{AB}=0\) for \(\delta\) gives
\[
\begin{aligned}
\delta=
\frac{12(1+x)^2-(1+x)(x+\lambda)-2(x+\lambda)^2}
{2(1-\lambda)(4+5x+\lambda)}-\frac{(4+3x-\lambda)^2\widehat\Psi_g(x,\lambda)}
{8(1+x)(1-\lambda)(4+5x+\lambda)D^2}.
\end{aligned}
\]
The right-hand side is strictly increasing in \(\lambda\): its derivative
has a positive denominator, and every coefficient of its numerator in powers
of \(x\) is positive for \(0<\lambda<1\). Thus, for fixed \((x,\delta)\),
there is at most one \(\lambda\in(0,1)\) satisfying
\(F_{AB}(x,\delta,\lambda)=0\). When it exists, let
\(\overline{\lambda}_{CS}^g\) denote this value; set the threshold equal to zero
when the positive region is empty and to one when it covers the entire
interval. Since \(F_{AB}\) is strictly increasing in \(\delta\), it is
positive below the threshold and negative above it. Consequently,
\(CS^g>CS^{n,AB}\) iff \(\lambda<\overline{\lambda}_{CS}^g\). Since
\(F_{AB}\) is increasing in \(\delta\), this threshold weakly increases in
\(\delta\).

When both on-platform prices fall, every consumer is weakly better off.
Since \(\underline{\lambda}_{A1}^{g}<\underline{\lambda}_{B1}^{g}\), it is
also possible that \(A\)'s price rises while \(B\)'s price falls.  Consumers
who previously purchased \(B\) are then weakly better off.  The consumers
who continue to purchase \(A\) and are worse off are
\[
\theta\in
\left[
\max\left\{
\frac{p_A^{n,AB}-p_B^{n,AB}}{\delta(q_A-q_B)},
\frac{p_A^g-p_B^g}{q_A-q_B}
\right\},\Theta
\right].
\]
Consumers who previously purchased \(A\) cannot switch to \(C\), because
\(\frac{p_B^g-p_C^g}{q_B-q_C}
<\frac{p_A^{n,AB}-p_B^{n,AB}}{\delta(q_A-q_B)}\).  Among those who switch from
\(A\) to \(B\), utility falls precisely when
\(\theta>\frac{p_A^{n,AB}-p_B^g}{q_A-q_B}\).  Hence their worse-off interval is
\[
\theta\in
\left[
\max\left\{
\frac{p_A^{n,AB}-p_B^{n,AB}}{\delta(q_A-q_B)},
\frac{p_A^{n,AB}-p_B^g}{q_A-q_B}
\right\},
\frac{p_A^g-p_B^g}{q_A-q_B}
\right].
\]
Figure~\ref{fig: agent_CS_appendix} illustrates these groups.

\begin{figure}[h!]
  \FIGURE
  {\includegraphics[width=0.78\linewidth]{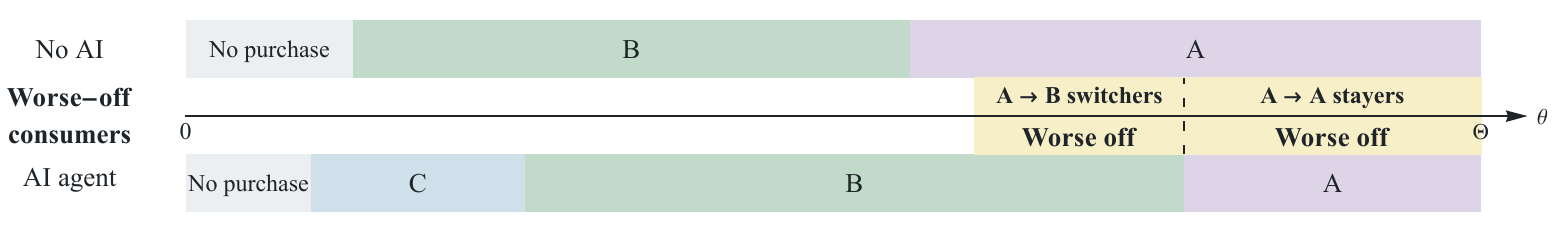}}
  {Individual Consumer Purchase and Welfare Changes under the AI Shopping
  Agent with \(S_P=\{A,B\}\)\label{fig: agent_CS_appendix}}
  {\scriptsize Figure follows the convention of
  Figure~\ref{fig: agent_CS}: outcomes above the axis show the no-AI
  benchmark, while those below show the agent-selected purchases. Switchers
  and stayers are defined as before.}
\end{figure}

It remains to consider the case in which both on-platform prices rise.
The two price-increase conditions from
Proposition~\ref{prop: agent_price and profits} are
\[
p_A^g>p_A^{n,AB}
\quad\text{iff}\quad \lambda>\underline{\lambda}_{A1}^g,
\qquad
p_B^g>p_B^{n,AB}
\quad\text{iff}\quad \lambda>\underline{\lambda}_{B1}^g.
\]
The sign characterization above implies
\(\max\{\underline{\lambda}_{A1}^g,\underline{\lambda}_{B1}^g\}
\geq \overline{\lambda}_{CS}^g\).
Hence, the two price-increase conditions imply
\(\lambda>\max\{\underline{\lambda}_{A1}^g,\underline{\lambda}_{B1}^g\}
\geq \overline{\lambda}_{CS}^g\), and therefore
\(CS^g-CS^{n,AB}=\Theta Q F_{AB}(x,\delta,\lambda)<0\).
Thus, consumer surplus strictly falls.  The individual welfare effects reinforce this result.
Consumers who continue to buy \(A\) or \(B\) pay higher prices and are strictly
worse off.  Moreover,
\(\frac{p_A^g-p_B^g}{q_A-q_B}
>\frac{p_A^{n,AB}-p_B^{n,AB}}{\delta(q_A-q_B)}\).  The higher
purchase threshold under the agent causes some consumers who previously bought
\(A\) to switch to the lower-quality \(B\), and these switchers are also
worse off.  The respective intervals for worse-off consumers who continue to
buy \(B\), continue to buy \(A\), and switch from \(A\) to \(B\) are
\[
\begin{aligned}
&\theta\in
\left[
\max\left\{\frac{p_B^{n,AB}}{\delta q_B},
\frac{p_B^g-p_C^g}{q_B-q_C}\right\},
\frac{p_A^{n,AB}-p_B^{n,AB}}{\delta(q_A-q_B)}
\right],\\
&\theta\in
\left[\frac{p_A^g-p_B^g}{q_A-q_B},\Theta\right],
\qquad
\theta\in
\left[
\frac{p_A^{n,AB}-p_B^{n,AB}}{\delta(q_A-q_B)},
\frac{p_A^g-p_B^g}{q_A-q_B}
\right].
\end{aligned}
\]
Finally, some consumers switch from \(B\) to \(C\).  Such a switcher is
worse off precisely when
\(\theta>\frac{p_B^{n,AB}-p_C^g}{q_B-q_C}\).  Combining this condition with
the switching interval gives
\[
\theta\in
\left[
\max\left\{\frac{p_B^{n,AB}}{\delta q_B},
\frac{p_B^{n,AB}-p_C^g}{q_B-q_C}\right\},
\frac{p_B^g-p_C^g}{q_B-q_C}
\right].
\]
Figure~\ref{fig: agent_CS}(d) illustrates these groups. \(\Halmos\)\vspace{3mm}

\noindent\textit{Proof of Proposition \ref{prop: agent_joint}:}
The result follows by combining
Propositions~\ref{prop: agent_price and profits} and~\ref{prop: agent_CS}.

\noindent\textbf{Assortment \(S_P=\{A,C\}\).}
The preceding propositions give
\[
\widehat{\Pi}^g>\widehat{\Pi}^{n,AC}
\quad\text{iff}\quad
Q>\underline Q^g\ \text{ and }\ \lambda<\overline{\lambda}_P^g,
\qquad
CS^g>CS^{n,AC}
\quad\text{iff}\quad
\lambda>\underline{\lambda}_{CS}^g.
\]
Thus both platform profit and consumer surplus increase when
\[
Q>\underline Q^g,
\qquad
\underline{\lambda}_{CS}^g
<\lambda<\overline{\lambda}_P^g.
\]
This win--win region is nonempty. For example, at
\((x,\delta,\lambda)=(1/4,1/2,1/2)\),
\[
\frac{\widehat{\Pi}^g-\widehat{\Pi}^{n,AC}}{\Theta Q}
=\frac{6013}{288800}>0,
\qquad
\frac{CS^g-CS^{n,AC}}{\Theta Q}
=\frac{761}{30400}>0.
\]
If \(\lambda>\underline{\lambda}_{CS}^g\) and either
\(Q\leq\underline Q^g\) or
\(\lambda\geq\overline{\lambda}_P^g\), consumer surplus increases while
platform profit weakly decreases. If \(Q>\underline Q^g\),
\(\lambda<\overline{\lambda}_P^g\), and
\(\lambda\leq\underline{\lambda}_{CS}^g\), platform profit increases while
consumer surplus weakly decreases. If
\(\lambda\leq\underline{\lambda}_{CS}^g\) and either
\(Q\leq\underline Q^g\) or
\(\lambda\geq\overline{\lambda}_P^g\), both weakly decrease. These cases
establish all four joint outcomes under \(S_P=\{A,C\}\).

\noindent\textbf{Assortment \(S_P=\{A,B\}\).}
Here, the preceding propositions give
\[
\widehat{\Pi}^g>\widehat{\Pi}^{n,AB}
\quad\text{iff}\quad
\lambda>\underline{\lambda}_P^g,
\qquad
CS^g>CS^{n,AB}
\quad\text{iff}\quad
\lambda<\overline{\lambda}_{CS}^g.
\]
At any interior root of \(F_{AB}=0\), direct substitution gives
\[
\begin{aligned}
\left.(\widehat{\Pi}^g-\widehat{\Pi}^{n,AB})\right|_{F_{AB}=0}
={}&-\frac{3\Theta Q(1-\lambda)^2x(x+\lambda)}
{8D^2(3x+4-\lambda)^2}\Bigl[\lambda(16+8\lambda-3\lambda^2)\\
&\qquad+x(12+31\lambda-\lambda^2)+x^2(15+6\lambda)\Bigr]<0.
\end{aligned}
\]
The bracketed expression is positive, so every interior consumer-surplus
threshold lies strictly below \(\underline{\lambda}_P^g\). If its positive
region is empty, \(\overline{\lambda}_{CS}^g=0\); if it covers the entire
interval, \(\delta=1\) and both thresholds equal one. Hence
\(\overline{\lambda}_{CS}^g\leq\underline{\lambda}_P^g\), so the win--win
case is empty. When \(\lambda>\underline{\lambda}_P^g\), platform profit
increases and consumer surplus decreases.  When
\(\lambda<\overline{\lambda}_{CS}^g\), consumer surplus increases and
platform profit decreases.  Finally, when
\(\overline{\lambda}_{CS}^g\leq\lambda\leq
\underline{\lambda}_P^g\), both weakly decrease. This establishes the
stated joint outcomes under \(S_P=\{A,B\}\). \(\Halmos\)\vspace{3mm}

\noindent\textit{Proof of Proposition \ref{prop: platform design}:}
We compare the platform's profit across assortments and AI modes on the common equilibrium-existence domain stated in the proposition, maintaining the notation introduced in Lemma~\ref{lem: assis_equil}.

\noindent\textbf{No-AI benchmark.}
The two relevant normalized platform profit differences are
\[
\begin{aligned}
\widehat{\Pi}^{n,AC}-\widehat{\Pi}^{n,AB}
&=\frac{\delta\Theta Q\lambda(x+1)^2
[9\lambda x+20\lambda+45x^2+72x+16]}
{(3x+4)^2(3x+4-\lambda)^2}>0,\\
\widehat{\Pi}^{n,AC}-\widehat{\Pi}^{n,BC}
&=\frac{\delta\Theta Q(1-\lambda)H_0(x,\lambda)}
{(3x+4)^2(3x+4\lambda)^2}>0.
\end{aligned}
\]
Here,
\[
\begin{aligned}
H_0(x,\lambda)={}&36\lambda^2x^2+96\lambda^2x+64\lambda^2
+81\lambda x^3+172\lambda x^2+96\lambda x\\
&+45x^4+81x^3+36x^2>0.
\end{aligned}
\]
Thus the unique optimal no-AI assortment is \(S_P=\{A,C\}\).

\noindent\textbf{AI agent.}
The corresponding comparisons are
\[
\begin{aligned}
\widehat{\Pi}^{g,AB}-\widehat{\Pi}^{g,AC}
&=\frac{\Theta Q\lambda(1-\lambda)(x+\lambda)
(\lambda x+4\lambda+3x)}{4D^2}>0,\\
\widehat{\Pi}^{g,AB}-\widehat{\Pi}^{g,BC}
&=\frac{\Theta Q(1-\lambda)H_g(x,\lambda)}{4D^2}>0.
\end{aligned}
\]
Here,
\[
H_g(x,\lambda)
=\lambda^3x+2\lambda^2x^2+7\lambda^2x+16\lambda^2
+5\lambda x^2+24\lambda x+9x^2>0.
\]
Therefore the optimal agent assortment is \(S_P=\{A,B\}\).

\noindent\textbf{Type-1 assistant equilibrium.}
For \(x>0\), \(1/2\leq\delta\leq1\), and \(0<\lambda<1\),
\[
\widehat{\Pi}^{s1,AB}-\widehat{\Pi}^{s1,AC}=\widehat{\pi}_B^{s1}-\widehat{\pi}_C^{s1}>0,
\qquad
\widehat{\Pi}^{s1,AB}-\widehat{\Pi}^{s1,BC}=\widehat{\pi}_A^{s1}-\widehat{\pi}_C^{s1}>0.
\]
Thus \(S_P=\{A,B\}\) is the assistant's best assortment.  The Type-1
assistant outcome is the agent outcome scaled by \(\delta\), so
\[
\widehat{\Pi}^{s1,AB}=\delta\widehat{\Pi}^{g,AB}\leq\widehat{\Pi}^{g,AB}.
\]
The inequality is strict when \(\delta<1\) and is an equality when
\(\delta=1\).

\noindent\textbf{Type-2a and Type-2b assistant equilibria.}
The normalized profit formulas give
\[
\begin{aligned}
\widehat{\Pi}^{s2a,AB}-\widehat{\Pi}^{s2a,BC}
&=\frac{4\Theta Q\delta(1-\lambda)(1+\delta x)
[4+4\delta x-\delta^2x+\delta^2\lambda x]}{E^2}>0,\\
\widehat{\Pi}^{s2b,AB}-\widehat{\Pi}^{s2b,BC}
&=\frac{\Theta Q\delta(1-\lambda)(1+\delta x)
[4+4\delta x-\delta^2x+\delta^2\lambda x]}{K^2}>0.
\end{aligned}
\]
Hence \(S_P=\{B,C\}\) cannot maximize platform profit under assistance in either equilibrium type;
only \(S_P=\{A,B\}\) and \(S_P=\{A,C\}\) remain to be compared with the
agent.

For \(j\in\{2a,2b\}\), recall that
\(s_B^j=\frac{p_C^{sj}q_B}{q_C}\) is the lower boundary of firm \(B\)'s
market-sharing deviation prices. The cross-configuration incentive
constraint therefore implies
\[
\widehat{\pi}_B^{sj}\geq\widehat{M}_{B,\times}^{j}
\geq\widehat{\pi}_B^1(s_B^j;p_A^{sj},p_C^{sj}).
\]
Direct substitution gives
\[
\begin{aligned}
\frac{\widehat{\pi}_B^1(s_B^{2a};p_A^{s2a},p_C^{s2a})-\widehat{\pi}_C^{s2a}}{Z}
&=\frac{4\delta(1-\lambda)
\left[\lambda(2-\delta+2\delta x+\delta^2x^2)
 +(1-\delta)x(2+\delta x)\right]}{E^2}>0,\\
\frac{\widehat{\pi}_B^1(s_B^{2b};p_A^{s2b},p_C^{s2b})-\widehat{\pi}_C^{s2b}}{Z}
&=\frac{\delta(1-\lambda)
\left[\lambda(2-\delta+2\delta x+\delta^2x^2)
 +(1-\delta)x(2+\delta x)\right]}{K^2}>0.
\end{aligned}
\]
Hence \(\widehat{\pi}_B^{sj}\geq\widehat{\pi}_B^1(s_B^j;p_A^{sj},p_C^{sj})>\widehat{\pi}_C^{sj}\), so \(\widehat{\Pi}^{sj,AB}>\widehat{\Pi}^{sj,AC}\).
Together with the comparison above, \(S_P=\{A,B\}\) is optimal for both Type 2a and Type 2b.

It remains to compare this assortment with the agent outcome. The defining
inequalities of the two local regions imply
\[
\mathcal L_{2a}\Longrightarrow
\widehat{\Pi}^{g,AB}\geq\widehat{\Pi}^{s2a,AB},
\qquad
\mathcal L_{2b}\Longrightarrow
\widehat{\Pi}^{g,AB}\geq\widehat{\Pi}^{s2b,AB}.
\]
Therefore every Type-2 assistant equilibrium yields weakly lower platform
profit than the agent outcome.

\noindent\textbf{Common boundary and conclusion.}
At \(\lambda=\underline{\lambda}_{2}^{\,s}\), the Type-2a and Type-2b price vectors, demands, firm
profits, and all three assortment profits coincide.  The equality is
assigned to Type 2b in Lemma~\ref{lem: assis_equil}, so the preceding
comparison includes the boundary once and leaves no gap.

Combining Types 1, 2a, and 2b, every assistant Nash equilibrium satisfies
\[
\widehat{\Pi}^{g,AB}\geq
\max_{S_P\in\{\{A,B\},\{A,C\},\{B,C\}\}}\widehat{\Pi}^{s,S_P}.
\]
These comparisons establish the preferred AI design on the common equilibrium-existence domain stated in the proposition.
On this domain, \(S_P=\{A,B\}\) with the agent yields weakly higher platform profit than the characterized assistant outcomes.
The comparison does not assign a continuation payoff to the assistant in \(\mathcal R_{\varnothing}\) and therefore does not identify the platform's preferred mode there.
The agent outcome remains characterized separately by Lemma~\ref{lem: agent_equil}.

For the cross-scenario comparison on that common domain,
\(\widehat{\Pi}^*=\widehat{\Pi}^{g,AB}\), \(CS^*=CS^g\),
\(\widehat{\Pi}^{n*}=\widehat{\Pi}^{n,AC}\), and \(CS^{n*}=CS^{n,AC}\).
Maintain \(x\) and \(D\) from Lemma~\ref{lem: assis_equil} and \(L\) from
Proposition~\ref{prop: agent_price and profits}. Comparing this selected AI outcome with
the no-AI benchmark \(S_P=\{A,C\}\) gives
\(\widehat{\Pi}^{g,AB}-\widehat{\Pi}^{n,AC}=\Theta QG_P(x,\delta,\lambda)\), where
\[
G_P(x,\delta,\lambda)
=\frac{(1-\lambda)[L^2+4\lambda(x+\lambda)^2]}{4D^2}
-\frac{\delta(x+1)(5x+4)}{(3x+4)^2}.
\]
To establish the threshold structure, direct differentiation gives
\[
\frac{\partial^2G_P}{\partial\lambda^2}
=-\frac{\mathcal K(x,\lambda)}{2D^4}<0,
\]
where
\[
\begin{aligned}
\mathcal K(x,\lambda)={}&
\lambda^5(x^2+16x+44)
+\lambda^4(35x^2+224x+112)
+\lambda^3(24x^3+378x^2+336x)\\
&
+\lambda^2(216x^3+334x^2)
+\lambda(27x^4+198x^3+44x^2)+81x^4+42x^3>0.
\end{aligned}
\]
Moreover,
\[
\left.\frac{\partial G_P}{\partial\lambda}\right|_{\lambda=0}
=\frac1{36}>0,
\qquad
\left.\frac{\partial G_P}{\partial\lambda}\right|_{\lambda=1}
=-\frac59<0,
\qquad
G_P(x,\delta,1)
=-\frac{\delta(x+1)(5x+4)}{(3x+4)^2}<0.
\]
Thus \(G_P\) is strictly concave in \(\lambda\) and has a unique interior
maximizer. Its positive set is therefore either empty, an interval beginning
at zero, or an interval bounded by two interior roots. Let
\(\underline{\lambda}_P^{*}\) and \(\overline{\lambda}_P^{*}\) denote the
lower and upper interior roots when they exist, allowing them to coincide at an interior double root.
Set \(\underline{\lambda}_P^{*}=0\) when only the upper interior root exists,
and set both thresholds to zero when \(G_P<0\) throughout \((0,1)\). Then
\[
\widehat{\Pi}^{g,AB}>\widehat{\Pi}^{n,AC}
\quad\text{iff}\quad
\lambda\in
(\underline{\lambda}_P^{*},\overline{\lambda}_P^{*}).
\]
Since
\[
\frac{\partial G_P}{\partial\delta}
=-\frac{(x+1)(5x+4)}{(3x+4)^2}<0,
\]
this platform-benefit interval weakly shrinks as \(\delta\) increases.

It remains to establish the unique quality-span threshold. At
\(\delta=1/2\),
\[
\frac{\partial G_P}{\partial x}
=-\frac{(13x+12)D^3
-\lambda^2(1-\lambda)^2(\lambda x+8\lambda+7x)(3x+4)^3}
{2D^3(3x+4)^3}.
\]
Expanding the numerator in powers of \(x\) and using \(0<\lambda<1\) gives
\[
\begin{aligned}
&(13x+12)D^3
-\lambda^2(1-\lambda)^2(\lambda x+8\lambda+7x)(3x+4)^3\\
&>4\lambda^3(64+17\lambda)
+\lambda^2(928+395\lambda^2)x
+1296\lambda x^2+324x^3+162x^4>0.
\end{aligned}
\]
Hence \(\partial G_P/\partial x<0\), so
\(G_P(x,1/2,\lambda)\) strictly decreases in \(x\) and, because
\(x=q_C/Q\), strictly increases in \(Q\).

At \(\delta=1/2\), let
\(M(Q)\equiv\max_{\lambda\in[0,1]}
G_P(q_C/Q,1/2,\lambda)\); a platform-profit gain is possible iff
\(M(Q)>0\). Continuity of \(G_P\) and compactness of \([0,1]\) imply that
\(M\) is continuous. Since its maximizer is interior and each
fixed-\(\lambda\) value strictly increases in \(Q\), \(M\) is strictly
increasing in \(Q\). The first-order condition for its unique maximizer
\(\lambda^*(x)\) gives
\[
\lambda^*(x)\longrightarrow-3+\frac{2\sqrt{21}}{3}
\quad\text{as }x\to\infty,\qquad
\frac{\lambda^*(x)}{x}\longrightarrow\frac{\sqrt{17}-3}{8}
\quad\text{as }x\downarrow0.
\]
Since \(x=q_C/Q\), substitution into \(G_P\) yields
\[
\lim_{Q\downarrow0}M(Q)=-\frac{29}{18}+\frac{28\sqrt{21}}{81}<0,\qquad
\lim_{Q\to\infty}M(Q)=\frac18>0.
\]
These limits imply that there is a unique
\(\underline Q_P^*>0\) satisfying \(M(\underline Q_P^*)=0\). At this point,
strict concavity implies that the zero set consists of a unique interior
double root. Above this threshold, at \(\delta=1/2\), the positive set
is a nonempty interval with an upper interior root; its lower endpoint
is either zero or a second interior root, under the convention above. Consequently,
\[
\begin{aligned}
Q\leq\underline Q_P^*&\ \Longrightarrow\ I_P^*=\varnothing
&&\text{for every admissible }\delta,\\
Q>\underline Q_P^*&\ \Longrightarrow\ I_P^*\neq\varnothing
&&\text{at }\delta=1/2.
\end{aligned}
\]

For consumers, Proposition~\ref{prop: agent_CS} gives
\(CS^{g}>CS^{n,AC}\) iff
\(\lambda>\underline{\lambda}_{CS}^{g}\).  Therefore, the joint outcome
is determined by the platform-benefit interval
\(I_P^*\equiv(\underline{\lambda}_P^*,\overline{\lambda}_P^*)\) and the
consumer-surplus threshold:
\[
\begin{array}{r@{\qquad}c@{\qquad}l}
Q>\underline Q_P^*,\ \lambda\in I_P^*,\
\lambda>\underline{\lambda}_{CS}^{g}
&\text{iff}&
\widehat{\Pi}^*>\widehat{\Pi}^{n*},\quad CS^*>CS^{n*},\\
Q>\underline Q_P^*,\ \lambda\in I_P^*,\
\lambda\leq\underline{\lambda}_{CS}^{g}
&\text{iff}&
\widehat{\Pi}^*>\widehat{\Pi}^{n*},\quad CS^*\leq CS^{n*},\\
\bigl(Q\leq\underline Q_P^*\ \text{or}\ \lambda\notin I_P^*\bigr),\
\lambda>\underline{\lambda}_{CS}^{g}
&\text{iff}&
\widehat{\Pi}^*\leq\widehat{\Pi}^{n*},\quad CS^*>CS^{n*},\\
\bigl(Q\leq\underline Q_P^*\ \text{or}\ \lambda\notin I_P^*\bigr),\
\lambda\leq\underline{\lambda}_{CS}^{g}
&\text{iff}&
\widehat{\Pi}^*\leq\widehat{\Pi}^{n*},\quad CS^*\leq CS^{n*}.
\end{array}
\]

All four strict cases are nonempty.  The following exact rational witnesses
list \((x,\delta,\lambda)\) and the corresponding normalized differences:
\[
\renewcommand{\arraystretch}{1.35}
\begin{array}{c|c|c}
(x,\delta,\lambda)&
(\widehat{\Pi}^*-\widehat{\Pi}^{n*})/(\Theta Q)&
(CS^*-CS^{n*})/(\Theta Q)\\ \hline
(\tfrac1{20},\tfrac12,\tfrac12)&
 \tfrac{17570753}{318326912}& \tfrac{33415}{1917632}\\
(\tfrac1{20},\tfrac12,\tfrac1{20})&
 \tfrac{1264609691}{10648189520}&-\tfrac{28813589}{256582880}\\
(\tfrac1{20},\tfrac12,\tfrac34)&
-\tfrac{26424821}{1574329392}& \tfrac{4537085}{37935648}\\
(\tfrac9{20},\tfrac45,\tfrac1{20})&
-\tfrac{1088921629}{350944823120}&
-\tfrac{666445863}{701889646240}
\end{array}
\]
Hence, on the common equilibrium-existence domain, the platform's selected AI design may benefit both the platform and consumers, hurt both parties, or benefit one party while hurting the other. \(\Halmos\)\vspace{3mm}

\end{APPENDICES}

\end{document}